%% file: so5.tex
\documentclass[aps,rmp,floatfix]{revtex4}

\include{macros}

\begin{document}
\title{SO(5) Theory of Antiferromagnetism and Superconductivity}
\author{Eugene Demler}
\address{Department of Physics, Harvard University, Cambridge MA 02138}
\author{Werner Hanke}
\address{Institute for Theoretical Physics, University of W\"urzburg Am Hubland, D-97074 W\"urzburg, Germany}
\author{Shou-Cheng Zhang}
\address{Department of Physics, Stanford University, Stanford, CA 94305}

\begin{abstract}
Antiferromagnetism and superconductivity are both
fundamental and common states of matter. In many strongly
correlated systems, including the high $T_c$ cuprates, the heavy
fermion compounds and the organic superconductors, they occur next
to each other in the phase diagram and influence each other's
physical properties. The $SO(5)$ theory unifies these two basic
states of matter by a symmetry principle and describes their rich
phenomenology through a single low energy effective model. In this
paper, we review the framework of the $SO(5)$ theory, and its
detailed comparison with numerical and experimental results.
\end{abstract}

\maketitle

\tableofcontents

\section{INTRODUCTION}
\label{sec:introduction} The phenomenon of superconductivity (SC)
is one of the most profound manifestations of quantum mechanics in
the macroscopic world. The celebrated Bardeen-Cooper-Schrieffer
(BCS) theory\cite{BARDEEN1957} of superconductivity provides a
basic theoretical framework to understand this remarkable
phenomenon in terms of the pairing of electrons with opposite spin
and momenta to form a collective condensate state. Not only does
this theory quantitatively explain the experimental data of
conventional superconductors, the basic concepts developed from
this theory, including the concept of spontaneously broken symmetry,
the Nambu-Goldstone modes and the Anderson-Higgs mechanism, provide
the essential building blocks for the unified theory of
fundamental forces. The discovery of high temperature
superconductivity (HTSC)\cite{BEDNORZ1986,WU1987} in the copper
oxide material poses a profound challenge to the theoretical
understanding of the phenomenon of superconductivity in the extreme
limit of strong correlations. While the basic idea of electron
pairing in the BCS theory carries over to HTSC, other aspects,
like the weak coupling mean field approximation and the phonon
mediated pairing mechanism, may not apply without modifications.
Therefore, the HTSC systems provide an exciting opportunity to develop
new theoretical frameworks and concepts for strongly correlated
electronic systems.

Since the discovery of HTSC, a tremendous amount of experimental
data has been accumulated on this material. In this theoretical
review it is not possible to give a detailed review of all the
experimental findings. Instead, we refer the readers to a number
of excellent review articles on the
subject\cite{ORENSTEIN2000,KASTNER1998,IMADA1998,MAPLE1998,TIMUSK1999,CAMPUZANO2002,YEH2002,DAMASCELLI2003}.
Below, we summarize the phase diagram of the HTSC cuprates and
discuss some of the basic and (more or less) universal properties
in each phase.

\begin{figure}[hp]
\begin{center}
        \includegraphics[scale=0.5]{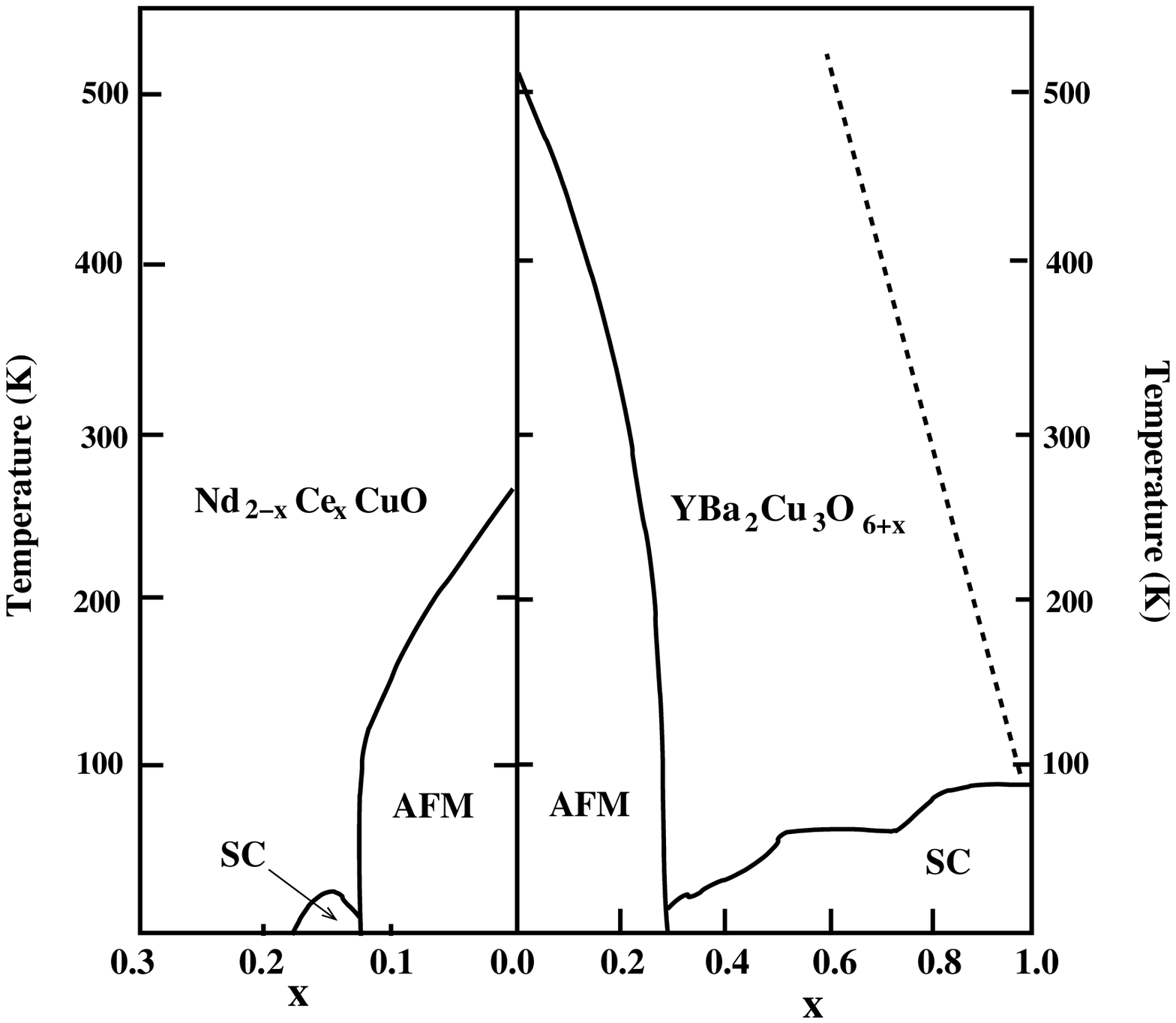}
        \caption{Phase diagram of the NCCO and the YBCO superconductors.}
\label{HTSC_Phase_Diagram}
\end{center}
\end{figure}

To date, a number of different HTSC materials have been
discovered. The most studied of these include the hole doped
$La_{2-x}Sr_xCuO_{4+\delta}$ (LSCO), $YBa_2Cu_3O_{6+\delta}$
(YBCO), $Bi_2Sr_2CaCu_2O_{8+\delta}$ (BSCO), and
$Tl_2Ba_2CuO_{6+\delta}$ (TBCO) materials and the electron doped
$Nd_{2-x}Ce_xCuO_4$ (NCCO) material. All these materials have two
dimensional (2D) $CuO_2$ planes and display an antiferromagnetic
(AF) insulating phase at half-filling. The magnetic properties of
this insulating phase are well approximated by the AF Heisenberg
model with spin $S=1/2$ and an AF exchange constant $J\sim 100
meV$. The Neel temperature for the three dimensional AF ordering
is approximately given by $T_N\sim 300 - 500 K$. The HTSC
material can be doped either by holes or by electrons. In the
doping range of $5\% \lesssim x \lesssim 15\%$, there is an SC
phase, which has a dome-like shape in the temperature versus doping
plane. The maximal SC transition temperature, $T_c$, is of the order
$100K$. The three doping regimes are divided by the maximum of the
dome and are called the underdoped, optimally doped, and
overdoped regimes, respectively. The generic phase diagram of HTSC
is shown in Fig.\ \ref{HTSC_Phase_Diagram}.

One of the main questions concerning the HTSC phase diagram is the
transition region between the AF and SC phases. Partly because
of the complicated material chemistry in this regime, there is no
universal agreement among different experiments. Different
experiments indicate several different possibilities, including
phase separation with an inhomogeneous density 
distribution \cite{HOWALD2001,LANG2002},
uniform mixed phase between AF and
SC\cite{BREWER1988,Miller2003}, and periodically ordered spin and
charge distributions in the form of stripes or
checkerboards\cite{TRANQUADA1995}.


The phase diagram of the HTSC cuprates also contains a regime with
anomalous behavior conventionally called the pseudogap phase.
This region of the phase diagram is indicated by the dashed line
in Fig.\ \ref{HTSC_Phase_Diagram}. In conventional
superconductors, a pairing gap opens up at $T_c$. In a large class
of HTSC cuprates, however, a gap, which can be observed in a variety 
of spectroscopic experiments, starts to open up at
a temperature $T^*$, much higher than $T_c$. Many experiments
indicate that the pseudogap ``phase" is not a true thermodynamical
phase but rather a precursor toward a crossover behavior. The
phenomenology of the pseudogap behavior is extensively reviewed
in\cite{TIMUSK1999,TALLON2001}.

The SC phase of the HTSC has a number of striking properties not
shared by conventional superconductors. First of all, phase
sensitive experiments indicate that the SC phase of most cuprates
has $d$-wave like pairing
symmetry\cite{VANHARLINGEN1995,TSUEI2000}. This is also supported
by the photoemission experiments, which show the existence of the
nodal points in the quasiparticle gap
\cite{CAMPUZANO2002,DAMASCELLI2003}. Neutron scattering
experiments find a new type of collective mode, carrying spin one,
lattice momentum close to $(\pi,\pi)$ and a resolution limited
sharp resonance energy around $20\sim 40 meV$. Most remarkably,
this resonance mode appears only below $T_c$ in the optimally doped
cuprates. It has been found in a number of
materials, including the YBCO, BSCO and the TBCO classes of
materials\cite{ROSSATMIGNOD1991A,MOOK1993,FONG1995,DAI1996,FONG1996,MOOK1998,DAI1998,FONG2000,FONG1999A,HE2001,HE2002}. 
Another property uniquely different from the conventional
superconductors is the vortex state. Most HTSC are type II
superconductors in which the magnetic field can penetrate into the SC
state in the form of a vortex lattice, with the SC order being
destroyed at the center of the vortex core. In conventional
superconductors, the vortex core is filled by normal metallic
electrons. However, a number of different experimental probes,
including neutron scattering, muon spin resonance ($\mu$sR), and
nuclear magnetic resonance (NMR) have shown that the vortex cores
in the HTSC cuprates are antiferromagnetic, rather than normal
metallic\cite{LEVI2002,KATANO2000,LAKE2001,LAKE2002,MILLER2002,MITROVIC2001,KHAYKOVICH2002,Mitrovic2002,Kakuyanagi2003,Kang2003,Fujita2003}.
This phenomenon has been observed in almost all HTSC materials,
including LSCO, YBCO, TBCO and NSCO; thus, it appears to be a
universal property of the HTSC cuprates.

The HTSC materials also have highly unusual transport properties.
While conventional metals have a $T^2$ dependence of resistivity,
in accordance with the predictions of the Fermi liquid theory, the
HTSC materials display a linear $T$ dependence of the resistivity
near optimal doping. This linear $T$ dependence extends over a
wide temperature window and seems to be universal among most of
the cuprates. When the underdoped or sometimes optimally doped SC
state is destroyed by applying a high magnetic field, the
resulting ``normal state" is not a conventional conducting
state\cite{ANDO1995,ANDO1996,BOEBINGER1996,HILL2001} but exhibits
insulating like behavior, at least along the $c$ axis, i.e. the axis perpendicular to the $CuO_2$ planes. This
phenomenon may be related to the insulating AF vortices mentioned
in the previous paragraph.

The HTSC materials attracted great attention because of the high
SC transition temperature. However, many of the striking
properties discussed above are also shared by other materials,
which have a similar phase diagram but typically with much
reduced temperature and energy scales. The 2D organic
superconductor $\kappa-(BEDT-TTF)_2X$ (X=anion) display a similar
phase diagram in the temperature versus pressure plane, where a
direct first order transition between the AF and  SC phases can
be tuned by pressure or magnetic
field\cite{LEFEBVRE2000,TANIGUCHI2003,SINGLETON2002}. In this
system, the AF transition temperature is approximately $T_N\sim
30K$, while the SC transition temperature is $T_c\sim 14K$. In
heavy fermion compounds
$CeCu_{2}(Si_{1-x}Ge_{x})_{2}$\cite{KITAOKA2001}, $CePd_2Si_2$ and
$CeIn_3$\cite{MATHUR1998}, the SC phase also appears near the
boundary to the AF phase. In all these systems, even though the
underlying solid state chemistries are rather different, the
resulting phase diagrams are strikingly similar and robust. This
similarity suggests that the overall feature of all these phase
diagrams is controlled by a single energy scale. Different classes
of materials differ only by this overall energy scale.
Another interesting example of competing AF and SC
can be found in quasi-one-dimensional Bechgaard salts.
The most well studied material from this family,
$(TMTSF)_2PF_6$, is an AF insulator
at ambient pressure and becomes a triplet SC
above a certain critical pressure \cite{JEROME1980,VULETIC2002,LEE1997,LEE2000}.

The discovery of HTSC has greatly stimulated the theoretical
understanding of superconductivity in strongly correlated systems.
Since the theoretical literature is extensive, the readers are
referred to a number of excellent reviews and representative
articles\cite{Anderson1997,INUI1988,SCALAPINO1995,SCHRIEFFER1989,ABRIKOSOV2000,Chubukov2002,Laughlin2002,SACHDEV2002A,ZAANEN1999,CHAKRAVARTY2001,KIVELSON2002B,FRANZ2002,IOFFE2002,VARMA1999,SENTHIL2001,WEN1996,DAGOTTO1994,NORMAN2003,BALENTS1998,SHEN2002,ANDERSON2003,FU2004}.
The present review article focuses on a particular theory, which unifies
the AF and SC phases of the HTSC cuprates based on an
approximate $SO(5)$ symmetry\cite{ZHANG1997}. The $SO(5)$ theory
draws its inspiration from the successful application of symmetry
principles in theoretical physics. All fundamental laws of Nature
are statements about symmetry. Conservation of energy, momentum
and charge are direct consequences of global symmetries. The form
of fundamental interactions are dictated by local gauge
symmetries. Symmetry unifies apparently different physical
phenomena into a common framework. For example, electricity and
magnetism were discovered independently and viewed as completely
different phenomena before the 19th century. Maxwell's theory and
the underlying relativistic symmetry between space and time
unified the electric field, $\vec E$, and the magnetic field, $\vec
B$, into a common electromagnetic field tensor, $F_{\mu\nu}$. This
unification shows that electricity and magnetism share a common
microscopic origin and can be transformed into each other by
going to different inertial frames. As discussed in the
introduction, the two robust and universal ordered phases of the
HTSC are the AF and SC phases. The central question of HTSC
concerns the transition from one phase to the other as the doping
level is varied. 

The $SO(5)$ theory unifies the three dimensional
AF order parameter $(N_x,N_y,N_z)$ and the two dimensional SC
order parameter $(Re \Delta, Im \Delta)$ into a single, five
dimensional order parameter called the {\it superspin}, in a way
similar to the unification of electricity and magnetism in
Maxwell's theory:
\begin{eqnarray}
F_{\mu\nu}\!=\! \left( \begin{array}{cccc}
               0 &   &   &   \\
               E_x & 0  &   &   \\
               E_y & B_z  & 0  &   \\
               E_z & -B_y  & B_x   & 0
               \end{array} \right)
               \ \ \ \Leftrightarrow \ \ \
n_a\!=\! \left( \begin{array}{c}
               Re \Delta\\
               N_x\\
               N_y\\
               N_z\\
               Im \Delta\\
               \end{array} \right).
\label{Maxwell}
\end{eqnarray}
This unification relies on the postulate that a common microscopic
interaction is responsible for both AF and SC in the HTSC cuprates
and related materials. A well-defined $SO(5)$ transformation
rotates one form of the order into another. Within this framework,
the mysterious transition from the AF to  SC phase as a
function of doping is explained in terms of a rotation in the five
dimensional order parameters space. Symmetry principles are not
only fundamental and beautiful, but they are also practically useful
in extracting information from a strongly interacting system,
which can be tested quantitatively. As seen in the examples 
applying the isospin $SU(2)$ and the $SU(3)$ symmetries to the
strong interaction, some quantitative predictions can be made and
tested even when the symmetry is broken. The
approximate $SO(5)$ symmetry between the AF and SC phases has
many direct consequences, which can be
tested both numerically and experimentally. We shall discuss a
number of these tests in this review article.

Historically, the $SO(5)$ theory concentrated on the
competition between AF and SC orders in the
high Tc cuprates.
The idea of some order competing with superconductivity
is common in several theories.
The staggered flux or the $d$-density
wave phase has been suggested in Refs.
\cite{AFFLECK1988,WEN1996,CHAKRAVARTY2001},
the spin-Peierls order has been discussed in
Refs. \cite{VOJTA1999,PARK2001}, and spin and charge density
wave orders have been considered in Refs.
\cite{ZAANEN1999A,KIVELSON2001,ZHANG2002}.
The $SO(5)$ theory extends simple consideration
of the competition between AF and SC 
in the cuprates by unifying the two order parameters
using a larger symmetry and examining
consequences of such symmetry.

The microscopic interactions in the HTSC materials are highly
complex, and the resulting phenomenology is extremely rich. The
$SO(5)$ theory is motivated by a confluence of the
phenomenological top-down approach with the microscopic bottom-up
approach, as discussed below.

{\it The top-down approach:} Upon first glance at the phase
diagram of the HTSC cuprates, one is immediately impressed by its
striking simplicity; there are only three {\it universal} phases
in the phase diagram of {\it all} HTSC cuprates: the AF, the SC
and the metallic phases, all with {\it homogeneous} charge
distributions. However, closer inspection reveals a bewildering
complexity of other possible phases, which may not be universally
present in all HTSC cuprates, and which may have inhomogeneous
charge distributions. Because of this complexity, formulating a
universal theory of HTSC is a formidable challenge. The strategy of the
$SO(5)$ theory can be best explained with an analogy:  we see a
colorful world around us, but the entire rainbow of colors is
composed of only three primary colors. In the $SO(5)$ theory, the
superspin plays the role of the primary colors. {\it A central
macroscopic hypothesis of the $SO(5)$ theory is that the ground
state and the dynamics of collective excitations in various phases
of the HTSC cuprates can be described in terms of the spatial and
temporal variations of the superspin}. This is a highly
constraining and experimentally testable hypothesis, since it
excludes many possible phases. It does include a homogeneous state
in which AF and SC coexist microscopically. It includes states with
spin and charge density wave orders, such as stripe phases,
checkerboards and AF vortex cores, which can be obtained from
spatial modulations of the superspin. It also includes quantum
disordered ground states and Cooper pair density wave, which can
be obtained from the temporal modulation of the superspin. The
metallic Fermi liquid state on the overdoped side of the HTSC
phase diagram seems to share the same symmetry as the high
temperature phase of the underdoped cuprates. Therefore, they can
also be identified with the disordered state of the superspin,
although extra care must be given to treat the gapless fermionic
excitations in that case. If this hypothesis is experimentally
proven to be correct, a great simplicity emerges from the
complexity: a full dynamical theory of the superspin field can be
the universal theory of the HTSC cuprates. Part of this review
article is devoted to describing and classifying phases which can
be obtained from this theory. This top-down approach focuses on
the low energy collective degrees of freedom and takes as the
starting point a theory expressed exclusively in terms of these collective
degrees of freedom. This is to be contrasted with the
conventional approach based on weak coupling Fermi liquid theory,
of which the BCS theory is a highly successful example. For an
extensive discussion on the relative merits of both approaches for
the HTSC problem, we refer the readers to an excellent, recent
review article in Ref. \cite{KIVELSON2002B}.

The $SO(5)$ theory is philosophically inspired by the
Landau-Ginzburg (LG) theory. The LG theory is a highly successful
phenomenological theory, in which one first makes observations of the
phase diagram, introduces one order parameter for each broken
symmetry phase and constructs a free energy functional by
expanding in terms of different order parameters 
(a review of earlier work based on this approach is given in Ref. 
\cite{VONSOVSKY1982}). However, given
the complexity of interactions and phases in the cuprates,
introducing one order parameter for each phase with unconstrained
parameters would greatly compromise the predictive power of
theory. 
The $SO(5)$ theory extends the LG theory
in several important directions. First, it postulates an
approximately symmetric interaction potential between the AF and
the SC phases in the underdoped regime of the cuprates, thereby
greatly constraining theoretical model building. Second, it
includes a full set of dynamic variables canonically conjugate to
the superspin order parameters, including the total spin, the
total charge and the so called $\pi$ operators. Therefore, unlike
the classical LG theory, which only contains the classical order
parameter fields without their dynamically conjugate variables,
the $SO(5)$ theory is capable of describing quantum disordered
phases and the quantum phase transitions between these phases.
Because the quantum disordered phases are described by the degrees
of freedom canonically conjugate to the classical order
parameters, a definite relationship, the so-called $SO(5)$
orthogonality relation, exists between them, which can give highly
constrained theoretical predictions. Therefore, in this sense, the
$SO(5)$ theory makes great use of the LG theory but also goes
far beyond in making more constrained and more powerful
predictions which are subject to experimental falsifications.

{\it The bottom-up approach:} Soon after the discovery of the HTSC
cuprates, Anderson\cite{ANDERSON1987} introduced the repulsive
Hubbard model to describe the electronic degrees of freedom in the
$CuO_2$ plane. Its low energy limit, the $t-J$ model, is defined
by\cite{ZHANG1988A}
\begin{eqnarray}
H= -t \sum_{\langle x,x'\rangle} c^\dagger_\sigma(x) c_\sigma(x')
+J \sum_{\langle x,x'\rangle} \vec S(x)\cdot \vec S(x'),
\label{t-J}
\end{eqnarray}
where the $t$ term describes the hopping of an electron with spin
$\sigma$ from a site $x'$ to its nearest neighbor $x$, with double
occupancy removed, and the $J$ term describes the nearest neighbor
spin exchange interaction. The main merit of these models does
not lie in the microscopic accuracy and realism but rather in the
conceptual simplicity. However, despite their simplicity, these
models are still very hard to solve, and their phase diagram 
cannot be compared directly with experiments.

The $t-J$ model certainly contains AF at half-filling. While it is
still not fully settled whether it has $d$-wave SC ground state
with a high transition temperature \cite{PRYADKO2003}, it is
reasonably convincing that it has strong $d$-wave pairing
fluctuations\cite{SORELLA2002}. Therefore, it is plausible that a
small modification could give a robust SC ground state. {\it The
basic microscopic hypothesis of the $SO(5)$ theory is that AF and
SC states arise from the same interaction with a common energy
scale of $J$}. This common energy scale justifies the treatment of AF and
SC on equal footing and is also the origin of an approximate
$SO(5)$ symmetry between these two phases. By postulating an
approximate symmetry between the AF and SC phases, and by
systematically testing this hypothesis experimentally and
numerically, the question of the microscopic mechanism of HTSC can
also be resolved. In this context, early numerical diagonalizations showed that the low-lying states of the $t - J$ model fit into irreducible representations of the $SO(5)$ symmetry group \cite{EDER1998}. If the $SO(5)$ symmetry is valid, then HTSC
shares a common microscopic origin with the AF, which is a well
understood phenomenon.

The basic idea is to solve these models by two steps. The first
step is a renormalization group transformation, which maps these
microscopic models to an {\it effective} superspin model on a
plaquette, typically of the size of $2a\times 2a$ or larger. This
step determines the form and the parameters of the effective
models. The next step is to solve the effective model either
through accurate numerical calculations or tractable analytical
calculations.

There is a systematic method to carry out the first step. Using
the contractor-renormalization-group (CORE)\cite{MORNINGSTAR1996}
approach, Altman and Auerbach\cite{ALTMAN2002} derived the
projected $SO(5)$ model from the Hubbard and the $t-J$ model.
Within the approximations studied to date, a simple and consistent
picture emerges. There are only five low energy states on
a coarse grained lattice site, namely a spin singlet state and a
spin triplet state at half-filling and a $d$-wave hole pair state
with two holes. These states correspond exactly to the local and
dynamical superspin degrees of freedom hypothesized in the
top-down approach. The resulting effective $SO(5)$ superspin
model, valid near the underdoped regime, only contains bosonic degrees
of freedom. This model can be studied by quantum Monte-Carlo
simulations up to very large sizes, and the accurate determination of
the phase diagram is possible (in contrast to the Hubbard and
$t-J$ models) because of the absence of the minus sign problems.
Once the global phase diagrams are determined, fermionic
excitations in each phase can also be studied by approximate
analytic methods. Within this approach, the effective $SO(5)$
superspin model derived from the microscopic physics can give a realistic description of the phenomenology and phase diagram of the HTSC cuprates and account for many of their physical properties \cite{DORNEICH2002A,DORNEICH2002B}. This agreement can be further tested, refined
and possibly falsified. This approach can be best summarized by
the following block diagram:

\begin{center}
\begin{tabular}{|c|c|c|ccc|c|}\cline{1-1}\cline{3-3}\cline{7-7}
Hubbard and t-J & {\it coarse graining}  &  Quantum SO(5)  & & {\it analytical and numerical} & & Phase diagram  \\
model  & $\Longrightarrow$  & model   & & $\Longrightarrow$ & & Collective modes  \\
of the electron        & {\it transformation}  & of the superspin  & & {\it calculations} & & other experiments... \\
\cline{1-1}\cline{3-3}\cline{7-7}
\end{tabular}
\end{center}

While the practical execution of the first step already introduces
errors and uncertainties, we need to remember that the Hubbard and
the $t-J$ models are effective models themselves, and they contain
errors and uncertainties compared with the real materials. The
error involved in our coarse grain process is not inherently
larger than the uncertainties involved in deriving the Hubbard and
the $t-J$ models from more realistic models. Therefore, as long as
we study a reasonable range of the parameters in the second step
and compare them directly with experiments, we could determine
these parameters.

This review is intended as a self-contained introduction to a
particular theory of the HTSC cuprates and related materials and
is organized as follows. Section \ref{sec:spin-flop} describes
three toy models which introduce some important concepts used in
the rest of the article. Section \ref{sec:so5effective} introduces
the concept of the $SO(5)$ superspin and its symmetry
transformation, as well as effective dynamical models of the superspin. 
The global phase diagram of the $SO(5)$
model is discussed and solved numerically in section
\ref{sec:so5global}. Section \ref{sec:microscopic} introduces
exact $SO(5)$ symmetric microscopic models, the numerical tests of
the $SO(5)$ symmetry in the $t-J$ and Hubbard models, and the
Altman-Auerbach procedure of deriving the $SO(5)$ model from
microscopic models of the HTSC cuprates. Section \ref{sec:pi}
discusses the $\pi$ resonance model and the microscopic mechanism
of HTSC. Finally, in section \ref{sec:experiment}, we discuss
experimental predictions of the $SO(5)$ theory and comparisons
with the tests performed so far. The readers are assumed to have
general knowledge of quantum many body physics and are referred
to several excellent textbooks for pedagogical introductions to
the basic concepts and theoretical
tools\cite{Anderson1997,ABRIKOSOV1993,NOZIERES1966,SCHRIEFER1964,TINKHAM1995,Auerbach1994,Sachdev2000A}.

\section{THE SPIN-FLOP AND THE MOTT INSULATOR TO SUPERCONDUCTOR TRANSITION}
\label{sec:spin-flop} Before presenting the full $SO(5)$ theory,
let us first discuss a much simpler class of toy models, namely
the anisotropic Heisenberg model in a magnetic field, the
hard-core lattice boson model and the negative $U$ Hubbard model.
The low energy limits of this class of models are all equivalent
to each other and can be described by a universal quantum field
theory, the $SO(3)$ quantum non-linear sigma model.
Although these models are simple to solve, they exhibit some of
the key properties of the HTSC cuprates, including strong
correlation, competition of different orders, low superfluid
density near the insulating phase, maximum of $T_c$, and the
pseudogap behavior.

The spin $1/2$ anisotropic AF Heisenberg model on a square lattice is described by the
following Hamiltonian:
\begin{eqnarray}
H= \sum_{\langle x,x'\rangle} S^z(x) V(x,x') S^z(x') + J
\sum_{\langle x,x'\rangle} (S^x(x) S^x(x) + S^y(x) S^y(x)) - B
\sum_{x} S^z(x). \label{Heisenberg}
\end{eqnarray}
Here, $S^\alpha=\frac{1}{2}\tau^\alpha$ is the Heisenberg spin
operators and $\tau^\alpha$ is the Pauli matrix. $J$ describes the
nearest-neighbor exchange of the $xy$ components of the spin,
while $V(x,x')$ describes the $z$ component of the spin
interaction. We shall begin by considering only the nearest
neighbor (denoted by $\langle x,x'\rangle$) spin interaction $V$.
$B$ is an external magnetic field. At the point of $B=0$ and
$J=V$, this model enjoys an $SO(3)$ symmetry generated by the
total spin operators:
\begin{eqnarray}
S^\alpha = \sum_x S^\alpha(x) \ \ , \ \ [S^\alpha,
S^\beta]=i\epsilon_{\alpha\beta\gamma} S^\gamma \ \ , \ \
[H,S^\alpha]=0. \label{total_spin}.
\end{eqnarray}
The order parameter in this problem is the Neel operator, which transforms according
to the vector representation of the $SO(3)$ group
\begin{eqnarray}
N^\alpha = \sum_x (-)^x S^\alpha(x) \ \ , \ \ [S^\alpha,
N^\beta]=i\epsilon_{\alpha\beta\gamma} N^\gamma. \label{Neel_spin}
\end{eqnarray}
Here $(-)^x=1$ if $x$ is on an even site and $(-)^x=-1$ if $x$ is
on an odd site. The symmetry generators and the order parameters
are canonically conjugate degrees of freedom, and the second part
of Eq. (\ref{Neel_spin}) is similar to the Heisenberg commutation
relation $[x,p]=i\hbar$ between the canonically conjugate position
and momentum. Just like $p$ can be expressed as
$\frac{\hbar}{i}\partial/\partial x$, one can express
\begin{eqnarray}
S^\alpha=i\epsilon^{\alpha\beta\gamma} N^\beta
\frac{\partial}{\partial N^\gamma} \ \ , \ \ N^\alpha S^\alpha =
0, \label{SO3_orthogonal}
\end{eqnarray}
where the second part of the equation, called the $SO(3)$
orthogonality relation, follows directly from the first. Both the
symmetry algebra, the canonical conjugation and the orthogonality
constraint are fundamental concepts important to the understanding
of the dynamics and the phase diagram of the model.

\begin{figure}[hp]
\begin{center}
        \includegraphics[scale=1.0]{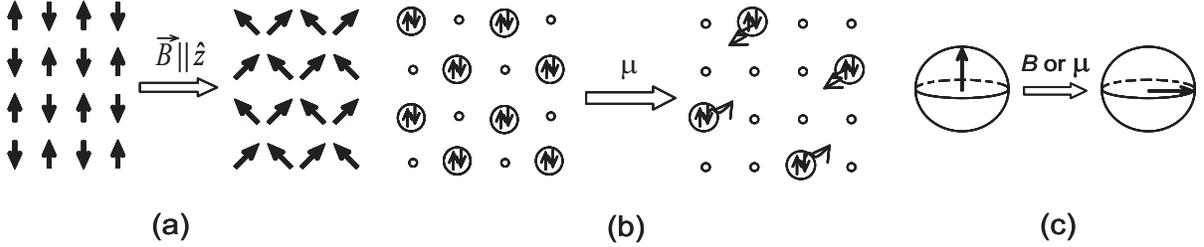}
        \caption{(a) The spin-flop transition of the XXZ Heisenberg model. (b) the
        Mott insulator to superfluid transition of the hard-core boson
        model or the $U<0$ Hubbard model. (c) Both can be described as the spin or
        the pseudospin flop transition in the $SO(3)$ non-linear $\sigma$ model, induced either
        by the magnetic field or the chemical potential.}
\label{Fig_spin_flop}
\end{center}
\end{figure}
\begin{figure}[hp]
\begin{center}
        \includegraphics[scale=0.8]{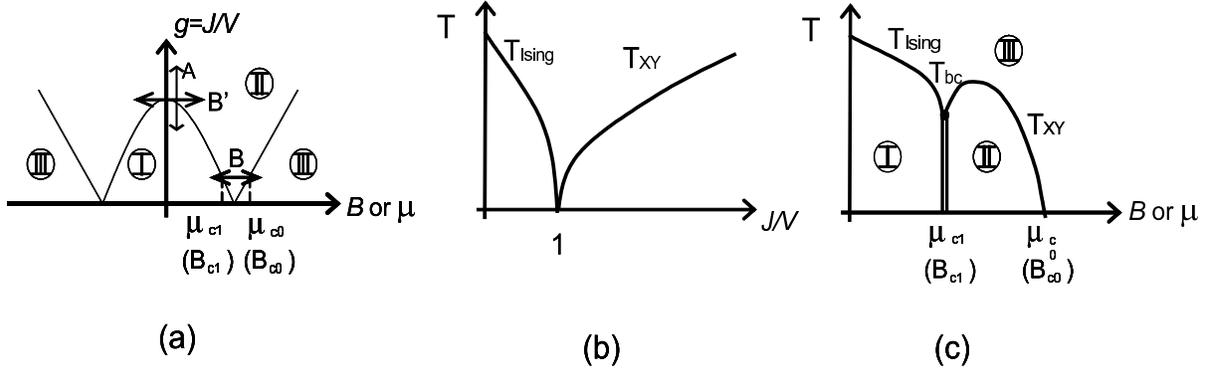}
        \caption{(a) Zero temperature phase diagram of the XXZ Heisenberg model,
        the hard-core boson model or the negative $U$ Hubbard model. Phase I
        is the Ising or the CDW phase, Phase II is the XY or the superfluid
        phase and phase III is the fully polarized or the normal
        phase. ``Class $A$" transition is induced by the
        anisotropy parameter $g=J/V$, while the ``Class $B$" transition is induced by the
        chemical potential or the magnetic field. (b) Finite
        temperature phase diagram for the ``class $A$" transition
        in $D=2$. Because of the $SO(3)$ symmetry at $J=V$ point,
        the transition temperature vanishes according to the
        Mermin-Wagner theorem. The dashed line denotes the mean
        field temperature. (c) Finite temperature phase diagram for the
        ``class $B$" transition in $D=3$. $T_{bc}$ denotes the $SO(3)$ symmetric
        bicritical point.}
\label{Fig_spin_flop_phase_diagram}
\end{center}
\end{figure}

Let us first consider the classical, mean field approximation to
the ground state of the anisotropic Heisenberg model defined in
Eq.(\ref{Heisenberg}). For $V>J$, the spins like to align
antiferromagnetically along the $z$ direction. In the Ising phase,
$S^z(x)=(-1)^x S$, the ground state energy per site is given by
$e_{Ising}(B) = -\frac{zV}{2} S^2$, where $z$ is the coordination
number, which is $4$ for the square lattice. Note that the
energy is independent of the $B$ field in the Ising phase. For
larger values of $B$, the spins ``flop" into the $XY$ plane, and
tilt uniformly toward the $Z$ axis. (See Fig.
\ref{Fig_spin_flop}a). Such a spin-flop state is given by
$S^z(x)=S cos\theta$ and $S^x(x)=(-1)^x S sin\theta$. The minimal
energy configuration is given by $cos\theta=B/zS(V+J)$, and the
energy per site for this spin-flop state is $e_{XY}(B) = -zJS^2/2
- B^2/2z(V+J)$. Comparing the energies of both states, we obtain
the critical value of $B$ where the spin-flop transition occurs:
$B_{c1}=zS\sqrt{V^2-J^2}$. On the other hand, we require
$|cos\theta|\leq 1$, which implies a critical field
$B_{c2}=zS(V+J)$ at which $|cos\theta|=1$, and the staggered order
parameter vanishes. Combining these phase transition lines, we
obtain the ``class $B$" transition in the ground state phase
diagram depicted in the $B$-$J/V$ plane (see Fig.
\ref{Fig_spin_flop_phase_diagram}a). Here and later in
the article, the ``class $B$" transition refers to the transition
induced by the chemical potential or the magnetic field. At the
$SO(3)$ symmetric point, $V=J$ and $B_{c1}=0$. For $V<J$, the
ground state has XY order even at $B=0$, and there is no spin-flop
transition as a function of the magnetic field $B$. The Ising to
XY transition can also be tuned by varying $J/V$ at $B=0$, and the
phase transition occurs at the special $SO(3)$ symmetric
Heisenberg point. This type of transition is also depicted in Fig.
\ref{Fig_spin_flop_phase_diagram}a and will be called ``class
$A$" transition in this paper.


The spin $1/2$ Heisenberg model can be mapped to a hard-core boson model, defined
by the following Hamiltonian:
\begin{eqnarray}
H= \sum_{\langle x,x'\rangle} n(x) V(x,x') n(x') - \frac{1}{2} J
\sum_{\langle x,x'\rangle} (b^\dagger(x) b(x') + h.c.) - \mu
\sum_{x} n(x). \label{hardcore_boson}
\end{eqnarray}
Here $b(x)$ and $b^\dagger(x)$ are the hard-core boson
annihilation and creation operators and $n(x)=b^\dagger(x) b(x)$
is the boson density operator. In this context, $V$, $J$ and $\mu$
describe the interaction, hopping and the chemical potential
energies, respectively. There are two states per site; $|1\rangle$
and $|0\rangle$ denote the filled and empty boson states,
respectively. They can be identified with the spin up
$|\uparrow\rangle$ and the spin down $|\downarrow\rangle$ states
of the Heisenberg model. The operators in the two theories can be
identified as follows:
\begin{eqnarray}
b(x)^\dagger = (-)^x (S^x(x) + i S^y(x)) \ \ \ b(x) = (-)^x (
S^x(x) - i S^y(x) ) \ \ \ n(x) = S^z(x) + \frac{1}{2}.
\label{mapping}
\end{eqnarray}
We see that these two models are identical to each other
when $\mu=B+zV/2$. From this mapping, we see that the spin-flop
phase diagram has another interpretation: the Ising phase is
equivalent to the Mott insulating phase of bosons with a
charge-density-wave (CDW) order in the ground state. The XY phase
is equivalent to the superfluid phase of the bosons. The two
paramagnetic states correspond to the full and empty states of the
bosons. While Heisenberg spins are intuitively associated with the
$SO(3)$ spin rotational symmetry, lattice boson models generically
have only a $U(1)$ symmetry generated by the total number
operator $N=\sum_x n(x)$, which transforms the boson operators by
a phase factor: $b^\dagger(x) \rightarrow e^{i\alpha}
b^\dagger(x)$ and $b(x) \rightarrow e^{-i\alpha} b(x)$. From this
point of view, it is rather interesting and non-trivial that the
boson model can also have an additional $SO(3)$ symmetry at the
special point $J=V$ because of its equivalence to the Heisenberg
model.

Having discussed the Heisenberg spin model and the lattice boson
models, let us now consider a fermion model, namely the negative
$U$ Hubbard model, defined by the Hamiltonian
\begin{eqnarray}
H= -t \sum_{\langle x,x'\rangle} (c^\dagger_\sigma(x) c_\sigma(x')
+ h.c.) + U \sum_x (n_\uparrow(x)-\frac{1}{2})
(n_\downarrow(x)-\frac{1}{2}) - \mu \sum_x n_\sigma(x),
\label{Hubbard}
\end{eqnarray}
where $c_\sigma(x)$ is the fermion operator and
$n_\sigma(x)=c^\dagger_\sigma(x) c_\sigma(x)$ is the electron
density operator at site $x$ with spin $\sigma$. $t$, $U$ and
$\mu$ are the hopping, interaction and the chemical potential
parameters respectively. The Hubbard model has a pseudospin
$SU(2)$ symmetry generated by the operators
\begin{eqnarray}
\eta^{-} = \sum_x (-)^x c_\uparrow(x) c_\downarrow(x) \ \ , \ \
\eta^{+} = (\eta^{-})^\dagger \ \ , \ \ \eta^{z} = \frac{1}{2}
\sum_\sigma (n_\sigma(x) -\frac{1}{2}) \ \ , \ \ [\eta^\alpha,
\eta^\beta]=i\epsilon_{\alpha\beta\gamma} \eta^\gamma. \label{eta}
\end{eqnarray}
where $\eta^{\pm}=\eta^x \pm i \eta^y$ and $\alpha=x,y,z$, as
before. Yang and Zhang\cite{YANG1989,YANG1990,ZHANG1990} pointed
out that these operators commute with the Hubbard Hamiltonian when
$\mu=0$ ( {\it i.e.} $[H,\eta^\alpha]=0$); therefore, they form the
symmetry generators of the model. Combined with the standard
$SU(2)$ spin rotational symmetry, the Hubbard model enjoys a
$SO(4)=SU(2)\otimes SU(2)/Z_2$ symmetry. This symmetry has
important consequences in the phase diagram and the collective
modes in the system.  In particular, it implies that the SC and
CDW orders are degenerate at half-filling. The SC and the CDW
order parameters are defined by
\begin{eqnarray}
\Delta^{-} = \sum_i c_{i\uparrow} c_{i\downarrow}  \ \ , \ \
\Delta^{+} = (\Delta^{-})^\dagger  \ \ , \ \ \Delta^{z} =
\frac{1}{2} \sum_{i\sigma} (-1)^i n_{i\sigma}   \ \ , \ \
[\eta^\alpha, \Delta^\beta]=i\epsilon_{\alpha\beta\gamma}
\Delta^\gamma, \label{Delta}
\end{eqnarray}
where $\Delta^{\pm}=\Delta^x \pm i \Delta^y$. 
The last equation above shows that the
$\eta$ operators perform the rotation between the SC and CDW
order parameters. Thus, $\eta^\alpha$ is
the pseudospin generator and $\Delta^\alpha$ is the
pseudospin order parameter. Just like the total spin and the Neel order parameter
in the AF Heisenberg model, they are canonically conjugate
variables. Since $[H,\eta^\alpha]=0$ at $\mu=0$, this exact
pseudospin symmetry implies the degeneracy of SC and CDW orders at
half-filling.

The phase diagram of the $U<0$ Hubbard model corresponds to a 1D
slice of the 2D phase diagram, as depicted in
Fig.\ref{Fig_spin_flop_phase_diagram}a. The exact pseudospin
symmetry implies that the ``class $B$" transition line for the
$U<0$ Hubbard model exactly touches the tip of the Mott lobe, as
shown by the $B'$ line in Fig.\ref{Fig_spin_flop_phase_diagram}a.
At $\mu=0$, SC and CDW are exactly degenerate, and they can be
freely rotated into each other. For $\mu\neq 0$, the system is
immediately rotated into the SC state. One can add additional
interactions in the Hubbard model, such as a nearest neighbor
repulsion, which breaks the $SU(2)$ pseudospin rotation symmetry
even at $\mu=0$. In this case, the pseudospin anisotropy either
picks the CDW Mott insulating phase or the SC phase at half-filling.
By adjusting the nearest neighbor interaction, one can move the
height of the ``class $B$" transition line.

We have seen that the hard-core boson model is equivalent
to the Heisenberg model because of the mapping (\ref{mapping}).
The $U<0$ model, on the other hand, is only equivalent to the
Heisenberg model in the low energy limit. In fact, it is 
equivalent to a $U>0$ Hubbard at half-filling in the presence
of a Zeeman magnetic field. The ground state of the half-filled
Hubbard model is always AF; therefore, its low energy limit is the
same as that of the Heisenberg model in a magnetic field. All
three models are constructed from very different microscopic
origins. However, they all share the same phase diagram, symmetry
group and low energy dynamics. In fact, these universal features
can all be captured by a single effective quantum field theory
model, namely the $SO(3)$ quantum non-linear $\sigma$ model. This
model can be derived as an effective model from the
microscopic models introduced earlier or it can be constructed
purely from symmetry principles and the associated operator
algebra as defined in Eq. (\ref{total_spin}) and
(\ref{Neel_spin}). The fact that both derivations yield the
same model is hardly surprising, since the
universal features of all these models are direct consequences of
the symmetry.

The $SO(3)$ non-linear $\sigma$ model is defined by the following Lagrangian density
for a unit vector field $n_\alpha$ with $n^2_\alpha =1$:
\begin{eqnarray}
{\cal L}= \frac{\chi}{2} \omega^2_{\alpha\beta} - \frac{\rho}{2}
(\partial_i n_\alpha)^2 - V(n) \ \ , \ \ \omega_{\alpha\beta} =
n_\alpha(\partial_t n_\beta -iB_{\beta\gamma} n_\gamma) - (\alpha
\leftrightarrow \beta), \label{so3L}
\end{eqnarray}
where the Zeeman magnetic field is given by $B_\alpha =
\frac{1}{2} \epsilon_{\alpha\beta\gamma} B_{\beta\gamma}$. Without
loss of generality, we pick the magnetic field $B$ to be along the
$z$ direction. $\chi$ and $\rho$ are the susceptibility and
stiffness parameters and $V(n)$ is the anisotropy potential, which
can be taken as $V(n)=-\frac{g}{2}n_z^2$. Exact $SO(3)$ symmetry
is obtained when $g=B=0$. $g>0$ corresponds to easy axis
anisotropy or $J/V<1$ in the Heisenberg model. $g<0$ corresponds
to easy plane anisotropy or $J/V>1$ in the phase diagram of
Fig.\ref{Fig_spin_flop_phase_diagram}. In the case of $g>0$, there
is a phase transition as a function of $B$. To see this, let us
expand  the first term in (\ref{so3L}) in the presence of the
$B$ field. The time independent part contributes to an effective
potential $V_{eff}=V(n) - \frac{B^2}{2}(n_x^2+n_y^2)$, from which
we see that there is a phase transition at $B_{c1} =
\sqrt{g/\chi}$. For $B<B_{c1}$, the system is in the Ising phase,
while for $B>B_{c1}$ the system is in the XY phase. Therefore,
tuning $B$ for a fixed $g>0$ traces out the ``class $B$"
transition line, as depicted in
Fig.\ref{Fig_spin_flop_phase_diagram}a. On the other hand, fixing
$B=0$ and varying $g$ traces out the ``class $A$" transition line,
as depicted in Fig.\ref{Fig_spin_flop_phase_diagram}a. Therefore,
we see that the $SO(3)$ non-linear $\sigma$ model has a similar
phase diagram as the microscopic models discussed earlier.
For a more detailed discussion of phase transitions
in $SO(3)$ non-linear $\sigma$ models we refer the readers to
an excellent review paper by Auerbach et al\cite{AUERBACH2000A}.

In $D=2$, both the XY and the Ising phase can have a finite
temperature phase transition into the disordered state. However,
because of the Mermin-Wagner theorem, a finite temperature phase
transition is forbidden  at the point $B=g=0$, where the system
has an enhanced $SO(3)$ symmetry. The finite temperature phase
diagram is shown in Fig. \ref{Fig_spin_flop_phase_diagram}b.
Approaching from the SC side, the Kosterlitz-Thouless transition
temperature $T_{XY}$ is driven to zero at the Mott to superfluid
transition point $J/V=1$. In the 2D XY model, the superfluid
density and the transition temperature $T_{XY}$ are related to
each other by a universal relationship\cite{NELSON1977};
therefore, the vanishing of $T_{XY}$ also implies the vanishing of
the superfluid density as one approaches the Mott to superfluid
transition. Scalettar et al\cite{SCALETTAR1989}, Moreo and
Scalapino\cite{MOREO1991} have performed extensive quantum Monte
Carlo simulation in the negative $U$ Hubbard model and have
indeed concluded that the superfluid density vanishes at the
symmetric point. The $SO(3)$ symmetric point leads to a large
regime below the mean field transition temperature where fluctuations dominate. The single particle spectral function of
the 2D attractive Hubbard model has been studied extensively by
Allen et al\cite{ALLEN1999} near half-filling. They identified the
pseudogap behavior in the single particle density of states within
this fluctuation regime. Based on this study, they argued that the
pseudogap behavior is not  only a consequence of the SC phase
fluctuations\cite{DONIACH1990,EMERY1995,UEMURA2002}
but also a consequence of the full $SO(3)$ symmetric fluctuations, which also
include the fluctuations between the SC and the CDW phases. Fig.
\ref{Fig_spin_flop_phase_diagram}c shows the generic finite
temperature phase diagram of these $SO(3)$ models. In this case,
the Ising and the XY transition temperatures meet at a single
bi-critical point $T_{bc}$, which has the enhanced $SO(3)$
symmetry. At the ``class $A$" transition point $g=B=0$, the quantum
dynamics is fully $SO(3)$ symmetric. On the other hand, at the
``class $B$" transition point $T=T_{bc}$, only the static
potential is $SO(3)$ symmetric. We shall return to a detailed
discussion of this distinction in section \ref{sec:so5bosons}.

The pseudospin $SU(2)$ symmetry of the negative $U$ Hubbard model
has another important consequence. Away from half-filling, the
$\eta$ operators no longer commute with the Hamiltonian, but they
are eigen-operators of the Hamiltonian, in the sense that
\begin{eqnarray}
[H,\eta^\pm]=\mp 2\mu\eta^\pm. \label{Heta}
\end{eqnarray}
Thus, the $\eta$ operators create well defined
collective modes in the system. Since they carry charge $\pm 2$,
they usually do not couple to any physical probes. However, in a
SC state, the SC order parameter mixes the $\eta$ operators with
the CDW operator $\Delta^z$, via Eq. (\ref{Delta}). From this
reasoning, Zhang\cite{ZHANG1990,ZHANG1991,DEMLER1996} predicted a
pseudo-Goldstone mode in the density response function at wave
vector $(\pi,\pi)$ and energy $-2\mu$, which appears only below
the SC transition temperature $T_c$. This prediction anticipated
the neutron resonance mode later discovered in the HTSC cuprates;
a detailed discussion shall be given in section \ref{sec:pi}.

From the toy models discussed in this section, we learned a few
very important concepts. Competition between different
orders can sometimes lead to enhanced symmetries at the multi-critical
point. Universal properties of very different microscopic models
can be described by a single quantum field theory constructed
from the canonically conjugate symmetry generators and order
parameters. The enhanced symmetry naturally leads to a small
superfluid density near the Mott transition. The pseudogap
behavior in the single particle spectrum can be attributed to the
enhanced symmetry near half-filling and new types of collective
Goldstone modes can be predicted from the symmetry argument. All
these behaviors are reminiscent of the experimental observations
in the HTSC cuprates. The simplicity of these models on the one
hand and the richness of the phenomenology on the other
inspired the $SO(5)$ theory, which we shall discuss in the
following sections.

\section{THE SO(5) GROUP AND EFFECTIVE THEORIES}
\label{sec:so5effective}
\subsection{Order parameters and SO(5) group properties}
\label{sec:so5group} The $SO(3)$ models discussed in the previous
section give a nice description of the quantum phase transition
from the Mott insulating phase with CDW order to the SC phase. 
However, these simple models do not have enough complexity
to describe the AF insulator at half-filling and the $d$-wave SC
order away from half-filling. Therefore, a natural step is to
generalize these models so that the Mott insulating phase with the
scalar CDW order parameter is replaced by a Mott insulating phase
with the vector AF order parameter. The pseudospin $SO(3)$
symmetry group considered previously arises from the combination of
one real scalar component of the CDW order parameter with one
complex or two real components of the SC order parameter. After
replacing the scalar CDW order parameter by the three components
of the AF order parameter and combining them with the two
components of the SC order parameters, we are naturally led to
consider a five component order parameter vector and the $SO(5)$
symmetry group which transforms it.

In section \ref{sec:spin-flop}, we introduced the crucial concept
of order parameter and symmetry generator. Both of these concepts
can be defined locally. In the case of the Heisenberg AF, at least
two sites, for instance, $\vec S_1$ and $\vec S_2$, are needed to define the
total spin $\vec S=\vec S_1+\vec S_2$ and the Neel vector $\vec
N=\vec S_1-\vec S_2$. Similarly, it is simplest to define the
concept of the $SO(5)$ symmetry generator and order parameter on
two sites with fermion operators $c_\sigma$ and $d_\sigma$,
respectively, where $\sigma=1,2$ is the usual spinor index. The AF
order parameter operator can be defined  naturally in terms of the
difference between the spins of the $c$ and $d$ fermions as
follows:
\begin{eqnarray}
N^\alpha = \frac{1}{2} (c^\dagger \tau^\alpha c - d^\dagger
\tau^\alpha d )\ \ ,\ \ n_2\equiv N_1  \ \ ,\ \ n_3\equiv N_2  \ \ ,\ \ n_4\equiv N_3.
\label{n234}
\end{eqnarray}
In view of the strong on-site repulsion in the cuprate problem,
the SC order parameter should be defined on a bond
connecting the $c$ and $d$ fermions. We introduce
\begin{eqnarray}
\Delta^\dagger = \frac{-i}{2} c^\dagger \tau^y d^\dagger
=\frac{1}{2}(-c_\uparrow^\dagger d_\downarrow^\dagger +
c_\downarrow^\dagger d_\uparrow^\dagger)\ \ , \ \ n_1  \equiv
\frac{\left(\Delta^{\dagger} + \Delta\right)}{2} \ \ , \ \ n_5
\equiv \frac{\left(\Delta^{\dagger} - \Delta\right)}{2i}.
\label{n15}
\end{eqnarray}
We can group these five components together to form a single
vector $n_a=(n_1, n_2, n_3, n_4, n_5)$, called the superspin since
it contains both {\it super}conducting and antiferromagnetic {\it
spin} components. The individual components of the superspin are
explicitly defined in the last parts of Eqs. (\ref{n234}) and
(\ref{n15}).

The concept of the superspin is useful only if there is a natural
symmetry group acting on it. In this case, since the order
parameter is five dimensional, it is natural to consider the most
general rotation in the five dimensional order parameter space
spanned by $n_a$. In three dimensions, three Euler angles are
needed to specify a general rotation. In higher dimensions, a
rotation is specified by selecting a plane and an angle of
rotation within this plane. Since there are $n(n-1)/2$ independent
planes in $n$ dimensions, the group $SO(n)$ is generated by
$n(n-1)/2$ elements, specified in general by antisymmetric
matrices $L_{ab}=-L_{ba}$, with $a=1,..,n$. In particular, the
$SO(5)$ group has ten generators. The total spin and the total
charge operators,
\begin{equation}
S_\alpha=\frac{1}{2} (c^\dagger \tau^\alpha c + d^\dagger
\tau^\alpha d ) \ \ , \ \ Q=\frac{1}{2}(c^\dagger c+d^\dagger
d-2),
\end{equation}
perform the rotation of the AF and SC order parameters within each
subspace. In addition, there are six so-called $\pi$ operators,
first introduced by Demler and Zhang\cite{DEMLER1995}, defined by
\begin{equation}
\pi^\dagger_\alpha = -\frac{1}{2}c^\dagger \tau^\alpha \tau^y
d^\dagger \ \ , \ \ \pi_\alpha = (\pi^\dagger_\alpha)^\dagger.
\label{pi}
\end{equation}
They perform the rotation from AF to SC and vice versa. These
infinitesimal rotations are defined by the commutation relations
\begin{equation}
[\pi^\dagger_\alpha, N_\beta] = i \delta_{\alpha\beta} \Delta^\dagger \ \ , \ \
[\pi^\dagger_\alpha, \Delta] = i N_\alpha.
\label{pi_n}
\end{equation}
The total spin components $S_\alpha$, the total charge $Q$, and the six $\pi$
operators form the ten generators of the $SO(5)$ group.

The superspin order parameters $n_a$, the associated $SO(5)$
generators $L_{ab}$, and their commutation relations can be
expressed compactly and elegantly in terms of the $SO(5)$ spinor
and the five Dirac $\Gamma$ matrices. The four component $SO(5)$
spinor is defined by
\begin{eqnarray}
\Psi_\mu = \left(
\begin{array}{c}
c_\sigma\\
d^\dagger_\sigma
\end{array}
\right).
\label{so5-spinor}
\end{eqnarray}
They satisfy the usual anti-commutation relations
\begin{eqnarray}
\{\Psi^\dagger_\mu,\Psi_\nu\}=\delta_{\mu\nu} \ \ , \ \
\{\Psi_\mu,\Psi_\nu\}= \{\Psi^\dagger_\mu,\Psi^\dagger_\nu\}=0.
\label{psi_commutation}
\end{eqnarray}
Using the $\Psi$ spinor and the five Dirac $\Gamma$ matrices (see
appendix A), we can express $n_a$ and $L_{ab}$ as
\begin{eqnarray}
n_a=\frac{1}{2}\Psi^\dagger_\mu \Gamma^a_{\mu\nu} \Psi_\nu \ \ , \
\ L_{ab}=-\frac{1}{2}\Psi^\dagger_\mu \Gamma^{ab}_{\mu\nu}.
\Psi_\nu \label{na_Lab}
\end{eqnarray}
The $L_{ab}$ operators forms the $SO(5)$ Lie algebra and satisfy
the commutation relation
\begin{eqnarray}
 \left[ L_{ab},  L_{cd} \right] =
 - i( \delta_{ac} L_{bd} + \delta_{bd} L_{ac}
   - \delta_{ad} L_{bc} -\delta_{bc} L_{ad}).
\label{LL}
\end{eqnarray}
The $n_a$ and the $\Psi_\mu$ operators form the vector and the
spinor representations of the $SO(5)$ group, satisfying the
equations
\begin{eqnarray}
 \left[ L_{ab},  n_c \right] =
  -i( \delta_{ac} n_b - \delta_{bc} n_a )
\label{Ln}
\end{eqnarray}
and
\begin{eqnarray}
\left[ L_{ab},  \Psi_\mu \right] =
  \frac{1}{2} \Gamma^{ab}_{\mu\nu} \Psi_\nu.
\label{LPsi}
\end{eqnarray}
If we arrange the ten operators $S_\alpha$, $Q$ and $\pi_\alpha$
into $L_{ab}$'s by the following matrix form:
\begin{eqnarray}
L_{ab} & = & \left(
\begin{array}{ccccc}
0 & \ & \ & \ & \ \\
\pi^\dagger_x\!+\!\pi_x & 0 & \ &\  & \ \\
\pi^\dagger_y\!+\!\pi_y & -S_z & 0 & \ & \ \\
\pi^\dagger_z\!+\!\pi_z & S_y & -S_x & 0 & \ \\
Q & \frac{1}{i}(\pi^\dagger_x\!-\!\pi_x )  &
\frac{1}{i}(\pi^\dagger_y\!-\!\pi_y
 ) &
    \frac{1}{i}(\pi^\dagger_z\!-\!\pi_z ) & 0
\end{array}
\right)
\label{Lab2}
\end{eqnarray}
and group $n_a$ as in Eqs. (\ref{n234}) and (\ref{n15}), we see
that Eqs. (\ref{LL}) and (\ref{Ln}) compactly reproduces all the
commutation relations presented previously. These equations show
that $L_{ab}$ and $n_a$ are the symmetry generators and the order
parameter vectors of the $SO(5)$ theory. The commutation relation
Eq. (\ref{Ln}) is the $SO(5)$ generalization of the $SO(3)$
communication relation as given in Eqs. (\ref{Neel_spin}) and
(\ref{Delta}).

In systems where the unit cell naturally contains two sites, 
such as the ladder and the bi-layer systems, the complete set
of operators $L_{ab}$, $n_a$ and $\Psi_\mu$ can be used to
construct model Hamiltonians with the exact $SO(5)$ symmetry, as
we will show in section \ref{sec:exact}. In these models, local
operators are coupled to each other so that only the total
symmetry generators, obtained as the sum of local symmetry
generators, commute with the Hamiltonian. For two dimensional
models containing only a single layer, grouping the lattice into
clusters of two sites would break lattice translational and
rotational symmetry. In this case, it is better to use a cluster
of four sites forming a square, which does not break rotational
symmetry and can lead naturally to the definition of a $d$-wave
pairing operator\cite{ZHANG1999,ALTMAN2002}. In this case, the
$L_{ab}$, $n_a$ and $\Psi_\mu$ operators are interpreted as the
effective low energy operators defined on a plaquette, which form
the basis for an effective low energy $SO(5)$ theory, rather than
the basis of a microscopic $SO(5)$ model.

Having introduced the concept of local symmetry generators and
order parameters based in real space, we will now discuss
definitions of these operators in momentum space. The AF and SC
order parameters can be naturally expressed in terms of the
microscopic fermion operators as
\begin{eqnarray}
N^\alpha  =  \sum_{p} c_{p+\Pi}^\dagger
\tau^\alpha c_p \ \ , \ \
\Delta^\dagger  =  \frac{-i}{2}\sum_{p} d(p)
c_{p}^\dagger \tau^y c_{-p}^\dagger \ \ ,\ \
d(p) \equiv cos(p_x)-cos(p_y),
\label{AFdSC}
\end{eqnarray}
where $\Pi\equiv (\pi,\pi)$ and $d(p)$ is the form factor for the $d$
wave pairing operator in two dimensions. They can be combined into
the five component superspin vector $n_a$ by using the same
convention as before.  The total spin and total charge operator
are defined microscopically as
\begin{eqnarray}
S_\alpha = \sum_{p} c_{p}^\dagger \tau^\alpha c_p\ \ ,\ \ Q =
\frac{1}{2} \sum_{p} (c_{p}^\dagger c_p -1), \label{SQ}
\end{eqnarray}
and the $\pi$ operators can be defined as
\begin{eqnarray}
\pi^\dagger_\alpha & = & \sum_{p} g(p) c_{p+\Pi}^\dagger
\tau^\alpha \tau^y c_{-p}^\dagger.
\label{pi_k}
\end{eqnarray}
Here the form factor $g(p)$ needs to be chosen appropriately to
satisfy the $SO(5)$ commutation relation (\ref{LL}). In the
original formulation of the $SO(5)$ theory, Zhang\cite{ZHANG1997}
chose $g(p)=d(p)$. In this case, the $SO(5)$ symmetry algebra
(\ref{LL}) only closes approximately near the
Fermi surface. Later, Henley\cite{HENLEY1998} proposed the choice
$g(p)=sgn(d(p))$(this construction requires introducing form
factors for the AF order parameter as well). 
When the momentum space operators $S_\alpha$,
$Q$ and $\pi^\dagger_\alpha$, as expressed in Eq. (\ref{SQ}) and
(\ref{pi_k}), are grouped into $L_{ab}$ according to Eq.
(\ref{Lab2}), the symmetry algebra (\ref{LL}) closes exactly.
However, the $\pi$-operators are no longer short
ranged.

The $SO(5)$ symmetry generators perform the most general rotation
among the five order parameters. The
quantum numbers of the $\pi$ operators exactly patch up the
difference in quantum numbers between the AF and SC order
parameters, as shown in the Table I.

\begin{table}[h]
\begin{center}
\begin{tabular}{|c|c|c|c|c|}    \hline
         &charge  & spin & momentum & internal angular momentum \\  \hline
$\Delta, \Delta^\dagger\ \textrm{or} \ n_1, n_5$ & $\pm 2$  & 0    & 0        & d wave  \\ \hline
$N^\alpha\ \textrm{or} \ n_{2,3,4}$ & 0  & 1    & $(\pi,\pi)$        & s wave   \\ \hline
$\pi_\alpha, \pi_\alpha^\dagger$ & $\pm 2$  & 1    & $(\pi,\pi)$        & d wave  \\ \hline
\end{tabular}
\caption{ Quantum numbers of the AF, the $d$-wave SC order
parameters, and the $\pi$ operator. Since the $\pi$ operator
rotates the AF and SC order parameters into each other, its
quantum numbers patch up the difference between the AF and SC
order parameters.}
\end{center}
\end{table}

With the proper choice of the internal form factors, the $\pi$
operators rotate between the AF and SC order parameters
according to (\ref{pi_n}). Analogous to the electro-magnetic
unification presented in the introduction, the $\pi$ operators
generate an infinitesimal rotation between the AF and SC order
parameters similar to the infinitesimal rotation between
the electric and the magnetic fields generated by the Lorentz
transformation. These commutation relations play a central role in
the $SO(5)$ theory and have profound implications on the
relationship between the AF and SC order -- they provide a
basis to unify these two different types of order in a single
framework. In the AF phase, the operator $N^\alpha$ acquires a
nonzero expectation value, and the $\pi$ and SC operators
become canonically conjugate variables in the sense of Hamiltonian
dynamics. Conversely, in the SC phase the operator $\Delta$
acquires a nonzero expectation value, and the $\pi$ and
AF operators become canonically conjugate variables. This canonical
relationship is the key to understanding the collective modes in
the $SO(5)$ theory and in HTSC.

The $SO(5)$ group is the minimal group to contain both AF and SC, the
two dominant phases in the HTSC cuprates. However, it is possible to
generalize this construction so that it includes other forms of
order. 
For example, in Ref. \cite{Podolsky2004}, it was  demonstrated
how one can combine AF and triplet SC using an 
$SO(4)$ symmetry \cite{ROZHKOV2002}. Such
a construction is useful for quasi-one-dimensional Bechgaard salts,
which undergo a transition from an AF insulating state to a triplet SC
state as a function of pressure\cite{JEROME1980,VULETIC2002,LEE1997,LEE2000}.  

To define an $SO(4)$ symmetry for a
one-dimensional electron system, we introduce
the total spin, total charge,
and $\Theta$ operators
\begin{eqnarray}
S_\alpha&=& \frac{1}{2} \sum_{k}
\left(
c_{+,k}^\dagger \tau^\alpha c_{+,k}
+ c_{-,k}^\dagger \tau^\alpha c_{-,k}
\right)
\nonumber\\
Q&=& \frac{1}{2} \sum_{k} \left( c_{+,k}^\dagger c_{+,k}
+ c_{-,k}^\dagger c_{-,k} - 2 \right) \nonumber\\
\Theta^\dagger&=&\frac{-i}{2}
\sum_k \left( c_{+,k}^\dagger \tau^y c_{+,-k}^\dagger -  
c_{-,k}^\dagger \tau^y c_{-,-k}^\dagger \right).
\label{quantumGenerators}
\end{eqnarray}
Here $c_{\pm,k}^\dagger$
creates right/left moving electrons of momentum $\pm k_f+k$. 
The spin operators $S_\alpha$ form an $SO(3)$ algebra of spin
rotations given by the second formula of equation (\ref{total_spin}).
We can also introduce isospin $SO(3)$ algebra by combining
the charge with the $\Theta$ operators
\begin{eqnarray}
I_x = \frac{1}{2}( \Theta^\dagger + \Theta),
\hspace{1cm}
I_y &=& \frac{1}{2i}( \Theta^\dagger - \Theta),
\hspace{1cm}
I_z = Q
\nonumber\\
\left[ I_a, I_b \right] &=& i \epsilon_{abc} I_c.
\end{eqnarray}
Spin and isospin operators together generate
an $SO(4) \approx SO(3) \times SO(3)$ symmetry,
which unifies triplet superconductivity
and antiferromagnetism. We define
the N\'eel vector and the TSC order parameter,
\begin{eqnarray}
N_\alpha &=& \frac{1}{2} \sum_{k} \left( c^\dagger_{+,k}
\tau^\alpha c_{-,k} + c^\dagger_{-,k} \tau^\alpha
c_{+,k} \right)
\label{AFOrderParameterII}\\
\Psi_\alpha &=& \frac{1}{i} \sum_{k} c_{+,k}
(\tau^y \tau^\alpha) c_{-,-k}\,\, , \nonumber
\end{eqnarray}
and combine them into a single tensor order parameter
\begin{eqnarray}
\hat{Q} &=& \left(\begin{array}{ccc}
\,\,({\rm Re} \vec{\Psi})_x\,\,\,\, ({\rm Im} \vec{\Psi})_x\,\,\,\, N_x \,\,\\
\,\,({\rm Re} \vec{\Psi})_y\,\,\,\, ({\rm Im} \vec{\Psi})_y\,\,\,\, N_y \,\,\\
\,\,({\rm Re} \vec{\Psi})_z\,\,\,\, ({\rm Im}
\vec{\Psi})_z\,\,\,\, N_z \,\,
\end{array}\right).
\label{QtensorII}
\end{eqnarray}
One can easily verify that $Q_{a\alpha}$ transforms as a vector under
both spin and isospin rotations
\begin{eqnarray}
\left[{S_\alpha,Q_{b\beta}}
\right] = i\epsilon_{\alpha\beta\gamma}Q_{b\gamma}
\hspace{1cm}
\left[
{I_a,Q_{b\beta}} \right]= i\epsilon_{abc}Q_{c\beta}.
\end{eqnarray}
One dimensional electron systems have been studied extensively using
bosonization and renormalization group analysis.  They have a line of
phase transitions between an antiferromagnetic and a triplet
superconducting phase at a special ratio of the forward and backward
scattering amplitudes.  Podolsky et al pointed out that anywhere on
this line the $\Theta$ operator commutes with the Hamiltonian of the
system.  Hence, one finds the $SO(4)$ symmetry at the AF/triplet SC
phase boundary without any fine tuning of the parameters. Consequences of
this symmetry for the Bechgaard salts are reviewed in
Ref. \cite{Podolsky2004}.

Other extensions and generalizations of $SO(5)$ are discussed in
Ref. \cite{MARKIEWICZ1998,NAYAK2000,WU2003,LIN1998,MURAKAMI1999,Schulz1998}.

\subsection{The SO(5) quantum nonlinear $\sigma$ model}
\label{sec:so5sigma} In the previous section, we presented the
concepts of local $SO(5)$ order parameters and symmetry
generators. These relationships are purely kinematic and do not
refer to any particular Hamiltonian. In section \ref{sec:exact},
we shall discuss microscopic models with exact $SO(5)$ symmetry,
constructed out of these operators. A large class of models,
however, may not have $SO(5)$ symmetry at the microscopic level,
but their long distance, low energy properties may be described in
terms of an effective $SO(5)$ model. In section
\ref{sec:spin-flop}, we saw that many different microscopic
models indeed have the $SO(3)$ non-linear $\sigma$ model as
their universal low energy description. Therefore, in order to
present a general theory of the AF and SC in the HTSC, we first
introduce the $SO(5)$ quantum non-linear $\sigma$ model.

The $SO(5)$ quantum non-linear $\sigma$ model describes the
kinetic and potential energies of coupled superspin degrees of
freedom. In the case of HTSC cuprates, the superspin degrees of
freedom are most conveniently defined on a coarse grained lattice,
with $2a\times 2a$ lattice spacing in units of the original
cuprate lattice spacing, where every super-site denotes a
(non-overlapping) plaquette of the original lattice (see Fig.
\ref{figALplaqlatt}). There are $4^4=256$ states on a plaquette in
the original Hubbard model, but we shall retain only the 6 lowest
energy states, including  a spin singlet state and three spin
triplet states at half-filling, and two paired states with two holes
or two particles away from half-filling (see Fig.
\ref{Fig_plaquette}). In sections \ref{sec:diagonal} and
\ref{sec:CORE}, we will show, with numerical calculations, that
these are indeed the lowest energy states in each charge sector.
Additionally, we will show explicitly that the local superspin degree of
freedom discussed in this section can be constructed from these
six low energy states. Proposing the $SO(5)$ quantum non-linear
$\sigma$ model as the low energy effective model of the HTSC
cuprates requires the following physical assumptions: 1) AF and SC
and their quantum disordered states are the only competing degrees
of freedom in the underdoped regime. 2) Fermionic degrees of
freedom are mostly gapped below the pseudogap temperature. For a
$d$-wave superconductor, there are also gapless fermion
degrees of freedom at the gap nodes. However, they do not play a
significant role in determining the phase diagram and collective
modes of the system. Our approach is to solve the bosonic part of
the model first, and then include gapless fermions
self-consistently at a later stage \cite{DEMLER1999,ALTMAN2002}.

From Eqs. (\ref{Ln}) and the discussions in sections
\ref{sec:so5group}, we see that $L_{ab}$ and $n_a$ are conjugate
degrees of freedom, very much similar to $[q,p]=i\hbar$ in quantum
mechanics. This suggests that we can construct a Hamiltonian from
these conjugate degrees of freedom. The Hamiltonian of the $SO(5)$
quantum non-linear $\sigma$ model takes the following form
\begin{eqnarray}
H = \frac{1}{2\chi}\sum_{x, a<b} L_{ab}^2(x) + \frac{\rho}{2}
\sum_{\langle x,x' \rangle, a} n_a(x) n_a(x') + \sum_{x, a<b}
B_{ab}(x) L_{ab}(x) + \sum_x V(n(x)), \label{Hsigma}
\end{eqnarray}
where the superspin $n_a$ vector field is subjected to the
constraint
\begin{eqnarray}
n_a^2=1. \label{n_constraint}
\end{eqnarray}
This Hamiltonian is quantized by the canonical commutation
relations (\ref{LL}) and (\ref{Ln}). Here, the first term is the
kinetic energy of the $SO(5)$ rotors, where $\chi$ has the
physical interpretation of the moment of inertia of the $SO(5)$
rotors. The second term describes the coupling of the $SO(5)$
rotors on different sites through the generalized stiffness
$\rho$. The third term introduces the coupling of external fields
to the symmetry generators, while $V(n)$ includes anisotropic
terms which break the $SO(5)$ symmetry to the conventional
$SO(3)\times U(1)$ symmetry. The $SO(5)$ quantum non-linear
$\sigma$ model is a natural combination of the $SO(3)$ non-linear
$\sigma$ model describing the AF Heisenberg model and the quantum
XY model describing the SC to insulator transition. If we restrict
the superspin to have only components $a=2,3,4$, 
then the first two terms describe the
symmetric Heisenberg model, the third term describes the coupling
to a uniform external magnetic field, while the last term can
represent easy plane or easy axis anisotropy of the Neel vector.
On the other hand, for $a=1,5$, the first term describes Coulomb
or capacitance energy, the second term is the Josephson coupling
energy, while the third term describes coupling to an external
chemical potential.

The first two terms of the $SO(5)$ model describe the competition
between the quantum disorder and classical order. In the ordered
state, the last two terms describe the competition between the
AF and SC order. Let us first consider the quantum
competition. The first term prefers sharp eigenstates of the
angular momentum. On an isolated site, $C\equiv \sum L_{ab}^2$ is
the Casimir operator of the $SO(5)$ group in the sense that it
commutes with all the $SO(5)$ generators. The eigenvalues of this
operator can be determined completely from group theory - they are
0, 4, 6 and 10, respectively, for the 1 dimensional $SO(5)$
singlet, 5 dimensional $SO(5)$ vector, 10 dimensional
antisymmetric tensor and 14 dimensional symmetric, traceless
tensors, respectively. Therefore, we see that this term always
prefers a quantum disordered $SO(5)$ singlet ground state, which
is also a total spin singlet. In the case where the effective
quantum non-linear $\sigma$ model is constructed by grouping the
sites into plaquettes, the quantum disordered ground state
corresponds to a plaquette ``RVB" state, as depicted in Figs.
\ref{Fig_plaquette}a and \ref{Fig_stripe_checker}a. This ground
state is separated from the first excited state, the five fold
$SO(5)$ vector state, by an energy gap of $2/\chi$. This gap will
be reduced when the different $SO(5)$ rotors are coupled to each
other by the second term. This term represents the effect of
stiffness, which prefers a fixed direction of the $n_a$ vector
to a fixed angular momentum. This competition is an
appropriate generalization of the competition between the number
sharp and phase sharp states in a superconductor and the
competition between the classical Neel state and the bond or
plaquette singlet state in the Heisenberg AF. The quantum phase
transition occurs near $\chi\rho \simeq 1$.

In the classically ordered state, the last two anisotropy terms
compete to select a ground state. To simplify the discussion, we
first consider the following simple form of the static
anisotropy potential:
\begin{eqnarray}
V(n) = -g (n_2^2+n_3^2+n_4^2).
\label{Vn}
\end{eqnarray}
At the particle-hole symmetric point with vanishing chemical
potential $B_{15}=\mu=0$, the AF ground state is selected by
$g>0$, while the SC ground state is selected by $g<0$. $g=0$ is
the quantum phase transition point separating the two ordered
phases. This phase transition belongs to ``class $A$" in the
classification scheme of section \ref{sec:spin-flop} and is
depicted as the ``$A1$" transition line in Fig. \ref{Fig_global}.
This point has the full quantum $SO(5)$ symmetry in the model
described above.

However, it is unlikely that the HTSC cuprates can be close to
this quantum phase transition point. In fact, we expect the
anisotropy term $g$ to be large and positive, making the AF phase
strongly favored over the SC phase at half-filling. However,
the chemical potential term has the opposite, competing effect
and favors SC. We can observe this by transforming 
the Hamiltonian into the Lagrangian density in the continuum limit:
\begin{eqnarray}
{\cal L} = \frac{\chi}{2}\omega_{ab}(x,t)^2 + \frac{\rho}{2}
(\partial_k n^a(x,t))^2 - V(n(x,t)), \label{La}
\end{eqnarray}
where
\begin{eqnarray}
\omega_{ab}=n_a (\partial_t n_b - i B_{bc} n_c) - (a\leftrightarrow b)
\label{omega}
\end{eqnarray}
is the angular velocity. We see that the chemical potential enters
the Lagrangian as a gauge coupling in the time direction.
Expanding the first term in the presence of the chemical potential
$\mu=B_{15}$, we obtain an effective potential
\begin{eqnarray}
V_{eff}(n) = V(n) - \frac{(2\mu)^2\chi}{2}(n_1^2+n_5^2),
\label{Veff}
\end{eqnarray}
from which we see that the bare $V$ term competes with the
chemical potential term. For $\mu<\mu_c=\sqrt{g/\chi}$, the AF
ground state is selected, while for $\mu>\mu_c$, the SC ground
state is realized. At the transition point -- even though each term
strongly breaks $SO(5)$ symmetry -- the combined term gives an
effective static potential which is $SO(5)$ symmetric, as we can
see from (\ref{Veff}). This quantum phase transition belongs to
``class $B$" in the classification scheme of section
\ref{sec:spin-flop}. A typical transition of this type is depicted
as the ``$B1$" transition line in Fig. \ref{Fig_global}. Even
though the static potential is $SO(5)$ symmetric, the full quantum
dynamics is not. This can be seen most easily  from the time
dependent term in the Lagrangian. When we expand out the square,
the term quadratic in $\mu$ enters the effective static potential
in Eq. (\ref{Veff}). However, there is also a $\mu$-dependent term
involving a first order time derivative. This term breaks the
particle hole symmetry and dominates over the second order
time derivative term in the $n_1$ and $n_5$ variables. In the
absence of an external magnetic field, only second order time
derivative terms of $n_{2,3,4}$ enter the Lagrangian. Therefore,
while the chemical potential term compensates the anisotropy
potential in Eq. (\ref{Veff}) to arrive at an $SO(5)$ symmetric
static potential, its time dependent part breaks the full quantum
$SO(5)$ symmetry. This observation leads to the concept of the
projected or static $SO(5)$ symmetry. A model with projected or
static $SO(5)$ symmetry is described by a quantum effective
Lagrangian of the form
\begin{eqnarray}
{\cal L} = \frac{\chi}{2} \sum_{\alpha=2,3,4} (\partial_t
n_\alpha)^2 -\chi \mu (n_1\partial_t n_5 - n_5\partial_t n_1) -
V_{eff}(n), \label{L_pso5}
\end{eqnarray}
where the static potential $V_{eff}$ is $SO(5)$ symmetric.

We see that ``class $A$" transition from AF to SC occurs at a
particle hole symmetric point, and it can have a full quantum
$SO(5)$ symmetry. The ``class $B$" transition from AF to SC is
induced by a chemical potential; only static $SO(5)$ symmetry can
be realized at the transition point. The ``class $A$" transition
can occur at half-filling in organic superconductors, where the
charge gap at half-filling is comparable to the spin exchange
energy. In this system, the AF to SC transition is tuned by
pressure, where the doping level and the chemical potential stay
fixed. The transition from the half-filled AF state to the SC
state in the HTSC cuprates is far from the ``class $A$'' transition
point, but static $SO(5)$ symmetry can be realized at the chemical
potential induced transition. However, as we shall see in section
\ref{sec:quantum}, there are also Mott insulating states with AF
order at fractional filling factors, for instance, at doping level
$x=1/8$. The insulating gap is much smaller at these fractional
Mott phases, and there is an effective particle-hole symmetry near
the tip of the Mott lobes. For these reasons, ``class $A$'' transition
with the full quantum $SO(5)$ symmetry can  be realized again near
the tip of fractional Mott phases, as in organic
superconductors. Transitions near the fractional Mott insulating
lobes are depicted as the ``$A2$" and ``$B2$" transitions in the
global phase diagram (see Fig. \ref{Fig_global}). In this case, a
transition from a fractional Mott insulating phase with AF order
to the SC state can again be tuned by pressure without changing
the density or the chemical potential.

The $SO(5)$ quantum nonlinear $\sigma$ model is constructed from
two canonically conjugate field operators $L_{ab}$ and $n_a$. In
fact, there is a kinematic constraint among these field operators.
In the case of the Heisenberg model, the total spin operator and
the AF Neel order parameter satisfy an orthogonality constraint,
as expressed in Eq. (\ref{SO3_orthogonal}). The $SO(5)$
generalization of this constraint can be expressed as follows:
\begin{eqnarray}
L_{ab} n_c + L_{bc} n_a + L_{ca} n_b=0. \label{so5_orthogonal}
\end{eqnarray}
This identity is valid for any triples $a$, $b$ and $c$, and can
be easily proven by expressing $L_{ab}=n_a p_b - n_b p_a$, where
$p_a$ is the conjugate momentum of $n_a$. Geometrically, this
identity expresses the fact that $L_{ab}$ generates a rotation of
the $n_a$ vector. The infinitesimal rotation vector lies on the
tangent plane of the four sphere $S^4$, as defined by Eq.
(\ref{n_constraint}), and is therefore orthogonal to the $n_a$
vector itself. Extending this geometric proof,
Wegner\cite{WEGNER2000} has shown that the $SO(5)$ orthogonality
relation also follows physically from maximizing the entropy. 
Taking the triple $a,b,c$ to be $2,3,4$, and recognizing that
$L_{\alpha\beta}=\epsilon_{\alpha\beta\gamma}S_\gamma$, this
identity reduces to the $SO(3)$ orthogonality relation in Eq.
(\ref{SO3_orthogonal}). This $SO(5)$ identity places a
powerful constraint on the expectation values of various
operators. In particular, it quantitatively predicts the value of
the $\pi$ order parameter in a mixed state between AF or SC.
For example, let's take the $a,b,c$ triple to be $1,2,5$. Eq.
(\ref{so5_orthogonal}) predicts that
\begin{eqnarray}
L_{15} n_2 + L_{52} n_1 + L_{21} n_5 =0 \ \ \Rightarrow \ \
\langle L_{25}\rangle = \langle Im \pi_x \rangle =
\frac{Q \langle n_2\rangle}{\langle n_1\rangle},
\label{mixed_ortho}
\end{eqnarray}
where we chose the SC phase such that $\langle
n_5\rangle=0$. Here, $Q=\langle L_{15}\rangle$ measures the hole
density. Since these four expectation values can easily be 
measured numerically and, in principle, experimentally, this
relationship can be tested quantitatively. Recently, Ghosal,
Kallin and Berlinsky\cite{GHOSAL2002} tested this relationship
within microscopic models of the AF vortex core. In this case, AF
and SC coexist in a finite region near the vortex core, so that
both $\langle n_1\rangle$ and $\langle n_5\rangle$ are
non-vanishing. They found that the $SO(5)$ orthogonality
constraint is accurately satisfied in microscopic models.

In this section, we presented the $SO(5)$ quantum non-linear
$\sigma$ model as a heuristic and phenomenological model. The key
ingredients of the model are introduced by observing the robust
features of the phase diagram and the low energy collective modes
of the HTSC cuprate system. This is the ``top-down" approach
discussed in the introduction. In this sense, the model has a
general validity beyond the underlying microscopic physics.
However, it is also useful to derive such a model directly from
microscopic electronic models. Fortunately, this ``bottom-up"
approach agrees with the phenomenological
approach to a large extent. A rigorous derivation of this quantum non-linear
$\sigma$ model from an $SO(5)$ symmetric microscopic model on a
bi-layer system will be given in section \ref{sec:exact}, while an
approximate derivation from the ``realistic" microscopic $t-J$ and
Hubbard model will be given in section \ref{sec:CORE}.

\subsection{The projected SO(5) model with lattice bosons}
\label{sec:so5bosons} In the previous section, we presented the
formulation of the $SO(5)$ quantum nonlinear $\sigma$ model. This
model is formulated in terms of two sets of canonically conjugate
variables - the superspin vector $n_a$ and the angular momentum
$L_{ab}$. The two terms which break the full quantum $SO(5)$
symmetry are the anisotropy term, $g$, and the chemical potential term,
$\mu$. Therefore, this model contains high energy modes,
particularly excitations of the order of the Mott insulating gap at
half-filling. For this reason, Greiter\cite{GREITER1997} and
Baskaran and Anderson\cite{BASKARAN1998} questioned
whether the effective $SO(5)$ symmetry can be implemented in the
low energy theory. In the previous section, it was shown that
these two symmetry breaking terms could cancel each other in the
{\it static} potential and the resulting effective potential
could still be $SO(5)$ symmetric. It was also pointed out that the
chemical potential term breaks the $SO(5)$ symmetry in the
{\it dynamic or time-dependent} part of the effective Lagrangian.
In response to these observations, Zhang et al. constructed the
{\it projected} $SO(5)$ models\cite{ZHANG1999}, which fully
project out the high energy modes, and obtained a low energy
effective quantum Hamiltonian, with an approximately $SO(5)$
symmetric static potential.

\begin{figure}[hp]
\begin{center}
        \includegraphics[scale=0.8]{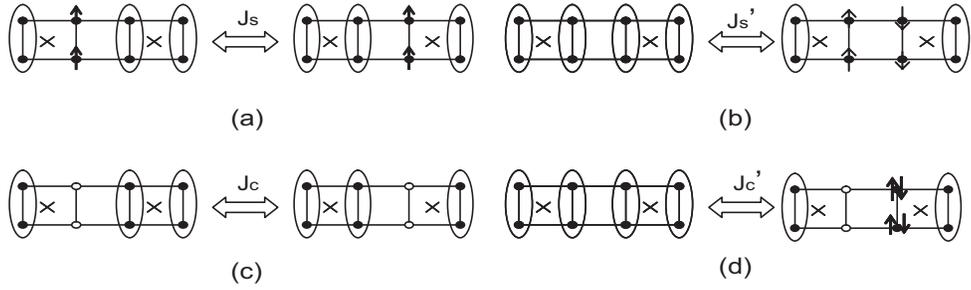}
        \caption{Illustration of hopping processes of the magnons and the hole
        pairs on a ladder. The cross denotes the center of a plaquette.
        An ellipse enclosing two sites denotes a spin singlet.
        (a) $J_s$ describes the magnon hopping, (b) $J'_s$ describes
        the spontaneous creation and annihilation of a magnon pair. (c) $J_c$ describes
        the hopping of a hole pair, (d) $J'_c$ describes the spontaneous
        creation and annihilation of a hole pair and a particle pair. In the
        full $SO(5)$ model, $J_s=J'_s$ and $J_c=J'_c$. In the projected $SO(5)$
        model, the particle pair states are removed and $J'_c=0$.}
\label{Fig_hopping}
\end{center}
\end{figure}

The first step is to perform a transformation from the $n_a$ and
$L_{ab}$ coordinates to a set of bosonic operators. We first
express the angular momentum operator as
\begin{equation}
L_{ab} = n_a p_b - n_b p_a, \label{Lnp}
\end{equation}
where $p_a$ is the canonical momentum conjugate to $n_a$,
satisfying the Heisenberg commutation relation:
\begin{equation}
[n_a, p_b] = i \delta_{ab}. \label{np}
\end{equation}
Furthermore, we can express the canonical coordinates and momenta
in terms of the boson operators as
\begin{equation}
n_a = \frac{1}{\sqrt{2}} (t_a + t_a^\dagger)  \ \ \ p_a =
\frac{1}{i\sqrt{2}} (t_a - t_a^\dagger),
\end{equation}
where the boson operators satisfy the commutation relation
\begin{equation}
[t_a, t_b^\dagger] = \delta_{ab},
\end{equation}
and the (half-filled) ground state is defined by $t_a
|\Omega\rangle=0$. There are five boson operators,
$t_\alpha=t_2,t_3,t_4$ are the boson operators for the magnetic
triplet excitations, also called the magnons, while
\begin{equation}
t_1 = \frac{1}{\sqrt{2}} (t_h + t_p)  \ \ \ t_5 =
\frac{1}{i\sqrt{2}}(t_h - t_p)
\label{p-h}
\end{equation}
are the linear combinations of the particle pair ($t_p$) and hole
pair ($t_h$) annihilation operators. In the $SO(5)$ quantum
non-linear $\sigma$ model formulation, there is an infinite number
of bosonic states per site. However, due to the first term in Eq.
(\ref{Hsigma}) (the angular momentum term), states with higher
angular momenta or, equivalently, higher boson number, are
separated by higher energies. Therefore, as far as the low energy
physics is concerned, we can restrict ourselves to the manifold of
six states per site, namely the ground state $|\Omega\rangle$ and
the five bosonic states $t^\dagger_a|\Omega\rangle$. This
restriction is called the hard-core boson constraint and can be
implemented by the condition $t^\dagger_a
t^\dagger_b|\Omega\rangle=0$. Within the Hilbert space of
hard-core bosons, the original $SO(5)$ quantum non-linear $\sigma$
model is mapped onto the following hard-core boson model:
\begin{eqnarray}
H &=& \Delta_s \sum_{x,\alpha=2,3,4} t_\alpha^\dagger t_\alpha(x)
+ \Delta_c \sum_{x,i=1,5} t_i^\dagger t_i(x)
+\mu \sum_x (t_p^\dagger t_p(x) - t_h^\dagger t_h(x)) \nonumber \\
&-& J_s \sum_{\langle xx' \rangle} n_\alpha(x) n_\alpha(x') - J_c
\sum_{\langle xx' \rangle} n_i(x) n_i(x'), \label{boson}
\end{eqnarray}
where $\Delta_c=2/\chi-2\mu^2\chi$ and $\Delta_s=2/\chi-g$ are the
creation energies for the charge pairs and the triplet magnons,
$\mu$ is the chemical potential, and $J_c$ and $J_s$ are the exchange
energies for SC and AF, respectively. In the isotropic case, they
are taken to be $\rho$ in the second term of Eq. (\ref{Hsigma}).
Expressing $n_i$ and $n_\alpha$ in terms of the bosonic operators,
we see that the $J_c$ and $J_s$ terms describe not only the
hopping, but also the spontaneous creation and annihilation of the
charge pairs and the magnons, as depicted in
Fig.\ref{Fig_hopping}.


When $\Delta_c=\Delta_s$, $J_c=J_s$ and $\mu=0$, the model
(\ref{boson}) has an exact quantum $SO(5)$ symmetry. In this case,
the energy to create charge excitations is the same as the energy
to create spin excitations. This situation could be realized in
organic and heavy fermion superconductors near the AF phase
boundary or the HTSC  near commensurate doping
fractions such as $x=1/8$, as we shall see in section
\ref{sec:quantum}. However, for HTSC systems near half-filling,
the energy to create charge excitations is much greater than the
energy to create spin excitations, {\i.e.} $\Delta_c \gg
\Delta_s$. In this case, the full quantum $SO(5)$ symmetry is
broken but, remarkably, the effective {\it static} potential can
still be $SO(5)$ symmetric. This was seen in the previous section
by the cancellation of the anisotropy potential $g$ term by the
chemical potential $\mu$ term. In a hard-core boson model
(\ref{boson}) with $\Delta_c \gg \Delta_s$, a low energy effective
model can be derived by retaining only the hole pair state while
projecting out the particle pair state. This can be done by
imposing the constraint
\begin{equation}
t^\dagger_p(x) |\Omega\rangle = 0 \label{constraint1}
\end{equation}
at every site $x$. The projected Hamiltonian takes the form
\begin{eqnarray}
H &=&
\Delta_s \sum_{ x } t_\alpha^\dagger t_\alpha(x) +
\tilde \Delta_c \sum_{ x } t^\dagger_h t_h (x)   \nonumber \\
&-& J_s \sum_{\langle xx' \rangle} n_\alpha(x) n_\alpha(x') - J_c
\sum_{ \langle xx' \rangle} n_i(x) n_i(x'), \label{pso5}
\end{eqnarray}
where $\tilde\Delta_c=\Delta_c-\mu$. The Hamiltonian (\ref{pso5}) has no
parameters of the order of $U$. To achieve the static $SO(5)$
symmetry, we need $\Delta_s\sim\tilde\Delta_c$ and $J_s\sim J_c$.
The first condition can always be met by changing the
chemical potential, whereas the second one requires certain fine
tuning. However, as we discuss in Section VD (see Fig.
\ref{fig5epars}), this condition emerges naturally when one
derives the model (\ref{pso5}) from the Hubbard model in the
relevant regime of parameters.

The form of the projected $SO(5)$ Hamiltonian hardly changes from
the unprojected model (\ref{boson}), but the definition of $n_1$
and $n_5$ is changed from
\begin{eqnarray}
n_1 &=& \frac{1}{\sqrt{2}} (t_1 + t_1^\dagger) =
\frac{1}{2} (t_h + t_p + t_h^\dagger + t_p^\dagger) \nonumber \\
n_5 &=& \frac{1}{\sqrt{2}} (t_5 + t_5^\dagger) =
\frac{1}{2i} (t_h - t_p - t_h^\dagger + t_p^\dagger)
\label{before}
\end{eqnarray}
to
\begin{equation}
n_1 = \frac{1}{2} (t_h + t_h^\dagger) \ \ \
n_5 = \frac{1}{2i} (t_h - t_h^\dagger).
\label{after}
\end{equation}
From Eq. (\ref{before}), we see that $n_1$ and $n_5$ commute with
each other before the projection. However, after the projection,
they acquire a nontrivial commutation relation, as can be seen
from Eq. (\ref{after}):
\begin{equation}
[n_1,n_5] = i/2.
\label{non-commute}
\end{equation}
Therefore, projecting out doubly occupied sites, commonly referred
to as the Gutzwiller projection, can be analytically implemented
in the $SO(5)$ theory by retaining the form of the Hamiltonian and
changing only the commutation relations.

\begin{figure}[hp]
\includegraphics[scale=0.3]{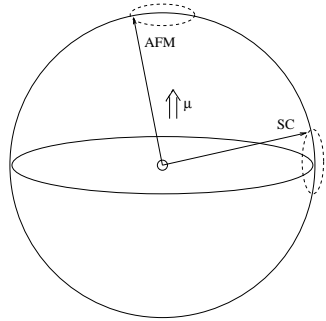}
\caption{The chiral $SO(5)$ sphere has an $SO(5)$ symmetric shape
but allows only one sense of the rotation in the SC plane
$(n_1,n_5)$. Small oscillations around the equator, or the $\pi$
triplet resonance, are unaffected by the chiral projection.
However, small oscillations around the north pole, or the $\pi$
doublet mode, are strongly affected: only one of the two such
modes is retained after the projection.} \label{Fig_chiral}
\end{figure}

\begin{figure}[hp]
\begin{center}
        \includegraphics[scale=0.8]{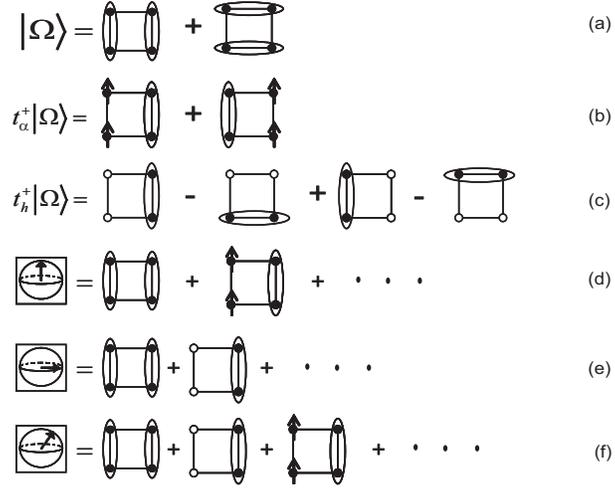}
        \caption{(a),(b) and (c) express the five bosonic states of the projected $SO(5)$ model
        in terms of the microscopic states on a plaquette. (d), (e) and (f) represent
        states with well-defined superspin directions, which can be obtained from the
        linear combinations of (a), (b) and (c). These states are analytically defined in
        Eq. (\ref{coherent}) and Table II.}
\label{Fig_plaquette}
\end{center}
\end{figure}

The Gutzwiller projection implemented through the modified
commutation relations between $n_1$ and $n_5$ is formally similar
to the projection onto the lowest Landau level in the physics of
the quantum Hall effect. For electrons moving in a 2D plane, the
canonical description involves two coordinates, $X$ and $Y$, and
two momenta, $P_X$ and $P_Y$. However, if the motion of the
electron is fully confined in the lowest Landau level, the
projected coordinate operators become non-commuting and are given
by $[X,Y]=il_0^2$, where $l_0$ is the magnetic length. In the
context of the projected $SO(5)$ Hamiltonian, the original rotors
at a given site can be viewed as particles moving on a four
dimensional sphere $S^4$, as defined by Eq. (\ref{n_constraint}),
embedded in a five dimensional Euclidean space. The angular
momentum term $\frac{1}{2\chi}L_{ab}^2$ describes the kinetic
motion of the particle on the sphere. The chemical potential acts
as a fictitious magnetic field in the $(n_1,n_5)$ plane. In the
Gutzwiller-Hubbard limit, where $\Delta_c \gg \Delta_s$, a large
chemical potential term is required to reach the limit
$\tilde\Delta_c \sim \Delta_s$. The particle motion in the
$(n_1,n_5)$ plane becomes quantized in this limit, as in the
case of the quantum Hall effect, and the non-commutativity of the
coordinates $(n_1,n_5)$ given by Eq. (\ref{non-commute}) arises as
a result of the projection. The projection does not affect the
symmetry of the sphere on which the particle is moving; however,
it restricts the sense of the kinetic motion to be chiral, {\it
i.e.}, only along one direction in the $(n_1,n_5)$ plane. (See
Fig. \ref{Fig_chiral}).
In this sense, the particle is moving on a {\it chiral} $SO(5)$
symmetric sphere. The non-commutativity of the $(n_1,n_5)$
coordinates is equivalent to the effective Lagrangian 
(see  Eq. (\ref{L_pso5}) of section \ref{sec:so5sigma}) containing
only the first order time derivative. In fact, from Eq. (\ref{L_pso5}),
we see that in this case the canonical momenta associated with the coordinates
$n_1$ and $n_5$ are given by
\begin{equation}
p_1=\frac{\delta L}{\delta \dot n_1}= \chi \mu n_5 \ \ , \ \
p_5=\frac{\delta L}{\delta \dot n_5}= - \chi \mu n_1.
\label{momenta}
\end{equation}
Applying the standard Heisenberg commutation relation for the
conjugate pairs $(n_1,p_1)$, or $(n_5,p_5)$ gives exactly the
quantization condition (\ref{non-commute}). Note that in Eq.
(\ref{momenta}) $\chi \mu$ plays the role of the Planck's
constant in quantum mechanics. We see that the projected $SO(5)$
Hamiltonian (\ref{pso5}) subjected to the quantization condition
(\ref{non-commute}) is fully equivalent to the effective
Lagrangian Eq. (\ref{L_pso5}), discussed in the last section.


Despite its apparent simplicity, the projected $SO(5)$ lattice
model can describe many complex phases, most of which are seen in
the HTSC cuprates. These different phases can be described in
terms of different limits of a single variational wave function
of the following product form:
\begin{equation}
|\Psi\rangle = \prod_x
(\cos\theta(x)+\sin\theta(x)(m_\alpha(x)t^\dagger_\alpha(x)
+\Delta(x) t^\dagger_h(x))) |\Omega\rangle. \label{coherent}
\end{equation}
where the variational parameters $m_\alpha(x)$ should be real,
while $\Delta(x)$ is generally complex. The normalization of the
wave function, $\langle\Psi|\Psi\rangle=1$, requires the
variational parameters to satisfy
\begin{equation}
\sum_\alpha |m_\alpha(x)|^2+|\Delta(x)|^2=1.\label{unit_length}
\end{equation}
Therefore, we can parameterize them as
$|m_\alpha(x)|^2=cos^2\phi(x)$ and
$|\Delta(x)|^2=sin^2\phi(x)$, which is similar to the $SO(5)$
constraint introduced in Eq. (\ref{n_constraint}). The expectation
values of the order parameters and the symmetry generators in this
variational state are given by
\begin{eqnarray}
\langle\Psi|n_\alpha(x)|\Psi\rangle &=& \frac{1}{\sqrt{2}}
sin2\theta(x)
Re(m_\alpha(x))\nonumber \\
\langle\Psi|n_1(x)|\Psi\rangle &=& \frac{1}{2} sin2\theta(x)
Re(\Delta(x))\nonumber \\
\langle\Psi|n_5(x)|\Psi\rangle &=& \frac{1}{2} sin2\theta(x)
Im(\Delta(x)) \label{order_parameter_expectation}
\end{eqnarray}
and
\begin{eqnarray}
\langle  \Psi | Q(x) | \Psi \rangle &=& \langle \Psi
t^\dagger_h(x) t_h(x) | \Psi \rangle
=\sin^2 \theta(x)~ |\Delta (x)|^2 \nonumber \\
\langle  \Psi | S_\alpha | \Psi \rangle &=& -\langle \Psi | i
\epsilon^{\alpha\beta\gamma} t^\dagger_\beta(x) t_\gamma(x) |\Psi
\rangle= -i \epsilon^{\alpha\beta\gamma} \sin^2 \theta~
m^*_\beta(x) m_\gamma(x)
\nonumber\\
\langle \Psi  | \pi_\alpha(x) | \Psi \rangle &=&\langle \Psi |
\frac{t^\dagger_\alpha(x) t_h(x)} {i\sqrt 2} | \Psi \rangle= \frac
{\sin^2 \theta ~m^*_\alpha(x) \Delta(x)}{i \sqrt 2}.
\label{symmetry_generator_expectation}
\end{eqnarray}

Initially, we restrict our discussions to the case where the
variational parameters are uniform, describing a translationally
invariant state. Evaluating different physical operators in this
state gives the result summarized in the following table:

\begin{table}[h]
\begin{center}
\begin{tabular}{|c|c|c|c|c|c|c|}    \hline
& &charge Q  & spin S & AF order $\langle n_\alpha \rangle$ & SC
order $\langle n_i \rangle$ & $\pi$ order $\langle \pi_\alpha
\rangle$  \\ \hline
(a) & ``RVB" state: $sin\theta=0$ & 0  & 0 & 0 & 0 & 0 \\
\hline
(b) & Magnon state: $cos\theta=0$ and $sin\phi=0$ & 0  & 1 & 0
& 0 & 0 \\ \hline
(c) & ``Hole pair" state: $cos\theta=0$ and
$cos\phi=0$ & -2 & 0 & 0 & 0 & 0 \\ \hline
(d) & AF state: $0<sin\theta<1$ and $sin\phi=0$ &
0  & indefinite & $\neq 0$ & 0 & 0 \\ \hline
(e) & SC state:
$0<sin\theta<1$ and $cos\phi=0$ & indefinite  & 0 &  0 & $\neq
0$ & 0 \\ \hline
(f) & Mixed AF/SC state: $0<sin\theta<1$ and $0 <
sin\phi < 0$ & indefinite  & indefinite & $\neq 0$ & $\neq 0$ &
$\neq 0$
\\ \hline
& $\pi$ state:
$cos\theta=0$ and
$0<cos\phi<1$ & indefinite & indefinite & 0 & 0 & $\neq 0$ \\
\hline
\end{tabular}
\caption{
Physical properties of various plaquette states classified according to
the $SO(5)$ order parameters and symmetry generators.
}
\end{center}
\end{table}


As we can see, this wave function not only describes classically
ordered states with spontaneously broken symmetry, but also
quantum disordered states which are eigenstates of spin and
charge. Generally, $\Delta_c$ and $\Delta_s$ favor
quantum disordered states, while $J_c$ and $J_s$ favor classically
ordered states. Depending on the relative strength of these
parameters, a rich phase diagram can be obtained.

\begin{figure}[hp]
\begin{center}
        \includegraphics[scale=0.5]{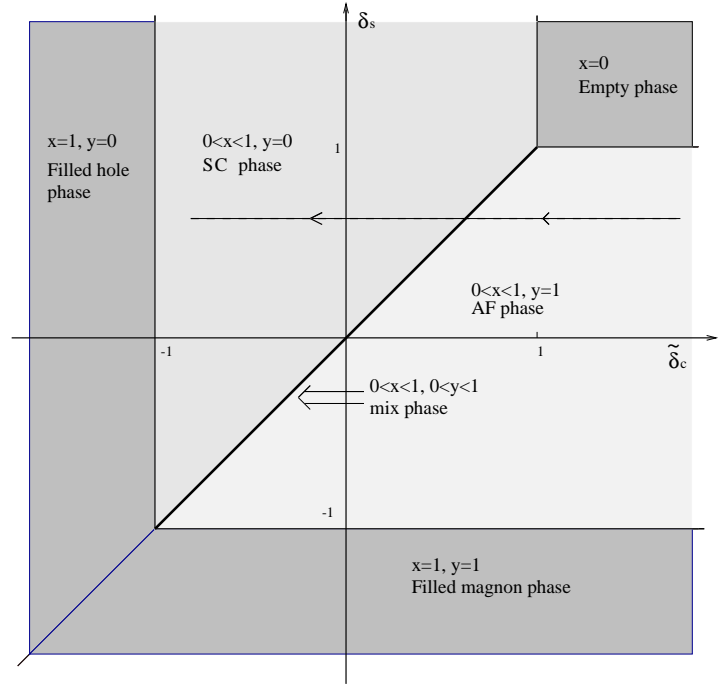}
        \caption{
Phase diagram of the projected $SO(5)$ model (\ref{pso5}) (for the
case $J_c=2J_s\equiv J$) as a function of $\delta_s=\Delta_S/4J$
and $\tilde\delta_c=\tilde{\Delta}_c/4J$ . Variation of the
chemical potential changes $\tilde{\Delta}_c$ and traces out a
one-dimensional trajectory as shown on the dotted line.
$x=sin^2\theta$ and $y=cos^2\phi$.} \label{Fig_pSO5PhaseDiagram}
\end{center}
\end{figure}
\begin{figure}[hp]
\begin{center}
        \includegraphics[scale=0.8]{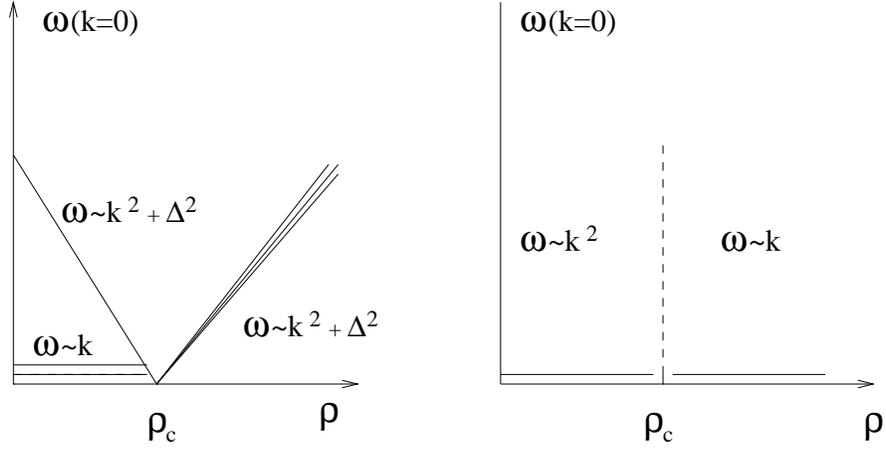}
        \caption{Spectra of the collective excitations of the
projected $SO(5)$ model as a function of density. The region
$0<\rho<\rho_c$ corresponds to the uniform mixed phase of SC and AF.
Region $\rho > \rho_c$ corresponds to the SC phase. The left panel
shows the spectra of the spin excitations. For $\rho<\rho_c$,
there are two gapless spin wave modes and one gapped spin
amplitude mode. For $\rho>\rho_c$, there is a spin triplet $\pi$
resonance mode.  The right panel shows the spectra of the gapless
charge excitations (in the absence of long range interactions).
For $\rho<\rho_c$ the charge mode has quadratic dispersion. The
dispersion relation changes from $\omega \propto k^2$ to  $\omega
\propto k$ for the $\rho>\rho_c$ regime.}
\label{Fig_pSO5Excitations}
\end{center}
\end{figure}

The phase diagram of the projected $SO(5)$ model with
$J_c=2J_s\equiv J$ is shown in Fig. \ref{Fig_pSO5PhaseDiagram}.
Changing the chemical potential modifies $\tilde{\Delta}_c$ and
traces out a one-dimensional path on the phase diagram. Along this
path the system goes from the AF state to the uniform AF/SC mixed phase
and then to the SC state. The mixed phase only corresponds
to one point on this trajectory (i.e. a single value of the
chemical potential $\mu_c$), although it covers a whole range of
densities $0 < \rho < \rho_c$. This suggests that the transition
may be thought of as a first order transition between the AF and
SC phases, with a jump in the density at $\mu_c$. The spectrum of
collective excitations shown in Fig. \ref{Fig_pSO5Excitations},
however, shows that this phase diagram also has important features
of two second order phase transitions. The energy gap to $S=1$
excitations inside the SC phase vanishes when the chemical
potential reaches the critical value $\mu_c$. Such a softening
should not occur for the first order transition but is required
for the continuous transition into a state with broken spin
symmetry. This shows that models with the $SO(5)$ symmetry have
intrinsic fine-tuning to be exactly at the border between a single
first order transition and two second order transitions; 
in subsequent sections this type
of transition shall be classified as type 1.5 transition. 
Further discussion of the phase diagram of
the projected $SO(5)$ model is given in section VC.
Note that effective bosonic Hamiltonians 
similar to (\ref{pso5}) have also been considered
in Refs. \cite{VANDUIN2000,PARK2001}.

\section{THE GLOBAL PHASE DIAGRAM OF SO(5) MODELS}
\label{sec:so5global}
\subsection{Phase diagram of the classical model}
\label{sec:classical} The two robust ordered phases in the HTSC
cuprates are the AF phase at half-filling and the SC phase away
from half-filling. It is important to ask how these two phases are
connected in the global phase diagram as 
different tuning parameters such as the temperature, the doping
level, the external magnetic field, etc, one varied. Analyzing the $SO(5)$
quantum nonlinear $\sigma$ model, Zhang has classified four
generic types of phase diagrams, presented as Fig. (1A)-(1D) in
reference \cite{ZHANG1997}. In the next section we are going to
investigate the zero temperature phase diagram where the AF and
the SC phases are connected by various quantum disordered states,
often possessing charge order. In this section, we first focus on
the simplest possibility, where AF and SC are the only two
competing phases in the problem, and discuss the phase diagram in
the plane of temperature and chemical potential, or doping level.

\begin{figure}[hp]
\begin{center}
        \includegraphics[scale=1.0]{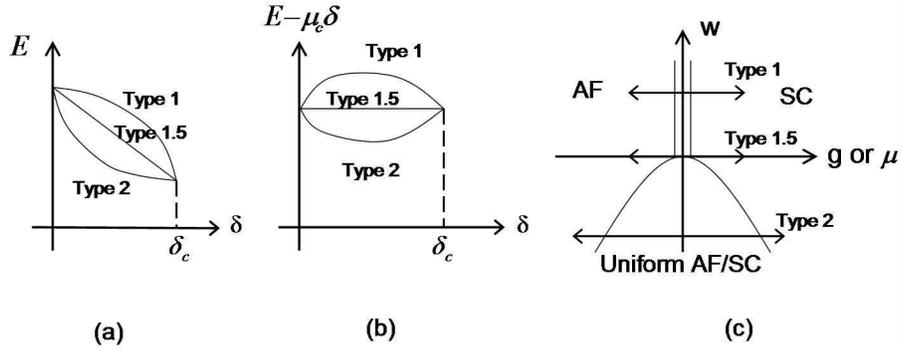}
        \caption{The energy (a) and the free energy (b) can depend on
        the density of a uniform AF/SC mixed state with a negative
        curvature when $u_{12}>\sqrt{u_1 u_2}$, (classified as ``type 1") or a positive
        curvature when $u_{12}<\sqrt{u_1 u_2}$ (classified as ``type 2").
        The $SO(5)$ symmetric limiting case of zero curvature, 
classified as ``type 1.5," is
        realized when $u_{12}=\sqrt{u_1 u_2}$. (c) The ``type 1" phase transition
        from the AF to SC state is a direct first order transition. There are two second
        order transitions from the AF to SC state in the ``type
        2" case. $SO(5)$ symmetry is realized at the intermediate 
case of ``type 1.5."}
\label{Fig_E_delta}
\end{center}
\end{figure}

Let us first discuss the general properties of phase transition
between two phases, each characterized by its own order parameter.
In particular, we shall focus on the phenomenon of the enhanced
symmetry at the multicritical point at which physically different
static correlation functions show identical asymptotic behavior.
In the case of CDW to SC transition discussed in section
\ref{sec:spin-flop}, the CDW is characterized by an Ising like
$Z_2$ order parameter, while the SC is characterized by a $U(1)$
order parameter. In the case of AF to SC transition, the order
parameter symmetries are $SO(3)$ and $U(1)$, respectively.
Generically, the phase transition between two ordered phases can
be either a single direct first order transition or two second
order phase transitions with a uniform mixed phase in between,
where both order parameters are non-zero. This situation can be
understood easily by describing the competition in terms of a LG
functional of two competing order parameters\cite{KOSTERLITZ1976},
which is given by
\begin{eqnarray}
F=r_1 \phi_1^2 + r_2 \phi_2^2 + u_1 \phi_1^4 +u_2 \phi_2^4 + 2
u_{12} \phi_1^2 \phi_2^2. \label{GL}
\end{eqnarray}
Here, $\phi_1$ and $\phi_2$ are vector order parameters with $N_1$
and $N_2$ components, respectively. In the case of current
interest, $N_1=2$ and $N_2=3$ and we can view
$\phi^2_1=n_1^2+n_5^2$ as the SC component of the superspin
vector, and $\phi^2_2=n_2^2+n_3^2+n_4^2$ as the AF component of
the superspin vector. These order parameters are determined by
minimizing the free energy $F$, and are given by the solutions of
\begin{eqnarray}
2u_1 \phi_1^2 + 2u_{12} \phi_2^2 +r_1 =0 \ \ , \ \
2u_{12} \phi_1^2 + 2u_2 \phi_2^2 +r_2 =0.
\label{GL_solution}
\end{eqnarray}
These equations determine the order parameters uniquely, except
in the case when the determinant of the linear equations
vanishes. At the point when
\begin{eqnarray}
u_1 u_2 = u_{12}^2
\label{GL_condition}
\end{eqnarray}
and
\begin{eqnarray}
\frac{r_1}{\sqrt{u_1}} = \frac{r_2}{\sqrt{u_2}},
\label{muc_condition}
\end{eqnarray}
the order parameters satisfy the relations
\begin{eqnarray}
\frac{\phi^2_1}{\sqrt{u_2}} + \frac{\phi^2_2}{\sqrt{u_1}} = const,
\label{ellipse}
\end{eqnarray}
but they are not individually determined. In fact, with the
re-scaling $\tilde\phi^2_1=\phi^2_1/\sqrt{u_2}$ and
$\tilde\phi^2_2=\phi^2_2/\sqrt{u_1}$, the free energy is exactly
$SO(5)$ symmetric with respect to the scaled variables, and 
Eq. (\ref{ellipse}) becomes identical to Eq. (\ref{n_constraint})
in the $SO(5)$ case. Since the free energy  only depends on the
combination $\tilde\phi^1_2+\tilde\phi^2_2$, one order parameter
can be smoothly rotated into the other without any energy cost.
Equation (\ref{GL_condition}) is the most important condition for
the enhanced symmetry. We shall discuss extensively in this paper
whether this condition is satisfied microscopically or close to
some multi-critical points in the HTSC cuprates. On the other hand,
equation (\ref{muc_condition}) can always  be tuned. In the case
of AF to SC transition, the chemical potential couples to the
square of the SC order parameter, as we can see from Eq.
(\ref{Veff}). Therefore, $r_1$ can be tuned by the chemical
potential, and equation (\ref{muc_condition}) defines the critical
value of the chemical potential $\mu_c$ at which the phase
transition between AF and SC occurs. At this point, the chemical
potential is held fixed, but the SC order parameter and the charge
density can change continuously according to Eq. (\ref{ellipse}).
Since the free energy is independent of the density at this point,
the energy, which differs from the grand canonical free energy by
a chemical potential term $\mu \delta$, can  depend only {\it
linearly} on the density. The linear dependence of the energy on
doping is a very special, limiting case. Generally, the energy
versus doping curve would either have a negative curvature,
classified as ``type 1," or a positive curvature, classified as
``type 2" (see Fig. \ref{Fig_E_delta}a). The special limiting case
of ``type 1.5" with zero curvature is only realized at the $SO(5)$
symmetric point. The linear dependence of the ground state energy
of a {\it uniform} AF/SC mixed state on the density is a
crucial test of the $SO(5)$ symmetry which can be performed
numerically, as we shall see in section \ref{sec:variational} and
\ref{sec:diagonal}. The constancy of the chemical potential and
the constancy of the length of the $SO(5)$ superspin vector
(\ref{ellipse}) as a function of density can be tested 
experimentally as well, as we shall discuss in section
\ref{sec:variational}.


The constancy of the chemical potential as a function of the
density in a {\it uniform system} is a very special situation
which only follows  from the enhanced symmetry at the phase
transition point. In a system with phase separation, the chemical
potential is also independent of the total density, but the local
density is non-uniform. The two phases are generally separated by
a domain wall. The $SO(5)$ symmetric case can be obtained from the
phase separation case in the limit where the width of the domain
wall goes to infinity and a uniform state is obtained. This
situation can be studied analytically by solving Eq.
(\ref{GL_solution}). Defining the parameters that characterize the
deviation from the symmetric point as $w=u_{12}-\sqrt{u_1 u_2}$
and $g=(\frac{r_1}{\sqrt{u_1}}-\frac{r_2}{\sqrt{u_2}})/2$, it is
obvious that the phase transition between the two forms of
order is tuned by $g$, while $w$ determines the nature of the
phase transition. The phase diagram in the $(g,w)$ plane is shown
in Fig. \ref{Fig_E_delta}c. For $w>0$, the two ordered phases are
separated by a first order line. This type of transition is
classified as ``type 1." On the other hand, when $w<0$, the two
ordered phases are separated by two second order phase transition
lines with an intermediate mixed phase where two orders coexist,
{\it i.e.} $\langle \phi_1 \rangle \neq 0$ and $\langle \phi_2
\rangle \neq 0$. This type of transition is classified as ``type
2." The limiting ``type 1.5" behavior corresponds to the symmetric
point $w=0$. Approaching this point from $w>0$, the first order
transition becomes weaker and weaker and the latent
heat associated with the first order transition becomes smaller
and smaller. Therefore, the symmetric point can be viewed as the
end point of a first order transition. On the other hand,
approaching the symmetric point from $w<0$, the width of the
intermediate mix phase becomes smaller and smaller, until
the two second order transition lines merge into a single
transition at $w=0$. From the above discussion, we learn
an important lesson: the phase transition between two ordered
phases can  be either a direct first order transition or two
second order transitions with an intermediate mixed phase.
Furthermore, the symmetric point realizes a limiting behavior
which separates these two scenarios. Balents, Fisher and
Nayak\cite{BALENTS1998}, Lee and Kivelson\cite{LEE2003A} pointed
out that the ``type 1" and ``type 2" transitions of a Mott
insulator induced by varying the chemical potential is analogous
to the two types of superconductor to normal state transitions
induced by a magnetic field. The magnetic field induces a direct
first order transition from the SC state to the normal state in
``type 1" superconductors, while it induces two second order
transitions with an intermediate mix state in the ``type 2"
superconductors. Indeed, the limiting ``type 1.5" behavior
separating the ``type 1" and the ``type 2" superconductors also
has a special symmetry, where the Bogomol'nyi's bound for the
vortex is satisfied as an equality. We note
that recent work of Senthil et al
discussed an alternative scenario for a direct second order
transition between two phases with different order
parameters and without a higher symmetry 
at the transition point. This was achieved by having
fractionalizated excitations at the quantum critical
point \cite{SENTHIL2004}.

Let us now turn to the finite temperature phase transitions. In
$D=3$, finite temperature phase transitions associated with
continuous symmetry breaking are possible. Therefore, the order
parameters $\phi_1$ and $\phi_2$ can each have their own phase
transition temperature, $T_c$ and $T_N$. The interesting question
is how these two second order lines merge as one changes the
parameter $g$ or, equivalently, the chemical potential $\mu$,
which interchanges the relative stability of the two ordered
phases. There are two generic possibilities. The ``type 1" phase
diagram is shown  in Fig. \ref{Fig_T_mu}a, where the two second
order phase transition lines intersect at a {\it bi-critical}
point, $T_{bc}$, which is also the termination point of the first
order transition line separating the two ordered phases. This type
of phase diagram is realized for $u_{12}>\sqrt{u_1 u_2}$. The
first order transition at $\mu_c$ separates the AF and SC
states with different densities; therefore, the $T$ versus
$\delta$ phase diagram shown in Fig. \ref{Fig_T_mu}b contains a
region of phase separation extending over the doping range
$0<\delta<\delta_c$. The ``type 2" phase diagram is shown in Fig.
\ref{Fig_T_mu}c, where $T_c$ and $T_N$ intersect at a {\it
tetra-critical point}, below which a {\it uniform} AF/SC
mixed phase separates the two pure phases by two second
order transition lines. This type of phase diagram is realized for
$u_{12}<\sqrt{u_1 u_2}$.


In contrast to the conventional superconductors with a long
coherence length, the HTSC cuprates have a short coherence length
and a large Ginzburg region. Thus, they have the possibility of
observing non-trivial critical behaviors. An interesting point
concerns the symmetry at the multi-critical point where $T_N$ and
$T_c$ (or, more generally, $T_1$ and $T_2$) intersect.
At the multi-critical point defined by $r_1=r_2=0$, the critical
fluctuations of the order parameters couple to each other and
renormalize the coefficients of the fourth order terms $u_1$,
$u_2$ and $u_{12}$. There are several possible fixed points. The
symmetric fixed point, also known as the Heisenberg fixed point, is
characterized by $u^*_1=u^*_2=u^*_{12}$. The $O(N_1)\times O(N_2)$
symmetry is enhanced at this point to the higher $O(N_1+N_2)$
symmetry. Another fixed point, called the biconical tetra-critical
point in the literature, has non-vanishing values of $u_1^*$,
$u_2^*$ and $u_{12}^*$ at the fixed point which deviates from the
$O(N_1+N_2)$ symmetry. The third possible fixed point is the
decoupled fixed point, where $u_{12}^*=0$ and the two order
parameters decouple from each other at the fixed point. The
relative stability of these three fixed points can be studied
analytically and numerically. The general picture is that there
are two critical values, $N_c$ and $N_c'$. For $N_1+N_2<N_c$, the
symmetric bi-critical point is stable, for $N_c<N_1+N_2<N_c'$, the
biconical point is stable, while for $N_1+N_2>N_c'$, the decoupled
point becomes stable. The renormalization group calculations based
on the $4-\epsilon$ expansion\cite{KOSTERLITZ1976} places the
value of $N_c$ close to 4 and the value of $N_c'$ close to 11. The RG flow
diagram is shown in Fig. \ref{Fig_RG_flow} for the case of $N_1=3$
and $N_2=2$. Initially, all RG trajectories flow towards the
symmetric fixed point. The manner in which
the trajectories diverge close to the symmetric point 
depends on the values of the initial
parameters. The trajectories flow to the symmetric point when $u_{12}^2 = u_1
u_2$,  to the biconical point when $u_{12}^2< u_1 u_2$, and
flow outside of the regime of weak coupling RG analysis when
$u_{12}^2 > u_1 u_2$. In the case of competition between AF and
SC, $N=N_1+N_2=5$ is very close to $N_c$, leading to two important
consequences. First, the biconical point breaks the $SO(5)$
symmetry weakly. The value of the interaction parameters at the
biconical fixed point is given by $(u_1^*; u_2^*; u_{12}^*) = 2
\pi^2 \epsilon (0.0905; 0.0847; 0.0536)$. Extrapolating to
$\epsilon=1$ gives the root mean square deviation from the
symmetric $SO(5)$ point to about $26\%$, indicating weak $SO(5)$
symmetry breaking. The second consequence is that the critical
exponent associated with the flow away from the symmetric, $SO(5)$
point is extremely slow. The first loop $4-\epsilon$ expansion
gives the value of $1/13$ for the exponent associated with the
flow away from the symmetric point. To get an estimate of the
order of magnitude, we take the initial value of the scaling
variable taking the flow away from the $SO(5)$ fixed point to be
$0.04$. This value is obtained by considering the quantum
corrections associated with a projected $SO(5)$
model~\cite{ARRIGONI2000}. In this case, the significant deviation
away from the symmetric point can only be observed when the
reduced temperature is $t=(T-T_{bc})/T_{bc}\approx 10^{-11}$,
making the departure away from the $SO(5)$ symmetric point
practically unobservable. Indeed, numerical
simulations of the $SO(5)$ models presented in section
\ref{sec:numerics} are consistent with the $SO(5)$ symmetric
behavior in a wide range of temperatures and in very large systems. 
However, it should be noted that they do not prove the
ultimate stability of the symmetric point.


The question of the stability of the $SO(5)$ symmetric bicritical
point has been raised and discussed extensively in 
literature\cite{BURGESS1998A,ARRIGONI2000,MURAKAMI2000,HU2000,HU2001,AHARONY2002,JOSTINGMEIER2003,CALABRESE2003}.
Because the possible flow away from the bicritical point is
extremely slow, experimental and numerical observation of the
$SO(5)$ symmetric bi-critical behavior should be possible in a
wide range of temperatures, if the starting {\it microscopic}
parameters are already close to the symmetric point
$u_{12}=\sqrt{u_1 u_2}$. The $SO(5)$ symmetric bi-critical point
has a distinct set of critical exponents, summarized in Ref.
\cite{HU2000}, which can be distinguished experimentally  from the
usual $SO(3)$ and $U(1)$ behavior. In this sense, the
experimental observation of the bi-critical behavior would
demonstrate that the microscopic model of the HTSC cuprates is
close to the $SO(5)$ symmetry. In section
\ref{sec:experiment_phase_diagram} we shall discuss the analysis
of Murakami and Nagaosa\cite{MURAKAMI2000} showing
bi-critical scaling behavior in the $\kappa$-BEDT organic
superconductors. If the microscopic parameters are far from the
the symmetric point $u_{12}=\sqrt{u_1 u_2}$, other critical
behaviors could be observed. Aharony\cite{AHARONY2002} proposed
the decoupled tetra-critical fixed point with $u^*_{12}=0$. As
previously discussed , this critical point can  be observed in
experiments only if the microscopic value of $u^*_{12}$ is already
close to zero (due to the extremely slow flows of parameters). For
the HTSC cuprates, the AF vortex core experiments discussed in
section \ref{sec:AFvortex} clearly show that the AF and SC
order parameters are strongly and repulsively coupled with
$u^*_{12}>0$. Therefore, the decoupled fixed point is unlikely to
be relevant for these materials. However, this behavior could be
realized in some heavy fermion systems where different bands are
responsible for the AF and SC phases separately. Kivelson et
al\cite{KIVELSON2001} and Calabrese et al\cite{CALABRESE2003} also
considered the possibility of tri-critical points, where some of
the quartic terms $u_1, u_2, u_{12}$ become negative and the
sixth order terms become important. In this case, the phase
diagram could have topologies different from those listed here,
and the readers are referred to the more extensive discussions given
in Ref. \cite{KIVELSON2001}, especially Fig. 1c
and 1d of that reference. Negative values of the quartic
coefficient in the free energy (\ref{GL}) may come from the
runaway flows shown in Fig. \ref{Fig_RG_flow}. A multi-critical
point most closely related to the bi-critical point is the
biconical tetra-critical point. Its relevance to the HTSC cuprates
has been discussed in Ref.\cite{ZHANG1997,ZHANG2002}.

\subsection{Phase diagram of the quantum model}
\label{sec:quantum} Having discussed the finite temperature phase
diagram of the classical model, we now present the global phase
diagram of the quantum model at zero temperature. The quantum
phase transitions in the $SO(5)$ model were discussed in Fig. (1C)
and (1D) in reference \cite{ZHANG1997}. The quantum critical
behavior of the $SO(5)$ models have also been  studied extensively
in Ref. \cite{KOPEC2001,KOPEC2003,ZALCSKI2000,ZALESKI2000}. This
section extends the original analysis to include quantum
disordered states with inhomogeneous charge distributions. The
analysis carried out in this section is based on the bosonic
projected $SO(5)$ model, which bears great similarities to the
phase diagrams of the hardcore boson model studied extensively in
Ref.
\cite{FISHER1989,BRUDER1993,SCALETTAR1995,VANOTTERLO1995,PICH1998,HEBERT2002,BERNARDET2002}.
The iterative construction of the global phase diagram of the
$SO(5)$ model is also inspired by the global phase diagram of the
quantum Hall effect constructed by Kivelson, Lee and
Zhang\cite{KIVELSON1992}.

The projected $SO(5)$ model given in Eq. (\ref{pso5}) contains the
creation energy and the hopping process of the magnons and hole
pairs. The variational wave function for this model has the
general form given in (\ref{coherent}), with variational
parameters $\theta(x)$, $m_\alpha(x)$ and
$\Delta(x)=m_1(x)+im_5(x)$. The expectation value of the energy in
this state is given by
\begin{eqnarray}
& & \langle\Psi|H|\Psi\rangle = E(\theta(x),m_a(x)) \nonumber \\
&=& -\frac{J_s}{2} \sum_{xx';\alpha=2,3,4} sin2\theta(x)
sin2\theta(x') m_\alpha(x) m_\alpha(x')
- \frac{J_c}{4}\sum_{xx';i=1,5} sin2\theta(x) sin2\theta(x') m_i(x) m_i(x') \nonumber \\
& & + \Delta_s \sum_{x;\alpha=2,3,4} \sin^2\theta(x) m_\alpha^2(x)
+ \tilde \Delta_c \sum_{x;i=1,5} \sin^2\theta(x) m_i^2(x).
\label{mean_field_energy}
\end{eqnarray}
The variational minimum is taken with respect to the normalization
condition (\ref{unit_length}). In the regime when the quantum
fluctuations are small, $\theta(x)$ can be taken to be fixed and
uniform. In this case, the variational energy is nothing but the
energy functional of a classical, generally anisotropic $SO(5)$
rotor model, which has been  studied extensively
numerically\cite{HU2001}. At the point $J_c=2J_s$ and $\tilde
\Delta_c=\Delta_s$ in parameter space, this rotor model is $SO(5)$
symmetric at the classical level. However, unlike the classical
$SO(5)$ rotor, the projected $SO(5)$ model also contains quantum
fluctuations and quantum disordered phases. The phase diagram of
the projected $SO(5)$ model has been studied extensively by
Quantum Monte Carlo
simulations\cite{CHEN2003A,JOSTINGMEIER2003,DORNEICH2002A,RIERA2002,RIERA2002A},
and the results will be reviewed in details in section
\ref{sec:numerics}. When the quantum fluctuations are not strong
enough to destroy classical order, the general topology of the
phase diagram is similar to that classified in section
\ref{sec:classical}.

In this section, we discuss the regime when quantum fluctuations
are non-negligible and focus on the global phase diagram when the
classical order competes with the quantum disorder and uniform
states compete with non-uniform states. In Fig.
\ref{Fig_plaquette} and table II, we see that the classically
ordered states are obtained from the linear superpositions of
quantum disordered states. The quantum disordered states are
realized in the regime where the kinetic energy of the superspin
$\Delta_s$ and $\Delta_c$ overwhelms the coupling energy of the
superspin $J_s$ and $J_c$, and the superspin vector becomes
disordered in the temporal domain. In this sense, the quantum
description of the superspin goes far beyond the classical LG
theory discussed in the previous section.


By arranging the six elementary states from Fig.
\ref{Fig_plaquette} into a spatially non-uniform patterns, we have
infinitely many possibilities. In addition to the classically ordered AF
and SC states, in Fig. \ref{Fig_stripe_checker} 
we illustrate some of the basic non-uniform states
and their associated wave functions, expressed in terms of
$\theta(x)$, $m_\alpha(x)$ and $\Delta(x)$. 
Stripe order was theoretically predicted
and experimentally observed in the HTSC
cuprates\cite{ZAANEN1989,TRANQUADA1995,KIVELSON1998,WHITE1998}. In
a typical stripe phase, a magnetic stripe of width $2a$ is
separated by a charge stripe of width $2a$, where $a$ is the
lattice constant. The stripe state come in two forms. For the
in-phase stripes, both the charge and the spin periodicity is $4a$
in the direction transverse to the stripe direction. For the
out-of-phase stripes, the charge periodicity is $4a$, while the
spin periodicity is $8a$. The charge stripe can either be
insulating, or superconducting. The SC stripes are defined by their
phase angle; the two nearby SC stripes can be either in-phase or
out-of-phase. The case when both the AF and the stripes are out
of phase can be viewed as a {\it superspin spiral}, in which the
superspin direction rotates continuously along the direction
transverse to the stripes. (See Fig. \ref{Fig_stripe_checker}c).
Both types of stripes discussed here have both AF and SC orders.
Another possibility is the checkerboard pair-density-wave
(PDW)\cite{CHEN2002B}, depicted in Fig. \ref{Fig_stripe_checker}d.
It can be obtained from the in-phase stripe by quantum disordering
the hole pairs in the SC stripe. This state is insulating with AF
and charge orders. We stress that all insulating states in the
$SO(5)$ theory are obtained from the quantum disordered states of
the hole pairs; therefore, they are paired insulators, in contrast
to ordinary band insulators or a Wigner crystal state of the
electrons.

Therefore, some of the inhomogeneous states observed in
the HTSC cuprates can be  described naturally in terms of the
temporal and spatial ordering of the superspin. The key question
is how they are energetically stabilized in the projected $SO(5)$
model. These spatially non-uniform states are usually realized
when extended interactions are considered. These extended
interactions take the form
\begin{eqnarray}
& & H_{ext} = (V_{c}\sum_{\langle xx'\rangle}+V'_{c}\sum_{\langle
\langle xx'\rangle\rangle}) n_h(x) n_h(x') \nonumber \\
& & + (V_{s}(S_T)\sum_{\langle xx'\rangle}+V'_{s}(S_T)\sum_{\langle
\langle xx'\rangle\rangle})  \sum_{S_T=0,1,2}
(t^\dagger(x)t^\dagger(x'))_{S_T} (t(x)t(x'))_{S_T}
\nonumber \\
& & + J_\pi \sum_{\langle xx'\rangle} (t_\alpha^\dagger(x)t_\alpha(x') t_h^\dagger(x')t_h(x) + H.c.)
+V_\pi \sum_{\langle xx'\rangle} (n_h(x) n_t(x')+n_h(x')n_t(x)) +...
\label{extended}
\end{eqnarray}
Here $\langle xx'\rangle$ and $\langle\langle xx'\rangle\rangle$
denote the summation over the nearest-neighbor and the
next-nearest-neighbor on a square lattice. $(t(x)t(x'))_{S_T}$
refers to the total spin $S_T=0,1,2$ combinations of two magnons
on sites $\langle xx'\rangle$. The $V_c$ and $V'_c$ terms describe
the interaction of the hole pairs, the $V_s$ and $V'_s$ terms
describe the interaction of the magnons, and the $J_\pi$ and
$V_\pi$ terms describe the mutual interaction of the hole pair and
the magnon. Since the projected $SO(5)$ model is defined on a
coarse-grained lattice, the density of the hole pairs, $n_h$, is
related to the hole doping density by $n_h=2\delta$. The model
Hamiltonian given by $H+H_{ext}$ has been studied extensively by
Chen et al\cite{CHEN2003A} by using both quantum Monte Carlo
methods and mean field theory. Here we summarize the basic
qualitative results. In order to study the phase diagram of this
model, we first focus on the charge sector. The charge sector of
the projected $SO(5)$ model is the same as the hard-core boson
model introduced in Eq. (\ref{hardcore_boson}) of section
\ref{sec:spin-flop}. This model has been much studied in the
context of superfluid-to-insulator
transition\cite{BRUDER1993,SCALETTAR1995,VANOTTERLO1995,PICH1998,HEBERT2002,BERNARDET2002}.
Without the extended interactions $V(x,x')$ in Eq.
(\ref{Heisenberg}), the phase diagram of the hard-core boson is
given in Fig. \ref{Fig_spin_flop_phase_diagram}a. Half-filling of
the original electron systems in the cuprates corresponds to the
vacuum state of the hole-pairs, or phase III in Fig.
\ref{Fig_spin_flop_phase_diagram}a. The chemical potential $\mu$
induces a transition into the SC state, labelled as phase II.
Further increase of the chemical potential induces a transition
into a checker-board ordered state, labelled as phase I. This is
the ``class $B$" transition shown in Fig.
\ref{Fig_spin_flop_phase_diagram}a. Phase I corresponds to
$n_h=1/2$ of the hole pair bosons, or $\delta=1/4$ of the original
electrons. When extended interactions in $V(x,x')$ are included, a
new insulating phase develops near the overlapping region of phase
I and phase III, with boson density of $n_h=1/4$ of the hole pair
bosons, or $\delta=1/8$ of the original electrons (see, for instance,
Fig. 2 of Ref. \cite{BRUDER1993}). This insulating phase can have
either stripe or checkerboard like charge order. Generally, stripe
type of insulating order is favored for $V'_{c}\gg V_{c}$, and the
checkerboard-type order is favored in the opposite
limit\cite{PICH1998,HEBERT2002}. With even more extended
interactions, additional phases develop at lower {\it rational}
densities. These Mott insulating phases at various rational
densities are shown in Fig. \ref{Fig_global}. The phase boundary
between the insulating phases with charge order and the SC phases
can be generally classified into ``type 1, 1.5 and 2," according
to the terminology developed in section \ref{sec:classical} and
Fig. \ref{Fig_E_delta}. In the last two cases, a mixed phase, called
the supersolid phase, develops near the phase boundary.
After understanding the generic phase diagram of the hard-core
lattice boson model, we are now in a position to discuss the full
global phase diagram of the $SO(5)$ model $H+H_{ext}$, depicted in
Fig. \ref{Fig_global}. Here $J/V$ denote the typical ratio of
$J_c/V_c$, but it can obviously be replaced by other similar
parameters. The $n_h=0$ phase corresponds to the AF state at
half-filling, where magnons condense into the singlet ground
state. For large values of $J/V$, a pure SC state is obtained
where the hole pairs condense into the singlet ground state.
However, besides these two robust, classically ordered phases, we
also see new insulating phases at $n_h=1/4$, $n_h=1/8$ and $n_h=3/8$,
which correspond to $\delta=1/8$, $\delta=1/16$ and $\delta=3/16$ in
the real system. These new insulating states are stabilized by the
extended interactions and have both AF and PDW order (see
example Fig. \ref{Fig_stripe_checker}d). As the chemical potential
or the doping level is varied, a given system
traces out a one dimensional slice in this phase diagram,
with typical slices $B1$, $B2$ and $B3$ depicted in Fig.
\ref{Fig_global}(we expect the quantum parameter $J/V$ to be independent
of $\mu$ for a given family of materials). 
The nature of the phase transition $B1$ is
similar to that of the classical model already discussed in
section \ref{sec:classical}. In this case, the phase transition
from the AF to SC state can be further classified into ``types
1, 1.5 and 2," as discussed in section \ref{sec:classical}, with
the two latter cases leading to an AF/SC mixed phase at the phase
transition boundary. For lower values of $J/V$, the trace $B3$
encounters the $\delta=1/8$ insulating phase. The key signature of
this type of phase transition is that the SC $T_c$ will display a
pronounced minimum as the doping variation traces through the
$\delta=1/8$ insulating state. At the same time, the AF ordering
(possibly at a wave vector shifted from $(\pi,\pi)$) will show
reentrant behavior as doping is varied. The phase transition
around the fractional insulating phases can again be classified
into types ``1, 1.5 and 2," with possible AF/SC, AF/PDW, SC/PDW
and AF/PDW/SC mix phases.

So far we have classified all quantum phase transitions in the
$SO(5)$ models according to two broad classes. ``Class $A$"
describes transitions at a fixed chemical potential, typically at an
effectively particle-hole symmetric point. ``Class $B$" describes
transitions in which the chemical potential or the density is varied.
Each broad class is further classified into three ``types, 1, 2
and 1.5," depending on whether the transition is a direct first
order, two second order, or an intermediate symmetric point in
between. The full quantum $SO(5)$ symmetry can only be realized in
the ``class $A$ type 1.5" quantum phase transition. The Heisenberg
point in the hard-core boson problem discussed in section
\ref{sec:spin-flop} is one such example. The $g=0$ point in the
$SO(5)$ quantum non-linear $\sigma$ model (\ref{Hsigma}) is
another example. On the other hand, the static, or projected
$SO(5)$ symmetry can be realized in ``class $B$ type 1.5"
transitions. We believe that the AF to SC transitions in the YBCO,
BCCO and  NCCO systems correspond to ``class $B1$" transition.
These systems only have an AF to SC transition, which can be
further classified as  types ``1, 1.5 and 2," but they do not
encounter additional statically ordered fractional insulating
phases. On the other hand, the phase transition in the LSCO
system, where $T_c$ displays a pronounced dip at $\delta=1/8$,
corresponds to the ``class B3" transition (see Fig. \ref{Fig_global}). 
In the HTSC cuprates,
the charge gap at half-filling is very large, of the order of
$U\sim 6 eV$; it is not possible to induce the ``class A1"
transition from the AF to SC state by conventional means.
However, the charge gap at the fractional insulating states is
much smaller, of the order of $J$, and it is possible to induce
the ``class A2" insulator to superconductor transition by applying
pressure\cite{LOCQUET1998,SATO2000,ARUMUGAM2002,Takeshita2003}. It
would be interesting to determine if this transition point could have
the full quantum $SO(5)$ symmetry.

Therefore, we see that the concept of the $SO(5)$ superspin indeed
gives a simple and unified organizational principle to understand
the rich phase diagram of the cuprates and other related systems.
This construction of the global phase diagram can obviously be
iterated {\it ad infinitum} to give a beautiful fractal structure
of self-similar phases and phase transitions. All of this
complexity can be simply reduced to the five elementary quantum
states of the superspin.

\subsection{Numerical simulations of the classical and quantum models}
\label{sec:numerics} In this section, we review 
essentially exact numerical studies of the classical $SO(5)$ model and the
quantum projected $SO(5)$ model on a
lattice\cite{CHEN2003A,JOSTINGMEIER2003,DORNEICH2002A,HU1999,HU2001,RIERA2002,RIERA2002A}.
In section \ref{sec:CORE} we shall discuss the transformation from
the microscopic models into the effective $SO(5)$ models and
determine the effective parameters. Once this is accomplished, the
phase diagram of the model can be determined reliably by bosonic
QMC simulations. These calculations can be carried out for system
sizes up to two orders of magnitude larger than fermionic QMC
simulations, with the latter being plagued, in the physically
interesting regime -- i.~e., close to half-filling -- by the
minus-sign problem \cite{VONDERLINDEN1992}. The effective models
can also be studied numerically in 3D; this is crucial, since
there exists no AF ordered phase in 2D at finite temperature (nor
long-range SC order). Thus, we are forced to study the 3D case
in order to determine the phase diagram and to show that the
scaling behavior is consistent with an $SO(5)$ symmetric critical
behavior within the parameter regime studied (temperature and
system size).  This was possible due to a major step forward in
the numerically accessible system sizes
\cite{SANDVIK1997,SANDVIK1999,DORNEICH2001}: in the bosonic
projected $SO(5)$ model $\sim$10.000 sites were included, in
contrast to just $\sim$100 sites in fermionic QMC calculations
\cite{DOPF1992,IMADA1998,DAGOTTO1994}. The numerical results,
obtained by the QMC technique of Stochastic Series Expansion (SSE)
\cite{SANDVIK1997,SANDVIK1999} and reviewed here, show that the
projected $SO(5)$ model can give a realistic description of the
global phase diagram of the HTSC cuprates and accounts for many of
their physical properties.

The form of the projected $SO(5)$ Hamiltonian is given in Eq.
(\ref{pso5}). The extended $SO(5)$ model also includes the
interactions expressed in Eq. (\ref{extended}). We shall discuss
the simple $SO(5)$ model first. In Ref.~\cite{ZHANG1999}, this
Hamiltonian was studied analytically within a mean-field approach.
At the special point $J_c=2J_s\equiv J$ and
$\Delta_s=\tilde\Delta_c$, the mean-field level of the ground-state
energy of Hamiltonian (~\ref{pso5}) depends on the AF and SC order
parameters $x=\langle t^{\dag}_x\rangle$ and $y=\langle
t^{\dag}_h\rangle$ only via their combination $x^2+y^2$, which
reflects the $SO(5)$ invariance of the mean-field approximation.
In the full model, however, quantum fluctuations modify the zero-point
energy of the bosons in Eq.~(~\ref{pso5}); therefore, giving a
correction to the ground-state energy which depends on $x$ and $y$
separately and destroys $SO(5)$ symmetry~\cite{ZHANG1999}. Hence
it is essential to study the full quantum-mechanical model
(\ref{pso5}), including all quantum fluctuations, which can only be
done by means of numerical simulations. We then compare the
properties of the projected $SO(5)$ model first in two dimensions
(2D) with a variety of salient features of the HTSC such as the
global phase diagram and the neutron-scattering resonance.
Finally, we review an extension of these studies to the 3D
projected $SO(5)$ model. In particular, we show that the scaling
behavior near the multi-critical point, within the parameter regime
studied (system size and temperature), is consistent with an
$SO(5)$ symmetrical behavior. The departure away from $SO(5)$
symmetric scaling can only occur in a narrow parameter regime which
is hardly accessible either experimentally or numerically.

After numerically solving the projected $SO(5)$ model, we obtain
Fig. \ref{Fig_2D_pso5}a, which gives the mean hole-pair and magnon
densities as a function of the chemical potential for
$T/J\!=\!0.03$ and their $T\!\to\!0$ extrapolations
\cite{JOSTINGMEIER2003}.  Similar to the mean-field results, a
jump in the densities can be clearly seen at $\mu_c\!=\!  -0.175$,
with a shift in respect to the mean-field value due to the
stronger fluctuations of hole pairs, as seen in the Gaussian
contributions \cite{ZHANG1999}. The nature of the phase transition
at $\mu\!=\!\mu_c$ can be determined by studying histograms of the
hole-pair distribution for fixed $\mu\!=\!\mu_c$. While in a
homogeneous phase the density peaks at its mean value, at
$\mu\!=\!\mu_c$ we obtain two peaks, which indicates a first-order
transition with a phase separation between (almost) hole-free
regions and regions with high hole-pair density.  From Fig.
\ref{Fig_2D_pso5}b we see that the transition is of first order for
$T\!<\!T_P\!=\!(0.20\pm 0.01)J$ at $\mu\!=\!\mu_P\!=\!(-0.168\pm
0.002)J$.  Above $T_P$, the histograms show strongly fluctuating
hole-pair densities, suggesting the presence of critical behavior.

\begin{figure}[hp]
\begin{center}
        \includegraphics[scale=1.2]{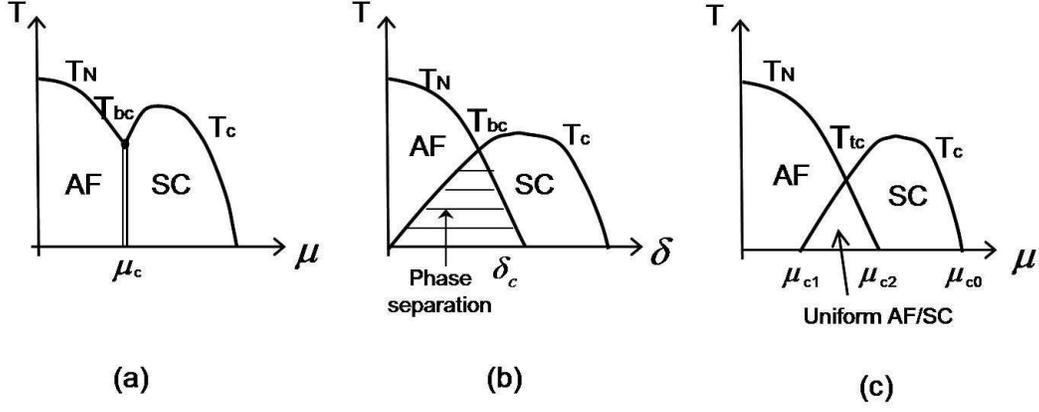}
        \caption{The finite temperature phase diagram in D=3 for the class
        ``B1" transition shown in Fig. \ref{Fig_global}. (a) and (b), correspond
        to a direct first order phase transition between AF and SC, as a function
        of the chemical potential and doping, respectively. This type of transition
        is classified as the ``type 1" transition. (c) corresponds to two
        second order phase transitions with a uniform AF/SC mix phase in
        between. This type of transition is classified as ``type 2"
        transition. The AF and  SC transition temperatures $T_N$ and $T_c$ 
merge into either
        a bi-critical $T_{bc}$ or a tetra-critical point $T_{tc}$.} \label{Fig_T_mu}
\end{center}
\end{figure}
\begin{figure}[hp]
\begin{center}
        \includegraphics[scale=0.6]{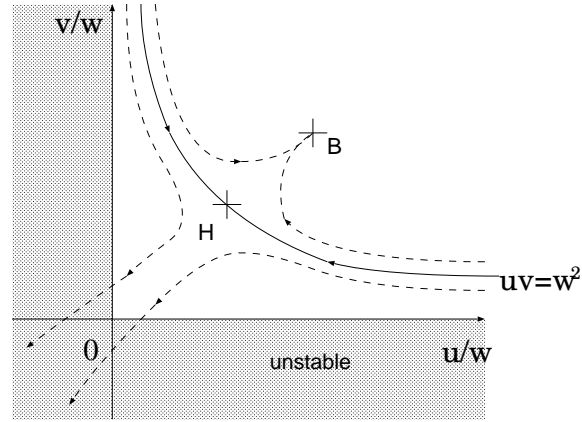}
        \caption{
Renormalization group flow in the $(u_1/u_{12}, u_2/u_{12})$
plane. (In this figure, $u=u_1$, $v=u_2$ and $w=u_{12}$.) The
renormalization group flow is initially attracted towards the
symmetric Heisenberg point labelled by H. The RG trajectories
diverge near the Heisenberg model, with a very small exponent.
Reproduced from Ref. \cite{MURAKAMI2000}.} \label{Fig_RG_flow}
\end{center}
\end{figure}
\begin{figure}[hp]
\begin{center}
        \includegraphics[scale=1.0]{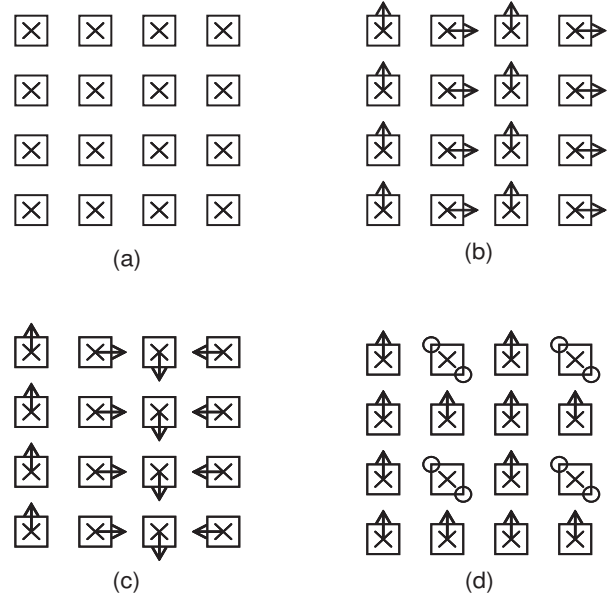}
        \caption{Some possible ground states of the projected $SO(5)$ model
 (see also Fig. \ref{Fig_plaquette}). The cross depicts an RVB like spin singlet state
 on a plaquette, the arrow denotes the direction of the superspin, and the open circles depict
 hole pairs. (a)
        The plaquette RVB state is described by $\theta(x)=0$ on every
        plaquette. (b) The in-phase SC stripe with
        $\alpha(x)=0,\pi/2,0,\pi/2$ on each stripe. (c) The superspin
        spiral with $\alpha(x)=0,\pi/2,\pi,3\pi/2$ on each stripe. (d)
        The hole pair checkerboard state with $\alpha(x)=0$
        everywhere, except on the hole pair plaquettes, where
        $\theta=\pi/2$ and $\alpha=\pi/2$.}
\label{Fig_stripe_checker}
\end{center}
\end{figure}

Based on these results, the phase diagram of the 2D projected
$SO(5)$ model is obtained in Fig. \ref{Fig_tricritical}. Unlike
the generic three dimensional phase diagrams presented in Fig.
\ref{Fig_T_mu}, there can be no finite temperature Neel transition
in D=2 because of the Mermin-Wagner theorem. On the other hand, a
continuous transition of the Kosterlitz-Thouless (KT) type is
possible for the SC to normal state transition at finite
temperature. The 3D phase diagram shown in Fig. \ref{Fig_T_mu}a
takes the form of Fig. \ref{Fig_tricritical} in D=2, where the
first order line separating the AF and  SC phases merges into
the continuous KT transition at a tricritical point P. The SC
phase with finite superfluid density $\rho_s$ is identified by a
power-law decay of the SC correlation function:
\[
C_h{(r)} = \big(t_h^\dagger{(r)} + t_h^{}{(r)}\big)
\big(t_h^\dagger{(0)} + t_h^{}{(0)}\big) \;.
\]
The KT transition line in Fig.~\ref{Fig_tricritical}a separates
power-law ($C_h{(r)} \propto r^{-\alpha}$) from rapid exponential
decay ($C_h{(r)} \propto \mbox{e}^{-\lambda r}$).  A reliable and
accurate distinction between these two decay behaviors requires a
finite-size scaling with large system sizes, as well as an
efficient QMC estimator for the Green functions appearing in the
correlation function.  With its non-local update scheme and with
our new estimators for arbitrary Green functions, SSE provides
both (for details see Ref.~\cite{DORNEICH2001}). An alternative
method for detecting a KT transition exploits the fact that the
superfluid density jumps from zero to a finite value at the KT
temperature $T_{KT}$\cite{NELSON1977}.  Within SSE the superfluid
density can be measured quite easily by counting winding
numbers\cite{HARADA1997}.
Numerically, this criterion is preferable to the arduous 
process of direct determination of decay
coefficients. Fig.~\ref{Fig_tricritical}a plots the phase diagram
obtained by applying both criteria independently.  The figure
shows that the projected $SO(5)$ model indeed has a KT phase with
quasi long-range order whose dome-like form in $\mu$-$T$ space
looks like that of the HTSC cuprates.  Both criteria 
produce the same clearly pronounced phase separation line.  It is
well known that a similar transition cannot occur for
antiferromagnets\cite{CHAKRAVARTY1988} and that the finite-$T$ AF
correlation length $\xi$ is always finite and behaves like $ \xi
\propto \mbox{e}^{2\pi\rho_s/k_B T}$, with $\rho_s$ being the spin
stiffness. This fact is confirmed by our numerical results.


One  condition required for an $SO(5)$ symmetric point is that the
formation energies of hole-pair bosons and of magnons are
identical. This condition is fulfilled along the line from $S$ to
the tricritical point $P$ in Fig. \ref{Fig_tricritical}.  Another
necessary condition is that hole pairs and magnons  behave in the
same way at long distances.  This condition is fulfilled on the
dashed line in Fig. \ref{Fig_tricritical}, where the AF and SC
correlation lengths $\xi$ become equal.  Interestingly, these two
conditions are met (within error bar accuracy) at the tricritical
point P.  Of course, the correlation length is still finite here;
however, we find relatively large $\xi$ values of order 10 to 15
in the immediate vicinity of point P, demonstrating the importance
of $SO(5)$ critical fluctuations in this region.

In addition, in realistic electron systems, the long-range part of
the Coulomb repulsion between the doubly-charged hole pairs
disfavors phase separation, while extended short ranged
interactions described by Eq. (\ref{extended}) could lead to the
formation of stripes and checkerboard types of states, as discussed
in section \ref{sec:quantum}. To study the effect of off-site
Coulomb interaction, we have added additional nearest-neighbor and
next-nearest-neighbor Coulomb repulsions $V_c$ and
$V_c'\!=\!0.67\!\;V_c$ to the projected $SO(5)$ model. Indeed, a
relatively modest Coulomb repulsion of $V_c/J\!\approx \! 0.2$ is
enough to completely destroy the phase separation.  Thus, one
interesting effect of Coulomb interaction in two dimensions is
to push down the tricritical point into a quantum-critical
point at $T\!=\!0$. In section \ref{sec:classical} and in Fig.
\ref{Fig_E_delta}c, we showed that the $SO(5)$ symmetric behavior
is recovered at the special point when a direct first order
transition changes into two second order transitions. Therefore,
the extended Coulomb interaction plays the role of the $w$
parameter in Fig. \ref{Fig_E_delta}c and could restore the
$SO(5)$ symmetry at the quantum critical point.

When larger values of extended interaction parameters in Eq.
(\ref{extended}) are considered, new insulating phases are
expected, following from the general discussions in section
\ref{sec:quantum} and Fig. \ref{Fig_global}. Indeed, Chen et al
\cite{CHEN2003A} have performed extensive QMC simulation of the
$SO(5)$ model and have determined its generic phase diagram, as 
shown in Fig. \ref{FIG-Phase-QMC}. In addition to the AF and SC
phases, there is an insulating pair-density-wave state around
doping range of $x=1/8$, where hole pairs form a checkerboard
state in the AF ordered background, as depicted in Fig.
\ref{Fig_stripe_checker}d. Near the phase boundaries between the
AF, PDW and SC phases, there are mixed phases with
coexisting order. The topology of the phase diagram obtained from
the QMC simulation agrees well with the mean field theory of the
extended $SO(5)$ model.
One of the main features of the $SO(5)$ theory is that it provides an
elegant explanation for the neutron resonance peak observed in
some HTSC cuprates at
$q\!=\!(\pi,\pi)$\cite{DEMLER1995,ZHANG1997}. We refer the reader
to the detailed discussion of the resonance mode in section
\ref{sec:pi}. Experiments show that the resonance energy
$\omega_{\mbox{\scriptsize res}}$ is an increasing function of
$T_c$, i.e.\ $\omega_{\mbox{\scriptsize res}}$ increases as a
function of doping in the underdoped region and decreases in the
overdoped region\cite{FONG2000}.  Fig. \ref{Fig_tricritical}b
plots the resonance frequency determined from the spin correlation
spectrum obtained for the projected $SO(5)$ model. As illustrated
in Fig. \ref{Fig_pSO5Excitations}, the spin-wave excitations are
massless Goldstone modes in the AF phase at $\mu\!<\!\mu_c$ (and
$T\!=\!0$) and become massive when entering into the SC phase.
$\omega_{\mbox{\scriptsize res}}$ increases monotonically up to the 
optimal doping $\mu_{\mbox{\scriptsize opt}}\!\approx\! 1$. In the
overdoped range of the simple $SO(5)$ model, however,
$\omega_{\mbox{\scriptsize res}}$ is increasing more, in
contrast to what happens in the cuprates. The resonance peak
continuously loses weight as $\mu$ increases, which is
consistent with experimental observations \cite{FONG2000}.

A comparison of the critical temperature $T_c$ obtained from
Fig.~\ref{Fig_tricritical} and the resonance frequency
$\omega_{\mbox{\scriptsize res}}$ at optimal doping yields the
ratio $T_c/\omega_{\mbox{\scriptsize res,opt}}\!=\!0.23$.  Again, this is
in qualitative accordance with the corresponding ratio for
$\mathrm{YBa_2Cu_3O_{6+x}}$, for which the experimentally
determined values $T_c\!=\!93\,\mbox{K}$ (thus $k_B T_C\!=\!8.02$
meV) and $\omega_{\mbox{\scriptsize res,opt}}\!=\!41$ meV yield
$T_c/\omega_{\mbox{\scriptsize res,opt}}\!=\!0.20$.




Now we turn to the numerical simulations of the $SO(5)$ models in
D=3. Two aims motivate our studies of the projected $SO(5)$ model
in three dimensions (3D). First, we expect to find an AF and  SC
phase with real long-range order. 
We need to determine which of the two types
of phase diagrams introduced in section \ref{sec:classical}
(see Fig. \ref{Fig_T_mu}) is realized in the numerical
simulations. Second, we would like to determine whether
the projected $SO(5)$ model has a certain multi-critical point at
which the $SO(5)$ symmetry is asymptotically restored. Since the
cuprates have a pronounced 2D layer structure with relatively weak
couplings between adjacent CuO$_2$ planes, the 2D and the
isotropic 3D model (discussed here) should be two extreme poles
for the possible range of properties of real HTSC materials. Most
numerical data reviewed here have been obtained by a
finite-size scaling with lattice sites up to 10,000.
\cite{DORNEICH2001,JOSTINGMEIER2003,DORNEICH2002A,DORNEICH2002B}




The phase diagram and the scaling behavior of the classical
$SO(5)$ model has been studied in detail by Hu in Ref.\
\cite{HU2001} by means of classical Monte Carlo simulations (MC).
Classical MC are by orders of magnitude easier to perform and less
resource demanding than QMC simulations; hence, very large system
sizes can be simulated and highly accurate data can be obtained. The
classical $SO(5)$ model can be obtained directly from the quantum
$SO(5)$ model by taking the expectation value of the Hamiltonian
in the variational state, as given by Eq.
(\ref{mean_field_energy}) and assuming a constant value of
$\theta(x)$. It takes the form:
\begin{equation}
\label{eq_H_Hu}
  H = -J \sum_{\langle x,x' \rangle} m_a(x) m_a(x') + g \sum_x m^2_\alpha(x)
     + w \sum_x m^2_\alpha(x) m^2_i(x),
\end{equation}
where $g=\Delta_s-\tilde\Delta_c$ is the quadratic symmetry
breaking term, and $w$ is an additional quartic symmetry breaking
term. Hu established the $T(g)$ phase diagram, which is of the
type illustrated in Fig. \ref{Fig_T_mu}a. The model has an AF and
SC phase which meet at a bicritical point
$(T_{bc},g_{bc}\!=\!0)$. The boundary lines between the
disordered and AF phases, and between the disordered and
SC phases merge tangentially at the bicritical point, which
is an important characteristics of $SO(5)$ symmetry \cite{HU2001}.
The following scaling properties were determined by Hu and will
be used to study the restoration of $SO(5)$ symmetry in the
projected $SO(5)$ model.

For an analysis of the crossover phenomenon, an Ansatz for the
behavior of the helicity modulus $\helic$ in the range
$T\!<\!T_c(g)$ and $g\!>\!0$ is used, which is suggested by
scaling theory\cite{HU2001}:
\begin{equation}
\label{scal_ansatz}
    \helic(T,g)\propto (g-g_{bc})_{}^{\nu_5/\phi} \times
    f\big( (T/T_{bc}-1) \big/ (g-g_{bc})_{}^{1/\phi} \big)\,.
\end{equation}
Here, $\nu_5$ is the {\em critical exponent for correlation
length} at $n\!=\!5$ and $\phi$ the {\em crossover exponent}.
Using (\ref{scal_ansatz}), the values of $\nu_5$ and $\phi$ can be
determined in two steps. First, performing a $g$ scan of $\helic
(T\!=\!T_{bc},g)$ returns the ratio $\nu_5/\phi$:
\begin{equation}
\label{eq_nu5_div_phi} \helic(T_{bc},g) / \helic(T_{bc},g')
   = \big( (g-g_{bc})/(g'-g_{bc}) \big)^{\nu_5/\phi}\,.
\end{equation}
Then, $\phi$ is obtained from the slopes $\frac{\partial}{\partial
T}(\helic \big(T,g)/\helic (T,g') \big)$ via
\begin{equation}
  \label{eq_phi}
  \phi = \ln\Big(\frac{g_2-g_{bc}}{g_1-g_{bc}} \Big) \bigg/
    \ln\bigg( \frac{\partial}{\partial T}\frac{\helic(T,g_1)}
    {\helic(T,g_1')} \Big|_{T=T_{bc}} \Big/
    \frac{\partial}{\partial T}\frac{\helic(T,g_2)}
    {\helic(T,g_2')} \Big|_{T=T_{bc}} \bigg)\,
  \end{equation}
if $g_1$, $g_1'$, $g_2$, and $g_2'$ are related by
$(g_1-g_{bc})/(g_1'-g_{bc}) \!=\! (g_2-g_{bc})/(g_2'-g_{bc}) > 0$.
From the scaling plots presented in Fig.\ref{Fig_Hu}, Hu finds the
values $\nu_5/\phi\!=\!0.523\pm 0.002$ and $\phi\!=\!1.387\pm
0.030$.

According to the scaling Ansatz in (\ref{scal_ansatz}), 
the transition lines between the
disordered and AF phases, and between the disordered and
SC phases near the  bicritical point should be of the form
\begin{equation}
 \label{eq_B2_B3}
 B_2\cdot (g-g_{bc})^{1/\phi} = \frac{T_c(g)}{T_{bc}}-1
   \quad\;\mbox{and}\quad\;
 B_3\cdot (g_{bc}-g)^{1/\phi} = \frac{T_N(g)}{T_{bc}}-1\,.
 \end{equation}
The ratio $B_2/B_3$ should be given by the inverse ratio between
the AF and SC degrees of freedom, i.e.
\begin{equation}
 \label{eq_B2_div_B3}
 B_2/B_3\!=\!3/2\,.
\end{equation}
The values numerically determined by Hu indeed have the correct
ratio: $B_2\!=\!1/4$ and $B_3\!=\!1/6$.




We now proceed to the phase diagram of the 3D quantum $SO(5)$
model \cite{DORNEICH2002A,DORNEICH2002B}. Figure \ref{Fig_3D_pso5} shows 
the AF and SC
phases, as expected. Furthermore, the two phase transition lines
merge tangentially into a multi-critical point (at
$T_{bc}\!=0.960\pm 0.005$ and $g_{bc}\!=\!-0.098 \pm 0.001$) just
as in the classical $SO(5)$ system \cite{HU2001}. The line of
equal correlation decay of hole-pairs and triplet bosons also
merges into this bicritical point $P$ -- a necessary condition 
at this point for the restoration of $SO(5)$ symmetry.  Unlike the
corresponding phase in the classical model \cite{HU2001}, the SC
phase only extends over a finite $g$ range due to the
hardcore constraint of the hole-pair bosons and agrees with
experimentally determined phase diagrams of the cuprates.
Obviously, the quantum mechanical $SO(5)$ model is `more physical'
in this aspect than the classical $SO(5)$ model. In real cuprates
the ratio between the maximum temperatures $T_c$ and $T_N$ is
about 0.17 to 0.25, whereas in the projected $SO(5)$ model we
obtain the values $T_c/J\!=\!1.465\pm 0.008$ at
$\mu_{\mathit{opt}}/J\!\approx\!1.7$ and $T_N/J\!=1.29\pm 0.01$ at
$\mu\to\-\infty$; hence, $T_c$ is slightly larger than $T_N$. In
order to obtain realistic values for the transition temperatures,
it is necessary to include the $J_\pi$ and $V_\pi$ terms in
Eq. (\ref{extended}). These terms represent the repulsion between
the magnons and the hole pairs. If we take the expectation values
of the hole pair operators, these terms effectively represent a
doping dependent $J_s$, which can produce a more realistic phase
diagram.  Such terms break the $SO(5)$ symmetry of the static
potential at T=0 (see Section \ref{sec:so5bosons} after equation
(\ref{boson})). However, the static symmetry may still be
recovered at the bicritical point, as discussed in Section
\ref{sec:classical}. At this point we are primarily concerned with the
multi-critical behavior, so we stay with the simple $SO(5)$ model.

A closer look at the phase transition line between the points $S$
and $P$ (see Fig.\ \ref{Fig_3D_pso5}) reveals that this line is
slightly inclined, unlike the vertical line seen in the classical $SO(5)$ model. 
This indicates that a finite latent heat is connected
with the AF-SC phase transition.  In addition, this means that  
$\mu$ is not a scaling variable
for the bicritical point $P$, as it is in the classical model. 
The result in Fig.\ref{Fig_3D_pso5}
shows a phase separation regime at $\mu=\mu_c$ on the entire
transition line from S to P.



We now review the results of a scaling analysis for the 3D quantum
$SO(5)$ model, similar to the one performed by Hu \cite{HU2001} in
a classical $SO(5)$ system \cite{DORNEICH2002B}. From
this analysis we also find that the $SO(5)$ symmetry is
restored in the region around the bicritical point
($T_{bc}\!=\!0.96$, $\mu\!=\!-0.098$).


We have determined the critical exponents for the onset of AF and
SC orders for various chemical
potentials as a function of temperature.  Far into the SC range, 
at $\mu\!=\!1.5$, we find that the SC helicity modulus
follows the scaling form~\cite{FISHER1973}
\[
\helic \propto (1-T/T_c)^{\nu}\quad\;\mbox{with}\quad\;\nu=0.66\pm
0.02\,,
\]
which agrees with the values obtained by both the
$\epsilon$-expansion and numerical analysis of a 3D XY model.
On the AF side, error bars are larger. For $\mu=-2.25$,
\[
{C_{AF}(\infty)} \propto (1-T/T_c)^{\beta_3} \quad\;
\mbox{with}\quad\;\beta_3 = 0.35 \pm 0.03,
\]
as expected for a 3D classical Heisenberg model.

To determine $\nu$ and $\phi$, we use Eqs.
(\ref{eq_nu5_div_phi}) and (\ref{eq_phi}), which express the scaling
behavior in the crossover regime (cf. Ref.~\cite{HU2001}).  
We obtain the ratio
\[
\nu_5/\phi = 0.52 \pm 0.01,
\]
which matches the results of the
$\epsilon$-expansion\cite{KOSTERLITZ1976,HU2000}. 
$\phi$ is then obtained by using
(\ref{eq_phi}). The result is
\[
\phi = 1.43 \pm 0.05\,
\]
which also agrees with the $\epsilon$-expansion for
an $SO(5)$ bicritical point and with the results of
Ref.~\cite{HU2001}.

Let us finally return to the comment by Aharony
\cite{AHARONY2002}, who, via a rigorous argument,
demonstrated that the decoupled fixed point is stable,
as opposed to the biconical and $SO(5)$ fixed points. 
However, he also commented
that the unstable flow is extremely slow for the $SO(5)$ case due
to the small crossover exponent.

The scaling analysis of the 3D projected $SO(5)$ model has
produced a crossover exponent which matches 
the value obtained from a classical $SO(5)$ model and
from the $\epsilon$-expansion.  This provides strong evidence
that the static correlation functions at the $SO(5)$ multicritical
point are controlled by a fully $SO(5)$ symmetric point, at least
in a large transient region.  However,
the isotropic $SO(5)$ and biconical fixed points  have very similar critical
exponents. Thus, given the statistical and finite-size errors, as well as
the errors due to the extrapolation of the $\epsilon$-expansion
value to $\epsilon=1$, we cannot exclude the possibility
that the multicritical point on the phase diagram
is actually  the biconical one.  On the other hand, the
biconical fixed point should be accompanied by a uniform AF/SC mixed
region (as a function of chemical potential), which was not
observed. The decoupled fixed point appears to be the least
compatible with the numerical results presented above. Even if the
bicritical point is fundamentally unstable, as suggested  by
Aharony in \cite{AHARONY2002}, one would have to come
unrealistically close to $T_{bc}$ to observe this. For example,
for the projected $SO(5)$ models Ref. \cite{ARRIGONI1999}
estimated that deviations from the $SO(5)$ behavior may  be
observed only when the reduced temperature becomes smaller than
$|T-T_{bc}|/T_{bc}<10^{-11}$. On the other hand, the other scaling
variables, although initially of the order of $1$, rapidly
scale to zero due to the large, negative exponents.  Therefore,
the $SO(5)$ regime starts to become important as soon as the AF
and SC correlation lengths become large and basically continues to affect
the scaling behavior of the system in the whole
accessible region.

Summarizing, the accurate QMC calculations show that the projected
$SO(5)$ model which combines the idea of $SO(5)$ symmetry with a
{\it realistic} treatment of the Hubbard gap, is characterized by
an $SO(5)$ symmetric bicritical point, at least within a large
transient region. Possible flow away from this symmetric fix point
occurs only within a narrow region in reduced temperature, making
it impossible to observe either experimentally or numerically. 
This
situation is common to many systems in condensed
matter physics. For example, due to the well-known Kohn-Luttinger
effect~\cite{KOHN1965}, the Fermi-liquid fixed point is always
unstable towards a SC state.  However, this effect is
experimentally irrelevant for most metals since it only works at
extremely low temperatures. Another example is the ``ordinary''
superconductor to normal-state transition at $T_c$.  Strictly
speaking, due to the coupling to the electromagnetic field this
fixed point is always unstable \cite{HALPERIN1974}.  However, this
effect is experimentally irrelevant since the associated critical
region is extremely small.  Similarly, irrespective of the
question of ultimate stability, the $SO(5)$ fixed point is a robust
one in a large transient regime, and it can control the
physics near the AF and SC transitions. For all practical purposes,
the multi-critical point is dominated by the initial flow toward
the $SO(5)$ symmetric behavior.

\section{MICROSCOPIC ORIGIN OF THE SO(5) SYMMETRY}
\label{sec:microscopic}

\subsection{Quantum lattice models with exact SO(5) symmetry}
\label{sec:exact} Soon after the general $SO(5)$ theory was
proposed, a class of microscopic fermion models with exact $SO(5)$
symmetry was
constructed\cite{RABELLO1998,HENLEY1998,BURGESS1998A,SCALAPINO1998,WU2003A}.
These models fall into three general classes. The first class
contains models with two sites per unit cell, such as the ladder
and the bi-layer models. In these models, a simple condition among
the local interaction parameters ensures the full quantum $SO(5)$
symmetry. The second class contains models with only one site in
the unit cell but with longer ranged interactions. The third
class contains higher spin fermion models, in particular the spin
$3/2$ Hubbard model. Remarkably, in this case the models are
always $SO(5)$ symmetric without any fine tuning of the local
interaction parameters and doping level.

The microscopic $SO(5)$ symmetric models in the ladder or
bi-layer models were first constructed by Scalapino, Zhang and
Hanke (SZH)\cite{SCALAPINO1998} and have been studied
extensively both analytically and numerically
\cite{LIN1998,ARRIGONI1999,FURUSAKI1999,Schulz1998,SHELTON1998,BOUWKNEGT1999,DUFFY1998,EDER1999,FRAHM2001,HONG1999}.
In these models, there are two sites and $4^2=16$ states in the
unit cell. In section \ref{sec:so5group}, we already discussed the
construction of $SO(5)$ symmetry operators in terms of the fermion
operators for two sites in the unit cell. Here we shall address
the question of whether the microscopic Hamiltonian commutes with
the $SO(5)$ symmetry generators. Three interaction parameters,
$U$, $V$ and $J$, fully characterize the most general local
interactions on the two sites, which takes the form
\begin{eqnarray}
H(x) =
U(n_{c\uparrow}-\frac{1}{2})(n_{c\downarrow}-\frac{1}{2})
+(c\rightarrow d) +  V (n_c-1)(n_d-1) + J \vec S_c \vec S_d
-\mu (n_c+n_d).
\label{rung-H}
\end{eqnarray}
This Hamiltonian can be solved easily for the $16$ states on two
sites and the 6 energy levels are given in Fig. \ref{Fig_ladder}.
Since the $SO(5)$ symmetry generators can be expressed in terms of
the microscopic fermion operators, we can easily determine the
transformation properties of these states under the $SO(5)$ group.
There are three $SO(5)$ singlet states, and two fermionic quartet
states, which form the fundamental spinor representations of
$SO(5)$. We see that the four fermionic states in each group are
always degenerate, without any fine tuning of the interaction
parameters. The three spin triplet states at half filling and the
two paired states away from half-filling form the five dimensional
vector representation, but they are only degenerate if we specify
one condition, namely
\begin{equation}
J = 4(U+V).
\label{so5_condition}
\end{equation}
This condition ensures the local $SO(5)$ symmetry within the unit
cell. Remarkably, under this condition, a global $SO(5)$ symmetry
is also obtained for a bi-partite lattice including
nearest-neighbor hopping. This is best demonstrated when we write
the model in a manifestly $SO(5)$ covariant manner. On a
bi-partite lattice, we introduce the four-component spinor
operator
\begin{eqnarray}
\Psi_\alpha(x\in even) = \left(
\begin{array}{c}
c_\sigma(x)\\
d^\dagger_\sigma(x)
\end{array}
\right) \ \ \
\Psi_\alpha(x\in odd) = \left(
\begin{array}{c}
d_\sigma(x)\\
c^\dagger_\sigma(x)
\end{array}
\right).
\label{even-odd-spinor}
\end{eqnarray}
The microscopic Hamiltonian including intra-rung hopping $t_\bot$
and inter-rung hopping $t_\parallel$ is given by
\begin{eqnarray}
H =  -2t_\parallel \sum_{\langle x,x' \rangle}
(c^\dagger_\sigma(x) c_\sigma(x')+d^\dagger_\sigma(x)
d_\sigma(x')) -2t_\bot \sum_x (c^\dagger_\sigma(x) d_\sigma(x) +
h.c) + \sum_x H(x). \label{ladder-H}
\end{eqnarray}
Under  condition (\ref{so5_condition}), this Hamiltonian can be
expressed in a manifestly $SO(5)$ invariant manner:
\begin{eqnarray}
H= 2t_\parallel \sum_{\langle x,x' \rangle} (\Psi_\alpha(x)
R^{\alpha\beta} \Psi_\beta(x') + h.c.) + t_\bot (\Psi_\alpha
R^{\alpha\beta} \Psi_\beta + h.c.) + \sum_{x} \frac{J}{4}
L_{ab}^2(x) + (\frac{J}{8}+\frac{U}{2})
(\Psi^\dagger_\alpha\Psi_\alpha-2)^2, \label{manifest-ladder-H}
\end{eqnarray}
where the $R$ matrix is defined in the Appendix. This model was
originally constructed for the two-legged ladder system, but it
works equally well for a two dimensional bi-layer system.

The phase diagram of this $SO(5)$ symmetric model has been studied
extensively in the literature. This simple model has a rich and
rather complex phase diagram, depending on the coupling strength and
doping. However, because of the constraints imposed by the $SO(5)$
symmetry, the phase diagram is much better understood compared to
other related models. In the strong coupling limit, three phase
boundary lines are determined from the level crossing of the
bosonic states on two sites. At $V=-2U$, the $E_0$ state becomes
degenerate with the $E_3$ states; at $V=-U$, the $E_0$ state
becomes degenerate with the $E_1$ states; finally, at $V=0$ the
$E_1$ states become degenerate with the $E_3$ states. The strong
coupling phase diagram at half-filling is shown in Fig.
\ref{Fig_ladder}b.

In the strong coupling $E_0$ phase, a robust ground state is
obtained as a product of $SO(5)$ singlets on the rungs. This type
of insulating state does not break any lattice translational or
internal rotational symmetry. Since there are two electrons per
unit cell, this insulating state is also adiabatically connected
to the band insulator state. This state is separated from the
excited $SO(5)$ quintet vector states by a finite energy gap,
$\Delta=E_1-E_0=J$. In this regime, we consider the low energy
manifold consisting of six states, namely one $E_0$ state
$|\Omega\rangle$ and five $E_1$ states $n_a|\Omega\rangle$ per
rung. The low energy effective Hamiltonian can be obtained easily 
by the second order strong coupling expansion and is exactly
given by the $SO(5)$ quantum non-linear $\sigma$ model Hamiltonian
given in Eq. (\ref{Hsigma}), with $\chi^{-1}= J$ and $\rho =
J_{\parallel} = t^2_\parallel/(U+J/2)$. The operators $L_{ab}$ and
$n_a$ act on the six states in the following way:
\begin{eqnarray}
&&L_{ab}(x) |\Omega(x)\rangle = 0 \ \ , \ \ L_{ab}(x)
|n_c(x)\rangle
= i\delta_{bc} |n_a(x)\rangle - i\delta_{ac} |n_b(x)\rangle \nonumber \\
&& n_a(x) |\Omega(x)\rangle = |n_a(x)\rangle \ \ , \ \ n_a(x)
|n_b(x)\rangle = \delta_{ab}|\Omega(x)\rangle.
\label{operator-action}
\end{eqnarray}
Since the quantum model is exactly $SO(5)$ symmetric, the
anisotropy term $V(n)$ vanishes identically. Therefore,
we see that the $SO(5)$ quantum non-linear $\sigma$ model,
phenomenologically introduced in section \ref{sec:so5sigma}, can
indeed be rigorously derived from the microscopic SZH model
defined on a ladder and on a bi-layer.

In the $E_0$ regime, the SZH model on the half-filled ladder has a
$SO(5)$ rung singlet ground state with a finite gap towards the
$SO(5)$ quintet excitations. A chemical potential term of the
order of the gap induces a second order quantum phase transition
into the SC phase. On the other hand, the SZH model on the
bi-layer has a quantum phase transition even at half-filling, when
$J_{\parallel}/J \sim 1$. For $J>J_{\parallel}$, the ground state
is a Mott insulator without any symmetry breaking with a finite
gap towards the quintet excitations. For $J<J_{\parallel}$, the
ground state is classically ordered and breaks the $SO(5)$
symmetry spontaneously by aligning the superspin in a particular
direction, which can be either AF or SC. Since the residual
symmetry is $SO(4)$, the Goldstone manifold of the $\sigma$ model
is the four dimensional sphere $SO(5)/SO(4)=S^4$.
Away from half-filling, the $SO(5)$ symmetry is broken by the
chemical potential term. According to Table I, the $\pi$ operators
carry charge $\pm2$, and we have $[H, \pi^\dagger_\alpha]=2 \mu
\pi^\dagger_\alpha $. However, although the Hamiltonian does not
commute with all the $SO(5)$ generators, it still commutes with
the Casimir operator $L_{ab}^2$. For this reason, all states are
still classified by $SO(5)$ quantum numbers and the $SO(5)$
symmetry makes powerful predictions despite a broken symmetry
away from half-filling. The phase diagram for the two dimensional
SZH bi-layer model is shown in Fig. \ref{Fig_bilayer}. For
$J_\parallel \gg J$, the ground state is classically ordered. The
chemical potential induces a quantum phase transition from the
$SO(5)$ uniform mixed AF/SC state to the SC state at $\mu=0$. This
transition is exactly the superspin flop transition discussed in
section \ref{sec:so5sigma}. For $J_\parallel \ll J$, the ground
state is quantum disordered at half-filling. A second order
quantum phase transition from the singlet Mott insulator state to
the SC state is induced at finite $\mu=\mu_c$. The exact $SO(5)$
bi-layer model offers an ideal theoretical laboratory to study the
collective modes, especially the $\pi$ resonance mode discussed in
section \ref{sec:pi}, since their sharpness is protected by the
exact $SO(5)$ symmetry. The Mott phase has five massive collective
modes, a doublet of charge modes and a triplet of spin modes. The
energy of the two charge modes splits at finite chemical
potential, and the energy of one of the charge modes vanishes at
the second order phase transition boundary. This charge mode
continues into the SC phase as the phase Goldstone mode. The spin
triplet mode of the Mott phase continues smoothly into the SC phase
and becomes the pseudo-Goldstone mode, or the $\pi$ resonance mode
of the SC phase. The ordered phase at half-filling has four
Goldstone modes. The direction of the order parameter can be
smoothly rotated from AF to SC at half-filling. When the order
parameter points in the AF direction, the four Goldstone modes
decompose into two spin wave modes and two charge modes. When the
order parameter is rotated into the SC direction, the four
Goldstone modes decompose into a spin triplet and a Goldstone
phase mode. The energy of the triplet Goldstone mode (the massive
$\pi$ mode) increases continuously with the chemical potential,
while the phase Goldstone mode remains gapless.

Having discussed the $E_0$ regime at length, let us now turn to
the $E_1$ regime, where the $SO(5)$ quintet state has the lowest
energy. In this case, we can restrict ourselves to the low energy
manifold of five states on each rung. The effective theory within
this low energy manifold can again be obtained by the strong
coupling second order perturbation theory, and is given by
\begin{eqnarray}
H = K\sum_{\langle x,x' \rangle} L_{ab}(x) L_{ab}(x'), \label{E1}
\end{eqnarray}
where $K=t^2_\parallel/(U/2-J/4)$. This effective Hamiltonian is
the $SO(5)$ generalization of the AF spin 1 Heisenberg model. Here
we must distinguish between the one dimensional ladder model
and the two dimensional bi-layer model. In one dimensional models,
the ground state is separated from the $SO(5)$ vector excitation
by a finite energy gap. In fact, an exact ground state can be
constructed for the $SO(5)$ vector model by generalizing the AKLT
model for the spin 1 chain. Such a state also preserves the
lattice translational and internal rotational symmetry. However,
in two dimensional bi-layer models, the effective exchange
coupling between the $SO(5)$ vectors will lead to a state with
spontaneously broken $SO(5)$ symmetry, with the $SO(5)$ adjoint
order parameter $\langle L_{ab} \rangle \neq 0$. This order
parameter is formed by the linear superposition of two $SO(5)$
vector states, $n_a$ and $n_b$. Without loss of generality, let us
consider the case where $\langle L_{15} \rangle \neq 0$. In this
case, the $SO(5)$ generators $L_{15}$, $\{L_{23},L_{24},L_{34}\}$
leaves the state invariant. These set of generators form a
$U(1)\times SU(2)$ symmetry group. Therefore, the Goldstone
manifold is the coset space
\begin{eqnarray}
SO(5)/\left(U(1)\times SU(2)\right)=CP_3, \label{Cp3}
\end{eqnarray}
where $CP_3$ is the six (real) dimensional complex projective
space, which can be described by the complex coordinates
$(z_1,z_2,z_3,z_4)$, satisfying
$|z_1|^2+|z_2|^2+|z_3|^2+|z_4|^2=1$ and with the points related
by a $U(1)$ gauge transformation $z_i\rightarrow e^{i\alpha}z_i$
identified. Since the $CP_3$ manifold is six dimensional, there
are six Goldstone bosons in this case. Here we see that there is
an important difference between the $SO(5)$ symmetric SZH model
and the $SO(3)$ symmetric Heisenberg model. In the Heisenberg
model, the vector representation is identical the adjoint
representation, there is only one type of classically ordered AF
state. In the $SO(5)$ case, the symmetry breaking can occur either
in the vector or the adjoint representations of the $SO(5)$ group,
which are inequivalent, and the resulting Goldstone manifolds are
$S^4$ and $CP_3$, respectively. The adjoint symmetry breaking
pattern has been used by Murakami, Nagaosa and Sigrist to unify
$p$ wave SC with ferromagnetism\cite{MURAKAMI1999}.

In the weak coupling limit, powerful renormalization group (RG)
analysis has been applied to study the $SO(5)$ 
symmetry in ladder models. The main
conclusions are similar to the strong coupling analysis;
therefore, we will only review the most remarkable and distinct
results. Lin, Balents and Fisher\cite{LIN1998}, Arrigoni and
Hanke\cite{ARRIGONI1999}, Schulz\cite{Schulz1998}, Shelton and
Senechal\cite{SHELTON1998} carried out detailed RG analysis and
showed that RG transformation always scales the most generic
ladder model towards an $SO(5)$ symmetric ladder model. This is a
remarkable result and showed that the quantum $SO(5)$ symmetry
does not need to be postulated at the microscopic level but could
emerge as a result of scaling in the long wave length and low
energy limit. More over, Lin, Balents and Fisher\cite{LIN1998}
showed that even the $SO(8)$ symmetry could emerge at
half-filling. Another interesting and remarkable result was
obtained recently. In the transition region between the singlet
$E_0$ phase and the charge ordered $E_3$ phase, the RG analysis of
the weak coupling limit showed the existence of a new phase,
called the staggered flux phase, or the DDW (d-density-wave)
phase, which has staggered circulating current on the
plaquettes\cite{MARSTON2002,Schollwoeck2002}. This phase has been
proposed to explain the pseudo-gap behavior in the HTSC
cuprates\cite{AFFLECK1988,CHAKRAVARTY2001}.

Exactly $SO(5)$ symmetric models can also be constructed for the
single layer model\cite{RABELLO1998,HENLEY1998,BURGESS1998A}. In
this case, there is no natural way to group two sites to form a
local, four-component $SO(5)$ spinor. However, one can introduce a
$SO(5)$ spinor in momentum space by defining
\begin{eqnarray}
\label{spinor} ^t\Psi_{{\bf p}} = \left\{
                  c_{{\bf p} \uparrow}   ,
                  c_{{\bf p} \downarrow} ,
        g ({\bf p}) c^{\dagger}_{{\bf -p+\Pi}, \uparrow}  ,
        g ({\bf p}) c^{\dagger}_{{\bf -p+\Pi}, \downarrow}
                  \right\},
\end{eqnarray}
where $g ({\bf p})={\rm sgn}(\cos p_x - \cos p_y) = \pm 1$ is the
form factor introduced by Henley\cite{HENLEY1998}. As discussed in
section \ref{sec:so5group}, this factor is needed to ensure the
closure of the $SO(5)$ algebra. Indeed, with this choice, the
$\Psi$ spinors form the canonical commutation relation
\begin{eqnarray}
\label{canonical} \{ \Psi^{\dagger}_{{\bf p}\alpha}, \Psi_{{\bf
p'}\beta} \}
     = \delta_{\alpha \beta}\delta_{{\bf p,p'}} , \hskip 2cm    \\
\{ \Psi^{\dagger}_{{\bf p}\alpha}, \Psi^{\dagger}_{{\bf p'}\beta}
\} = \{ \Psi_{{\bf p}\alpha}, \Psi_{{\bf p'}\beta} \}
     = -g({\bf p}) R_{\alpha \beta} \delta_{{\bf p+p',\Pi}} .
\end{eqnarray}
If we restrict  both ${\bf p}$ and ${\bf p'}$ to  be inside the
magnetic Brillouin zone, the right hand side of the second
equation vanishes and the $\Psi_{{\bf p} \alpha}$ spinors commute
in the same way as the $c_{{\bf p}\sigma}$ spinors. Any
Hamiltonian constructed by forming singlets of the the basic
spinors would be manifestly $SO(5)$ symmetric.

Because of the non-analyticity associated with the function $g
({\bf p})$, this class of $SO(5)$ symmetric models contain long
ranged interactions in real space. However, similar kinds of long
ranged interactions are also present in the original BCS model
due to the truncation of interactions in momentum space.
Therefore, this class of $SO(5)$ models can be best viewed as low
energy effective models resulting from integrating out states far
from the Fermi surface. These models may address an important
issue in the field of HTSC, which concerns the nature of the
quasi-particle spectrum at the $d$ wave SC to AF transition. In
the pure $d$ wave SC state, the SC order parameter is described by
the form factor $d(p)=(\cos p_x - \cos p_y)$. When the system is
rotated into a uniform mixed AF/SC state, the form factor of
the resulting AF order parameter is given by $g(p)d(p)=|\cos p_x -
\cos p_y|$, which contains nodes at the same positions as in the
pure $d$ wave SC state. When doping is further reduced, a uniform
component of the AF gap develops across the Fermi surface, filling
the $d$ wave nodes. This uniform AF gap gradually evolves into the
AF Mott insulating gap at half-filling. See Fig. \ref{Fig_d_wave}.
Based on this scenario, Zacher et al \cite{ZACHER2000} explained
the $d$-wave like dispersion of the quasi-particle in the
insulating state\cite{RONNING1998}. Filling the $d$ nodes with the
uniform AF gap also naturally explains the ``small gap" observed
in the photoemission experiments in the lightly doped
cuprates\cite{Shen2004}.  This theory of the quasi-particle
evolution is also similar to the scenario of quantum disordering
the nodal quasi-particles developed in
\cite{BALENTS1998,BALENTS1999,FRANZ2002,HERBUT2002}.
Recent studies have found that the generalized Hubbard model for spin
3/2 fermions enjoys an {\it exact and generic} $SO(5)$ symmetry
without any fine tuning of model parameters and filling factors
\cite{WU2003A}. Such a model can be accurately realized in systems
of ultra-cold atoms on optical lattices, where the interaction is
local and s wave scattering dominates 
\cite{JAKSCH1998,GREINER2002,HOFSTETTER2002}. 
In the Hubbard model with spin
1/2 fermions, two fermions on the same site can only form a total
spin $S_T=0$ state; the $S_T=1$ state is forbidden by the Pauli
principle. Therefore, only one local interaction parameter
specifies the on-site interaction. By a similar argument, two spin
3/2 fermions on the same site can only form the total spin
$S_T=0,2$ states; the $S_T=1,3$ states are forbidden by the Pauli
principle. Therefore, the generalized Hubbard model for spin 3/2
fermions is given by
\begin{eqnarray}
\label{hm1} &&H=-t\sum_{\langle ij\rangle ,\sigma} \big \{
c^\dagger_{i\sigma}
 c_{j\sigma} +h.c.\big \}-\mu \sum_{i\sigma} c^\dagger_{i\sigma} c_{i\sigma}
 \nonumber \\
&&+U_0 \sum_i P_0^\dagger(i) P_0(i) +U_2 \sum_{i,m=\pm2,\pm1,0}
P_{2m}^\dagger(i) P_{2m}(i),~~~
\end{eqnarray}
where $t$ is the hopping integral,  $\mu$ is the chemical
potential, and $P^\dagger_0, P^\dagger_{2m}$ are the singlet
($S_T=0$) and quintet ($S_T=2$) pairing operators, defined as
\begin{eqnarray} &&P^\dagger_{0}(i) (P^\dagger_{20}(i))=
{1\over \sqrt{2}} \{ c^\dagger_{3\over2} c^\dagger_{-{3\over2}}
\mp c^\dagger_{1\over2} c^\dagger_{-{1\over2}}\},
\nonumber \\
&&P^\dagger_{2,2}(i)= c^\dagger_{3\over2}
c^\dagger_{1\over2},\hspace{10mm}
P^\dagger_{2,1}(i)= c^\dagger_{3\over2} c^\dagger_{-{1\over2}}, \nonumber \\
&&P^\dagger_{2,-1}(i)= c^\dagger_{1\over2}
c^\dagger_{-{3\over2}},\hspace{5mm} P^\dagger_{2,-2}(i)=
c^\dagger_{-{1\over2}} c^\dagger_{-{3\over2}}.
\end{eqnarray}
Remarkably, this generalized Hubbard model for spin 3/2 fermions
is always $SO(5)$ symmetric, without any fine tuning of parameters
and filling factors. This can be seen easily from the energy level
diagram of a single site, which contains 16 states and 6 energy
levels for spin 3/2 fermions, as depicted in Fig.
\ref{Fig_spin32}. The $E_{1,4,6}$ levels are non-degenerate, the
degeneracy of the $E_{2,5}$ levels is four-fold, and the
degeneracy of the $E_{3}$ level is five-fold. We see that without
any fine tuning of interaction parameters, this pattern of
degeneracy exactly matches the singlet, the quartet (fundamental
spinor) and the quintet (fundamental vector) representations of
the $SO(5)$ group. It can also be easily verified that the hopping
term also preserves the global $SO(5)$ symmetry. In fact, it
preserves an even larger symmetry group, namely $SO(8)$. The
$SO(8)$ symmetry is always broken by interactions; however, under
special circumstances, its subgroups, $SO(7)$, $SO(6)$ and
$SO(5)\times SU(2)$ can be realized in addition to the generic
$SO(5)$ symmetry.
In this article we mainly focus on application of the
$SO(5)$ theory to the AF/SC systems. However, from the above
discussions, we see that ultra-cold atoms on optical lattices also
provide a fertile ground for investigating higher symmetries in
strongly correlated systems, because the higher spins of the
atoms and the accuracy of local interaction approximation. In the
case of the spin 3/2 systems, the generic $SO(5)$ symmetry makes
powerful predictions on the symmetries at quantum phase
transition lines, spectrum degeneracies, topology of the ground
state manifolds and low energy effective theories of the Goldstone
bosons. With the emerging convergence between the atomic and
condensed matter physics, we expect symmetry concepts and its
multiple manifestations to play an ever increasing role in these
fields.

Fermions in exact $SO(5)$ models have a beautiful non-abelian
holonomy associated with them\cite{DEMLER1999}. The four
components of an $SO(5)$ spinor represent four states but only two
energy levels, each being doubly degenerate. As one varies some
adiabatic parameters and returns to the same starting value,
the states inside a doublet can be rotated into each other by a
unitary transformation. This interesting mathematical property has
been used to predict $SO(5)$ generalization of the Andreev effect
and the non-abelian Aharonov-Bohm effect\cite{DEMLER1999}.

\subsection{Variational wave functions}
\label{sec:variational} In this section we shall discuss a crucial
test of the $SO(5)$ symmetry by investigating the microscopic wave
functions of the $t-J$ model. In section \ref{sec:classical}, we
showed that the transition from the AF state at half-filling to a
pure $d$-wave SC state away from half-filling can generally be
classified into three types. Within the general form of the static
potential as given in Eq. (\ref{GL}), the ``type 1" first order
transition is realized for $u_{12}^2>u_1 u_2$. For $u_{12}^2<u_1
u_2$, the ``type 2" transition involves two second order
transitions with an intermediate mixed phase where the AF and the
$d$-wave SC order coexist uniformly. Only for $u_{12}^2=u_1 u_2$ is
an intermediate ``type 1.5" transition realized, where the
potential can be re-scaled to take an $SO(5)$ symmetric form and a
smooth rotation between the AF and the $d$-wave SC states is
possible. If we only investigate states with uniform densities,
these three possibilities can be distinguished easily by curvature
in the plot of the ground state energy as a function of doping
$\delta$. The curvature would be negative (concave), positive
(convex), or zero (flat) for these three possibilities, as shown
in Fig. (\ref{Fig_E_delta}). In the concave case, the uniform
phase would be thermodynamically unstable, and a Maxwell's
construction leads to a phase separated ground state, where each
phase has a distinct density.

This interesting prediction can be tested numerically in the $t-J$
model. At this moment, exact diagonalization of the $t-J$ model
with large system size is not possible due to the exponential
growth of the Hilbert space, and reliable Monte Carlo simulation
cannot be carried out due to the fermion minus sign problem. A
successful method employs the variational Quantum-Monte-Carlo
(VMC) method (see, in particular,
\cite{GROS1989,HIMEDA1999,CALANDRA2000} and references therein).
Historically, the VMC method was first used to investigate the RVB
type of variational wave functions proposed by
Anderson\cite{ANDERSON1987}. By investigating various variational
wave functions, this method can address the issue of $d$-wave
pairing in the ground state and the possibility of a uniform mixed
phase with AF and $d$-wave SC order for the 2D $t-J$ model.

The first question is whether the uniform mixed state has a lower
energy than the pure $d$-wave SC or AF state near half-filling. In
earlier work by Zhang et al. \cite{ZHANG1988} and by Yokoyama and
Ogata \cite{YOKOYAMA1996}, it was shown that the Gutzwiller
approximation (GA) gives a reliable estimate for the variational
energies for the pure $d$-wave SC state. However, if the AF order
parameter is taken into account in the GA, there exists no region
in the phase diagram where the AF state is stabilized. On the
other hand, in Ref.~\cite{HIMEDA1999}, it was shown that when the
variational parameters $\Delta_d$, $\Delta_{AF}$ and $\mu$ were
determined from a VMC simulation, where {\it the double occupancy
prohibition is rigorously treated}, then the Gutzwiller-projected
trial wave function of the uniform mixed state has a lower energy
than the pure $d$-wave SC state with $\Delta_{AF} = 0$, in the
doping range $0<\delta<10\%$. Using Green's-function Monte Carlo
with stochastic reconfiguration (GFMCSR), Calandra and
Sorella\cite{CALANDRA2000} also concluded that the AF correlations
coexists with SC and persists up to $\delta=10\%$. Himeda and
Ogata used the following Gutzwiller projected trial wave function:
\begin{equation}
  \label{eq:himeda2.2}
  \left| \psi\right>=P_G \left| \psi_0 \left( \Delta_\mathrm{d},
      \Delta_\mathrm{AF},\mu\right) \right> \;,
\end{equation}
where $\Delta_d$, $\Delta_{AF}$ and $\mu$ are the variational
parameters relating to $d$-wave SC and AF order and $\mu$ is the
chemical potential. $P_G = \prod_i \left( 1 -\hat n_{i\uparrow}
    \hat n_{i \downarrow} \right) $ stands for the Gutzwiller projection
operator. The wave function $\left| \psi_0 \left(
\Delta_\mathrm{d}, \Delta_\mathrm{AF},\mu\right) \right>$ is a
mixed BCS/Spin-Density-Wave function, i.~e.
\begin{eqnarray}
  \label{eq:himeda2.3}
  \left| \psi_0 \left( \Delta_\mathrm{d}, \Delta_\mathrm{af},\mu\right)
  \right>
  &=& \prod_{k,s(=\pm)} \left(u^{(s)}_k +v^{(s)}_k
    d^{(s) \dagger }_{k\uparrow} d^{(s) \dagger }_{-k\downarrow} \right)
 \left| 0 \right> \;,
 \end{eqnarray}
where the index $s=\left\{\pm\right\}$ takes care of the electron
operators acting on the $A(B)$ sublattice in the AF state. The
$u_k$'s and $v_k$'s contain the variational parameters $\Delta_d$,
$\Delta_{AF}$ and $\mu$ and are defined in detail in
\cite{HIMEDA1999}. Fig.~\ref{figHimeda2} is reproduced from this
paper and plots the ground-state energy and the staggered
magnetization as a function of doping $\delta$.


We see that in the Himeda and Ogata variational QMC work the
uniform mixed phase of AF and $d$-wave SC has a lower energy than the
pure $d$-wave SC state up to a doping of about 10\%. At
half-filling, the energy was found to be close to the best
estimated value in the Green's function MC method (-0.1994 to
-0.20076), which provides support for the wave-function
Ansatz \eqref{eq:himeda2.3}.

The second point of interest is that, according to the Himada and
Ogata results in Fig.~\ref{figHimeda2}, the ground-state energy is
a {\it linear function of doping $\delta$} in this region, with
essentially zero curvature. This implies that the chemical
potential $\mu$ is constant. Since the wave function of Himada and
Ogata describes a mixed state with uniform density, the
energy versus doping plot can generally have three distinct
possibilities, as enumerated in Fig. (\ref{Fig_E_delta}).
Therefore, from the fact that the curvature is nearly flat we
determine that the condition $u_{12}^2=u_1 u_2$ is fulfilled,
which places the $t-J$ model at $J/t=0.3$ into the domain of
attraction of the $SO(5)$ fixed
point\cite{ARRIGONI2000,MURAKAMI2000}.

\subsection{Exact diagonalization of the t-J and the Hubbard model}
\label{sec:diagonal} In the previous section we discussed the test
of the $SO(5)$ symmetry through the variational wave functions in
the $t-J$ model. In this section, we shall describe numerical
calculations of the dynamic correlation functions and the exact
diagonalization of the spectrum, which also tests the $SO(5)$
symmetry of the microscopic $t-J$ and Hubbard models. A
microscopic model has a symmetry if its generators $G$ commute
with the Hamiltonian $H$, {\it i.e.} $[H,G]=0$. In the $SO(5)$
theory, the $\pi_\alpha$ operators are the non-trivial generators
of the symmetry. In models constructed in section \ref{sec:exact},
the $\pi_\alpha$ operators indeed commute with the Hamiltonian.
However, there are models where the symmetry generators do not
commute with the Hamiltonian, but they satisfy a weaker condition,
$[H,G^\pm]=\pm\lambda G^\pm$, where $\lambda$ is a $c$ number
eigenvalue (see e.g. Eq. (\ref{Heta}) in Section
\ref{sec:spin-flop}). These operators are called eigen-operators
of the Hamiltonian. In this case, from one eigenstate of the
Hamiltonian, one can still generate a multiplet of eigenstates by
the repeated actions of $G^\pm$. However, the eigenstates within a
multiplet are not degenerate, but their energies are equally
spaced by $\lambda$. A classic example is the precession of a spin
in a magnetic field, where
\begin{equation}
H = \omega_0 S_z; \quad \quad [H, S_{\pm}] = \pm \omega_0
S_{\pm}\,.
\end{equation}
and $\omega_0$ is the Lamor frequency of the spin precession.
Although in this case the spin-rotational symmetry is broken
explicitly by the magnetic field in the $z$-direction and the
eigenstates within the multiplets are no longer degenerate, the
multiplet structure of the symmetry is still visible in the
spectrum and can be sampled by the ladder operators. If one
calculates the dynamical response function of $S_\pm$, only a
single $\delta$-peak is present at $\omega = \omega_0$.

The $\pi_\alpha$ operators defined in equation
(\ref{pi_k}) do not commute with the Hubbard or
$t-J$ model Hamiltonian, but analytical and numerical calculations
show that they are approximate eigen-operators of these model, in
the sense that
\begin{equation}
[H, \pi^\dagger_\alpha] \approx \omega_\pi \pi^\dagger_\alpha
\label{approximate_eigen}
\end{equation}
is satisfied in the low energy sector. This means that the dynamic
auto-correlation function of the $\pi_\alpha$ operators contains a sharp
pole at $\omega_\pi$, with broad spectral weight
possibly istributed at higher energies. Using a T-matrix approximation,
Demler and Zhang \cite{DEMLER1995} verified this approximate
equation with $\omega_\pi = J (1-n)/2 - 2 \mu$. This calculation
will be reviewed in section \ref{PiFermiLiquid}. The first
numerical test for a low-energy $SO(5)$ symmetry in a microscopic
model has been performed by Meixner {\it et al.}
\cite{MEIXNER1997} using the Lanczos \cite{LANCZOS1950} exact
diagonalization technique. Analysis presented in this paper showed
that the dynamical correlation function of the $\pi$-operator
\begin{equation}
     \pi^{\dagger}_{\alpha}(\omega)= -\frac{1}{\pi} {\rm Im}
      \langle \Psi_0^{N}|\pi_{\alpha}
      \frac{1}{\omega - H+E_0^{N+2}+ i\eta}
        \pi_{\alpha}^\dagger|\Psi_0^N\rangle.
\label{pi-correlation}
\end{equation}
(with $H$ being the standard Hubbard Hamiltonian,
$|\Psi_0^N\rangle$ its ground state with $N$ electrons and
$E_0^{N}$ the corresponding ground state energy) yielded a single
sharp excitation peak at low energy $\omega_\pi$, accompanied by
an incoherent background at higher energies. The large separation
between the peak and the continuum and the large relative spectral
weight of the peak demonstrated that indeed the $\pi$-operator is
an {\it approximate} eigenoperator of the Hamiltonian (see
Fig.~\ref{figMZ322meixneru8}).  Also in accordance with the
perturbative result of Ref.\cite{DEMLER1995}, the ``precession
frequency'' $\omega_\pi$ decreases for decreased doping.
Furthermore, a comparison with a bubble approximation for this
correlation function showed that the sharp peak near $\omega_\pi$
originated solely from vertex corrections (i.e. collective
behavior).

Not only can the dynamic correlation function of the $\pi_\alpha$ operators
(\ref{pi-correlation}) be measured numerically for
microscopic models, thus providing a test of the approximate
$SO(5)$ symmetry, but they can also be directly measured in neutron
scattering experiments in the SC state. We shall discuss these
experiments in section \ref{sec:pi}.

Exact numerical diagonalization of the $t-J$ and Hubbard
models gives eigenstates and eigenvalues on a finite size cluster,
whose degeneracy pattern can be used directly to test the $SO(5)$
symmetry. In order to explain the main idea, let us first examine
the variational wave function of the projected $SO(5)$ model
given in Eq. (\ref{coherent}). This wave function describes a
broken symmetry state formed by a linear superposition of states
with different spin or charge quantum numbers. This type of state
can only be realized in infinite systems. On a finite size system,
all eigenstates must have definite spin and charge quantum
numbers. Denoting $t^\dagger(x)=m_\alpha(x)t^\dagger_\alpha(x)
+\Delta(x) t^\dagger_h(x)$, we can expand the coherent state
described by Eq. (\ref{coherent}) as
\begin{equation}
|\Psi\rangle = \{\cos\theta^N + \cos\theta^{N-1} \sin\theta \sum_x
t^\dagger(x)+\cos\theta^{N-2} \sin\theta^2 \sum_{x\neq y}
t^\dagger(x)t^\dagger(y)+\cos\theta^{N-3} \sin\theta^3 \sum_{x\neq
y\neq z} t^\dagger(x)t^\dagger(y)t^\dagger(z)+...\} |\Omega\rangle.
\label{coherent-expand}
\end{equation}
For $\Delta(x)=0$, we see that the AF ordered state can be
expressed as a linear superposition of states with different
numbers of magnons, forming states with different total spins.
While states with different total spins are separated by finite energy
gaps in a finite size system, these energy gaps could
vanish in the thermodynamic limit, allowing magnons to ``condense"
into the ground state. For $m_\alpha(x)=0$, we see that the SC
state can be expressed as a linear superposition of states with
different numbers of hole pairs, with different total charge. A
smooth rotation from the AF state to the SC state becomes
possible if one can freely substitute each magnon by a hole pair
without energy cost. This places a powerful requirement on the
spectrum. The $\sum_x t^\dagger(x)|\Omega\rangle$ term in
(\ref{coherent-expand}) contains a single magnon state with
($S=1$, $Q=0$) or a single hole pair state with ($S=0$, $Q=-1$).
$SO(5)$ symmetry requires them to be degenerate. This can
be easily achieved by tuning the chemical potential, which changes
the energy of the hole pair state without changing the energy of
the magnon state. Once the chemical potential is fixed, there are
no additional tuning parameters. The $\sum_{x\neq y}
t^\dagger(x)t^\dagger(y)|\Omega\rangle$ term in
(\ref{coherent-expand}) contains a two-magnon state with ($S=2$,
$Q=0$), a one-magnon-one-hole-pair state with ($S=1$, $Q=-1$) and
a two-hole-pair state with ($S=0$, $Q=-2$). $SO(5)$ symmetry again
requires them to be degenerate, which is a highly non-trivial
test. We can easily perform this analysis for states with
different numbers of magnons and hole pairs.

This pattern of the energy levels has been tested directly in the
exact diagonalization of the $t-J$ model by Eder, Hanke and Zhang
\cite{EDER1998}. The $t$$-$$J$ model, because of its more limited
Hilbert space (no double occupancies), allows the exact
diagonalization of larger systems (18, 20 and more sites). Since
the $t$$-$$J$ model explicitly projects out the states in the
upper Hubbard band, some of the questions about the compatibility
between the Mott-Hubbard gap and $SO(5)$ symmetry can also be
answered explicitly. In the exact diagonalization studies, total
energy, momentum, angular momentum, spin and the charge quantum
numbers of the low energy states can be determined explicitly.
These quantum numbers are summarized in Fig.
(\ref{Fig_multiplet})a.

%

Eigenstates obtained from the $t-J$ or Hubbard Hamiltonian can
always be interpreted as multi-particles states of the
underlying electron. However, it would be highly non-trivial if
the low energy states could also be interpreted as multi-particle
states formed from the collective degrees of freedom such as the
magnons and the hole pairs. The first non-trivial finding of
Ref.~\cite{EDER1998} is that this is indeed the case.
Fig.~\ref{figFKPirreps} shows the first four $(\nu=0$ to $\nu=3)$
sets of low-lying states of an 18-site $t-J$ model
\cite{EDER1998}. We see that the lowest energy state in the $S=1,
Q=0$ sector indeed has $s$ wave like rotational symmetry and total
momentum $(\pi,\pi)$, as expected from a magnon; the lowest energy
state in the $S=0, Q=-1$ sector indeed has a $d$-wave like
rotational symmetry and total momentum $0$. Similarly, states with
higher $S$ and $Q$ have quantum numbers expected from multiple
magnons and hole pairs. This finding confirms the basic assumption
of the $SO(5)$ theory, that the low energy collective degrees of
freedom can be described by the superspin alone.

At the next level, the pattern of symmetry can itself be tested.
The level $\nu$ of a given multiplet indicates the total number of
magnons and hole pairs. If $SO(5)$ symmetry is realized at a given
chemical potential $\mu_c$, we would expect the free energy to
depend only on $\nu$, the total number of magnons and hole pairs,
but not on the difference between the number of magnons and hole
pairs. As shown in Fig. \ref{Fig_multiplet}, the energy can depend
on $Q$ with three generic possibilities, similar to the
discussions we presented in section \ref{sec:classical} and Fig.
\ref{Fig_E_delta}. Only when the energy depends linearly on $Q$ can
the free energy be independent of $Q$ at a given critical
value of the chemical potential. From (\ref{figFKPirreps}) we see
that the energy levels indeed have this remarkable structure:
states whose total charge differ by $\Delta Q=-1$ have nearly the
same difference in energy. Therefore, the energy is approximately
a linear function of $Q$ or doping, similar to the situation
discussed in section \ref{sec:variational}. To be more
precise, the mean-level spacing within each multiplet (up to $Q=-2$) is $%
-2.9886$ with a standard deviation of $0.0769$. This standard
deviation is much smaller $\left( \sim J/8\right) $ than the
natural energy scale $J$ of the $t$$-$$J$ model and comparable to
or even smaller than the average SC gap. Therefore, if one now
adds the chemical potential term $H_{\mu} =-2\mu Q$, and chooses
$\mu=\mu _{c}$ equal to the mean-level spacing, the superspin
multiplets are nearly degenerate. At $\mu=\mu_c$, magnons can be
smoothly converted into hole pairs without free energy cost. This
means that in each term of the expansion in
(\ref{coherent-expand}) one can freely substitute
$t^\dagger_\alpha$ or $t^\dagger_h$ for $t^\dagger$, and the
direction of the superspin vector can be freely rotated from the
AF to the SC direction. The smallness of the standard deviation
indicates the flatness in the energy versus doping plot discussed
in the previous section. If the standard deviation is
significantly different from zero, this would indicate significant
curvature in the energy versus doping plot. Therefore, the
smallness of the standard deviation obtained by the exact
diagonalization is consistent with the flatness of the energy
versus doping plot obtained from the variational wave function
discussed in the previous section.

Another important aspect of the $SO(5)$ symmetry is the
Wigner-Eckart theorem \cite{GEORGI1982}. This theorem provides a
selection rule for the matrix elements of the operators based on
the $SO(5)$ symmetry of the system. It implies, for example, that
the $\pi$ operators (see equation (\ref{pi_k})) can only move us
within a given multiplet, since they are symmetry generators.  On
the other hand, AF and $d$-wave SC order parameters (see equations
(\ref{AFdSC})) should move us between different multiplets. Both
features have been verified in the numerical calculations in
\cite{EDER1998}.

We conclude this subsection with a general remark. Exact
diagonalizations (e.g.~\cite{DAGOTTO1994}) commonly study
ground-state correlations, but their spatial decay is often
inconclusive as a test of order due to small system size.
Discussions in this section show that it is possible that the
(excited) eigenstates reveal a well-defined structure characteristic
of a particular symmetry. Our strategy is to use the finite size
calculations as input for effective models describing the
collective degrees of freedom such as the superspin, or the
magnons and the hole pairs. Since quantum Monte Carlo calculations
can be performed for these models in a large size systems, the
question of long range order and their competition can be firmly
established.




\subsection{Transformation from the microscopic model to effective
SO(5) models} \label{sec:CORE}

From the two previous sections we have learned that both the
variational wave function and the exact diagonalization of the
$t-J$ model show that the ground state and low lying excited
states in the low doping range can be completely described in
terms of the superspin degree of freedom, with an approximate
$SO(5)$ symmetry. Altman and Auerbach\cite{ALTMAN2002} pioneered a
systematic procedure in which they derived the effective bosonic
$SO(5)$ model directly from the microscopic Hubbard and $t-J$
models through a renormalization group transformation called the
Contractor Renormalization (``CORE") method\cite{MORNINGSTAR1996}.
This mapping has several distinct advantages. First, this approach
helps to visualize clearly which processes and which excitations
dominate the low energy physics of the system. Second, their work
directly determines the parameters of the effective models defined
in Eq. (\ref{pso5}) and (\ref{extended}) in terms of the
microscopic parameters. The bosonic systems are often much easier
to analyze numerically, as one does not have to worry about Pauli
principles, Slater determinants, and the infamous sign problem in
the quantum Monte Carlo algorithms. In this section, we shall
describe their work.

Since we want to construct {\em bosonic} quasiparticles, we have
to divide the lattice into effective sites containing an {\em
even} number of elementary sites (with one electron per site). In
order to conserve the symmetry between the $x$ and $y$-direction in
the system, the original projected $SO(5)$ model was formulated on
a plaquette of $2\!\times\!2$ elementary sites. First we begin
with the low energy eigenstates of the Heisenberg plaquette, which
are determined easily. We find the nondegenerate ground state
$|\Omega\rangle$ (see Fig. \ref{Fig_plaquette} for a real-space
representation in terms of the microscopic states on a plaquette)
with energy $E_0\!=\!-2J$ and total spin $S\!=\!0$. This singlet
state will be the vacuum state of the effective bosonic projected
$SO(5)$ model. The next energy eigenstates are three triplet
states $t^\dagger_\alpha|\Omega\rangle$ with energy $E_t\!=\!-J$
and spin quantum numbers of $S\!=\!1$. All other energy
eigenstates of the Heisenberg plaquette have higher energies and
can be neglected in the low energy effective model. It should be
noted that the quasiparticles $t_\alpha$, which carry spin 1 and
charge 0, are {\em hardcore bosons} because one cannot create more
than one of them simultaneously on a single plaquette.

In their CORE study of the 2D Hubbard model, Altman and Auerbach
\cite{ALTMAN2002} started from the spectrum of lowest-energy
eigenstates of the 2 x 2 plaquette for 0, 1 and 2 holes,
respectively. The corresponding lowest spectrum of the triplet
$(t_\alpha^\dagger)$, pair boson $(t_h^\dagger)$ and fermionic
excitations is presented in Fig.~\ref{fig5espectrum}. The ground
state of two holes, also depicted in Fig.~\ref{Fig_plaquette}, is
described by
\begin{equation}
        \label{5e5}
        t_h^\dagger \left| \Omega \right> = \frac{1}{\sqrt{Z_b}}
        \left( \sum_{ij} d_{ij} c_{i\uparrow} c_{j \downarrow}+ \ldots \right)
        \left| \Omega \right>,
\end{equation}
where $d_{ji}$ is +1 (-1) on vertical (horizontal) bonds within a
plaquette and $\ldots$ stands for higher-order (U/t)-operators.
$Z_b$ is the wavefunction normalization. We note that
$t_h^\dagger$ creates a ``Cooper''-like hole pair with internal
$d$-wave symmetry with respect to the vacuum. The crucial point
here is that while there is no hole pair binding for the Hubbard
model on a dimer, there is binding in the range of $U/t \in (0,5)$
for a plaquette, a rather well-known fact (see, for instance,
Ref.\cite{HIRSCH1988}). However, this does not guarantee the integrity
of the pair binding on the infinity lattice, documented by the
fact that the hopping energy $t$ is much larger than the pair
binding energy, nor does it guarantee long-range SC order. To get
more insight into these questions, one has to construct
$H_{eff}$ via a CORE procedure.

In order to understand how the triplets and pair bosons behave on
the infinite lattice, we must determine the boson hopping energies
and the corresponding effective Hamiltonian. A suitable approach
for this has been suggested by Morningstar and Weinstein on the
basis of the CORE technique, which has been shown for the 2D
Hubbard model \cite{ALTMAN2002}, $t-J$ ladders \cite{CAPPONI2002}
and earlier for Heisenberg chains and ladders
\cite{MORNINGSTAR1996} to be extremely accurate. For example,
Morningstar and Weinstein obtained a very accurate 1D Heisenberg
model ground-state energy. This is even more impressive considering
the latter model has long-range, power-law decaying spin
correlations.


In order to implement the CORE technique, the lattice is
decomposed in small block units, as shown in Fig.~\ref{fig5eMJ},
where $H_0$ is the intra-block Hamiltonian and $V$ is the part
describing the coupling between the two neighboring blocks.  The
$M$ low-energy states $\{|\alpha^0\rangle\}_i^M$ are kept in each
block $i$ (here $M=4$ in the $2 \times 2$ plaquette $i$) to define
a reduced Hilbert space. The full Hamiltonian is then diagonalized
on $N$ connected units (in our example in  Fig.~\ref{fig5eMJ},
$N=2)$, i.~e. for the (superblock) Hamiltonian $H_S$. The $M^N$
(in our case, $M^N=4^2$) lowest energy states $|\Psi_n\rangle$
with energy $\epsilon_n$, $n=1 \ldots M^N$ are retained. These
true eigenstates of the $N(=2)$ block problem,
$\{|\Psi_n\rangle\}$, are then projected onto the reduced Hilbert
space spanned by the tensorial product $|\alpha^0_1 \ldots
\alpha^0_N\rangle$ of the $M(=4)$ states of each block, i.~e.
\begin{equation}
        \label{5e7}
        \left| \Psi'_n\right> = \sum_{\alpha_1 \ldots \alpha_N} \left<
        \alpha^0_1 \ldots \alpha^0_N\mid \Psi_n\right> \left| \alpha^0_1
        \ldots \alpha^0_N \right>,
\end{equation}
and Gram-Schmidt orthonormalized, finally yielding  the states $\{|
\tilde \Psi_n\rangle\}$. Then, the new superblock (renormalized)
Hamiltonian is defined as
\begin{equation}
        \label{5e9}
        \tilde H_S = \sum^{M^N}_n \epsilon_n \left|\tilde \Psi_n  \right>
        \left< \tilde \Psi_n \right|.
\end{equation}
By construction $ \tilde {H}_S$, has the same eigenvalues
$\epsilon_n$ as $H_S$ for $n=1, \ldots ,M^N$. Having constructed
the new superblock or renormalized Hamitonian $ \tilde {H}_S$, one
can write (in our $N=2$ example)
\begin{equation}
        \label{5e11}
        \tilde H_S = \tilde H_0 \otimes \mathrm{I} + \mathrm{I} \otimes
        \tilde H_0 +\tilde V,
\end{equation}
where $ \tilde {H}_0$ is simply the projected block Hamiltonian:
\begin{equation}
        \label{5e12}
          \tilde H_0 = \sum^{M}_{n=1} \epsilon_n \left|\alpha^0_n
              \right>
        \left< \alpha^0_n  \right|.
\end{equation}
The above equation (\ref{5e11}) gives the new renormalized
interblock coupling $\tilde{V}$, restricted to the reduced Hilbert
space. In the next step, one repeats the above procedure,
replacing $H_0$ and $V$ in the original superblock Hamiltonian
$H_S$ by  $\tilde{H}_0$ and $\tilde{V}$, and so on.

The projection onto the original plaquette product basis in the
Eq.~(\ref{5e7}) expresses, of course, the above-discussed
proliferation and possibly spatial decay of the block excitations.
More generally, this is incorporated within the CORE method, in a
superblock consisting of $N$ blocks and a corresponding
Hamiltonian containing $N$-body interactions. The construction to
obtain $\tilde{V}$ (Eq.~(\ref{5e11})) is different and, obviously,
one also obtains  $\tilde{V}$-terms, connecting $N$ clusters
instead of just $N=2$ (called range-$N$-approximation). It has
been shown in the above-cited various applications that 
the above range-2 approximation ($N=2$) and at most $N=3$
interactions already yield very accurate results. Thus, with a proper and
physically motivated choice of the truncated basis, range-N
interactions decay rapidly with $N$.

In Ref.~\cite{ALTMAN2002}, the CORE calculation was limited to
range-2 boson (triplets, hole pair boson) interactions leaving out
the fermion state. From the above Fig.~\ref{fig5eMJ}, i.~e. it is
clear that this amounts to diagonalizing two coupled ($2 \times
2$) plaquettes, for instance, an 8-site Hubbard cluster, which is very
straightforward by the Lanczos technique. The resulting effective
Hamiltonian for this range-2 four-boson model is exactly the
projected $SO(5)$ model defined in Eq. (\ref{pso5}) plus more
extended interactions defined in Eq. (\ref{extended}). Following
Altman and Auerbach, we compare in Fig.~\ref{fig5epars} the
magnitudes of the magnon hopping $J_s$ (denoted as $J_t/2 \simeq
J_{tt}/2$ in Ref. \cite{ALTMAN2002}) and the hole pair hopping
$J_c$ (denoted as $J_b$ in Ref. \cite{ALTMAN2002}) for a range of
$(U/t)$-values.


The first observation is that $J_t \sim J_{tt} \sim 0.6J$;
therefore, the triplet terms have a similar magnitude as those previously
(see also our simple pedagogical Heisenberg example) obtained
\cite{SACHDEV1990,GOPALAN1994}.

The second finding is crucial. The region of equal $J_t$
$(J_{tt})$ and $J_b$, equal magnon- and pair-boson hopping,
occurs very close to $U/t=8$. Thus, the value of the {\it
projected $SO(5)$ model with $J_t=J_b$} occurs in the {\it
physically relevant regime}: It is known from a large body of
numerically essentially exact (for example QMC) evaluations of the
2D Hubbard model that it reproduces salient features of the HTSC
cuprates precisely in this regime (see for example
\cite{DAGOTTO1994,IMADA1998}. This gives yet another piece of
evidence, in addition to those discussed in sections
\ref{sec:variational} and \ref{sec:diagonal}, that realistic
microscopic models can be described effectively by the projected
$SO(5)$ model close to the symmetric point.

Altman and Auerbach\cite{ALTMAN2002}, Capponi and
Poilblanc\cite{CAPPONI2002} also calculated the coefficient and
terms on $H^{int}$ in Eq.~(\ref{extended}), which contains
triplet-triplet, pair-pair and pair-triplet interactions. These
interaction terms were found to be small compared to $H^b$ and
$H^t$, but their influence has yet to be studied in detail. They
also estimated the truncation error of discarding range-3 terms
which, for physically relevant $U$-values, was found to be very
small (1\%).

An issue still left open is the role of fermions. Altman and
Auerbach have extended the above 4-boson model to a boson-fermion
model by augmenting the bosons with single-hole fermions ``by
hand.'' This is certainly a first step. However, a consistent
low-energy theory has to treat bosons and fermions within the CORE
procedure on equal footings. It should, however, be noted that the
short-range effects of the fermions on the effective boson
couplings were included in the above range-2 calculation. Altman
and Auerbach estimated the fermion-boson interaction by including
the hole fermions dispersion ``adhoc'', i.~e. using the
single-hole band-structure extracted previously by various groups
for large clusters \cite{DAGOTTO1994}.

In summary, the application of the CORE algorithm to the Hubbard model
has demonstrated two features which are of immediate relevance for
the $SO(5)$ theory: the $d$-wave hole pairs already present in the
2D Hubbard model on a single 2 x 2 plaquette maintain their
integrity in the ``infinite'' square lattice. The low energy
degrees of freedom are indeed described solely by the superspin.
Secondly, the hole-pair and magnon (triplet) hopping
fulfills the projected $SO(5)$ condition in the physically relevant
(U/t)-range.

\section{PHYSICS OF THE $\pi$ RESONANCE AND THE MICROSCOPIC MECHANISM OF SUPERCONDUCTIVITY}
\label{sec:pi} A key experimental manifestation of a higher
symmetry is the emergence of new particles or new
collective modes. Historically, this line of reasoning has led to
important discoveries in particle physics. For example, 
Gell-Mann used the $SU(3)$ symmetry of the strong interaction  to predict the
$\Omega^-$ resonance. Similarly, the electro-weak unification
based on the $SU(2)\times U(1)$ symmetry has led to the
prediction of the $W^\pm$ and the $Z$ bosons. In a strongly
interacting system, whether in particle physics or in condensed
matter physics, typical excitations have short life times and
broad line shapes. However, if higher symmetries are present, the
selection rules associated with the symmetry prevents the
excitation from decaying. The $SO(5)$ symmetry of
antiferromagnetism and superconductivity naturally predicts a new
class of collective excitations, called the $\pi$ resonance or
$\pi$ mode for short, which are the (pseudo-) Goldstone modes of
the spontaneous symmetry breaking. The $\pi$ resonance can be
identified naturally with the neutron resonance observed in the
HTSC cuprates \cite{ROSSATMIGNOD1991,MOOK1993,FONG1995}. 
In this section we will review basic experimental
facts about such resonances and discuss a theoretical scenario in
which they originate from the pseudo Goldstone modes associated
with the $SO(5)$ symmetry. The operator of the $\pi$-mode is a
symmetry generator of the $SO(5)$ symmetry, so the appearance of
the low lying resonance tells us about a small energy difference
between the $d$-wave SC and AF ground states of the doped
cuprates. The idea of the near degeneracy of the $d$-wave SC and
AF states lies at the heart of the $SO(5)$ approach, which assumes
that fluctuations between these two  states exhaust the low energy
sector of the system. Hence, experimental observation of the low
lying resonances provide a key foundation to the $SO(5)$ approach
to competing AF and SC in the cuprates. In this section we provide
several perspectives on the $\pi$-excitations. First, we use the $SO(5)$
non-linear sigma model to describe them as pseudo-Goldstone modes
of the approximate $SO(5)$ symmetry of the system. Second, we show that
the Fermi liquid analysis of the weakly interacting electron gas
in a two dimensional tight binding lattice produces the $\pi$-mode
as a sharp collective mode and gives a simple picture of this
excitation as an anti-bound state of two electrons with the total
spin $S=1$ and with the center of mass momentum $\Pi=(\pi,\pi)$.
Such excitation contributes to the spin fluctuation spectrum,
measured by the inelastic neutron scattering, only in the SC state
when there is a condensate of Cooper pairs. Finally, we discuss an
important role that the $\pi$-resonance plays in stabilizing the
SC state.

\subsection{Key experimental facts}
Resonant peak in the SC state of the cuprates was first observed
in optimally doped $YBa_2Cu_3O_7$
\cite{ROSSATMIGNOD1991,ROSSATMIGNOD1991A,ROSSATMIGNOD1992}.
Further experiments \cite{MOOK1993,FONG1995} established that this
is a magnetic resonance (spin $S=1$) at the AF wavevector
$\Pi=(\pi,\pi)$ which appears in the SC state. It has a constant
energy $\omega_0=41~{\rm meV}$ at all temperatures and intensity
that is strongly temperature dependent and vanishes at $T_c$.
Similar resonances have then been found in underdoped
$YBa_2Cu_3O_{6+x}$
\cite{DAI1996,FONG1996,MOOK1998,DAI1998,FONG2000,STOCK2003} and in
$Bi_2Sr_2CaCu_2O_{8+\delta}$ \cite{FONG1999A,HE2001} and
$Tl_2Ba_2CuO_{6+\delta}$ \cite{HE2002}.

An important feature of magnetic scattering in underdoped
$YBa_2Cu_3O_{6+x}$ \cite{DAI1996,DAI1998,MOOK1998,FONG2000} is
that the resonance precursors are detectable above $T_c$. Magnetic
correlations, however, are strongly enhanced in the SC state, and
there is a cusp in the temperature dependence of the resonant
scattering intensity at $T_c$ \cite{FONG2000}. Doping dependence
of the resonance energy and intensity indicate a strong
enhancement of magnetic fluctuations as we approach half-filling:
for underdoped $YBa_2Cu_3O_{6+x}$ the resonance energy decreases
with decreasing doping, and the intensity increases \cite{FONG2000}.
For overdoped $Bi_2Sr_2CaCu_2O_{8+\delta}$ it was found that the
energy decreased \cite{HE2001,HE2002}, which led to a suggestion
that the resonance energy follows the SC transition temperature
\cite{HE2001}.

The presence of the magnetic resonance in the SC state of many
cuprates suggests that it is closely related to the SC pairing.
This idea was reenforced by the experiments of Dai {\it et.al}
\cite{DAI2000}, in which the SC coherence in $YBa_2Cu_3O_{6.6}$
was suppressed by applying a magnetic field. It was found that the resonance
intensity decreased without any noticeable change in the resonance
energy. Finally, in \cite{DAI1999} it was demonstrated that the
exchange energy associated with the resonance has the right
magnitude, the temperature and doping dependences to
describe the SC condensation energy of $YBa_2Cu_3O_{6+x}$
materials.

\subsection{Contribution of the $\pi$ resonance
to the spin correlation function}

\label{PiFromINSExperiments}

Resonance that appears in the SC state suggests that what gets
scattered is Cooper pairs which are only present below $T_c$.
Based on this idea it was proposed \cite{DEMLER1995} that the
resonance observed in inelastic neutron scattering experiments is
due to the presence of the $\pi$-mode, a sharp collective mode in
the particle-particle  channel at momentum $\Pi =(\pi,\pi)$ with
spin $S=1$. In the normal state such excitation does not
contribute to the magnetic spectrum, since the latter is
determined by fluctuations in the particle-hole  channel.  Below
$T_c$, on the other hand, condensed Cooper pairs couple the
particle-hole and particle-particle channels
\cite{DEMLER1995,DEMLER1998A} and cause the $\pi$ excitation to
appear as a sharp resonance in the magnetic spectrum with
intensity set by the strength of mixing of the two channels,
$|\Delta(T)|^2$, where  $\Delta(T)$ is the amplitude of the SC
order parameter. Such a scenario provides a natural explanation for
the key properties of the observed resonance: its energy is
essentially the energy of the $\pi$ mode in the normal state and
is temperature independent \cite{DEMLER1998A}, whereas the
intensity of the resonance is set by $|\Delta(T)|^2$ and vanishes
at $T_c$.

Coupling of the particle-particle $\pi$-channel and the
particle-hole AF channel may be understood using the commutation
relations between the operators $\pi_\alpha$ and $N_\beta$ given
in equation (\ref{pi_n}).  In the SC state the $d$-wave SC order
parameter that enters the right hand side of equation (\ref{pi_n})
can be replaced by its expectation value in the ground state.
Hence, the commutator of $\pi$ and $N$ becomes a ${\rm c}$-number,
and the two fields become conjugate variables, just as coordinate
and momentum are conjugate to each other in elementary quantum
mechanics. The result of such coupling is that the $\pi$-mode
appears as a sharp resonance in the spin fluctuation spectrum. To
demonstrate this we consider the spin-spin correlation function at
wavevector $\Pi$
\begin{eqnarray}
\chi(\Pi,\omega) = -i \int e^{-i \omega t} \langle T N_\alpha(t)
N_\alpha(0) \rangle dt = \sum_n  |\langle n | N_\alpha | 0 \rangle|^2
\left\{ \frac{1}{\omega-E_n+i0}-\frac{1}{\omega+E_n-i0}
 \right\}.
\end{eqnarray}
Here $| 0 \rangle$ is the ground state and $n$-summation goes over
all excited states of the system.  One of the excited states is
created by the $\pi$-operator defined in equation (\ref{pi_k})
\begin{eqnarray}
|\pi_\alpha \rangle = \frac{1}{{\cal N}} \pi_\alpha^\dagger | 0 \rangle,
\end{eqnarray}
where ${\cal N}$ is the normalization factor.

It is useful to realize that if $\pi^\dagger$ acting on the ground
state creates an excited state, then $\pi$ should annihilate it
(otherwise it would create a state of lower energy than the ground
state \cite{NOZIERES1966}). Then we have
\begin{eqnarray}
1= \langle \pi_\alpha | \pi_\alpha \rangle = \frac{1}{{\cal N}^2} \langle 0
| \pi_\alpha \pi_\alpha^\dagger | 0 \rangle =
\frac{1}{{\cal N}^2} \langle 0
| \left[\pi_\alpha, \pi_\alpha^\dagger \right] | 0 \rangle \approx
\frac{(1-n)}{{\cal N}^2},
\end{eqnarray}
where $n$ is the filling fraction ($n=1$ corresponds to
half-filling). In writing the last equality we assumed $\langle
(g(p))^2 \rangle =1$ when averaged around the Fermi surface.

If we separate the contribution of the $\pi$ state to
$\chi(\Pi,\omega)$ we have
\begin{eqnarray}
\chi(\Pi,\omega) = | \langle \pi_\alpha | N_\alpha | 0 \rangle |^2
\frac{1}{(\omega-\omega_\pi+i0)} +{\rm part\;
regular\;at\;\omega_\pi}.
\end{eqnarray}
The resonant contribution to $\chi(\Pi,\omega)$ can be expressed
as
\begin{eqnarray}
\chi^{\rm res}(\Pi,\omega) &=& \frac{1}{{\cal N}^2} | \langle 0|
\pi_\alpha  N_\alpha | 0 \rangle |^2
\frac{1}{(\omega-\omega_\pi+i0)} = \frac{1}{{\cal N}^2} | \langle
0| \left[ \pi,  N \right] | 0 \rangle |^2
\frac{1}{(\omega-\omega_\pi+i0)}
\nonumber\\
&\approx&\frac{|\langle 0 | \Delta | 0 \rangle |^2}{(1-n)}
\frac{1}{(\omega-\omega_\pi+i0)}. \label{ExpressionChiRes1}
\end{eqnarray}
The expectation value in the numerator of the last expression is
simply the amplitude of the superonducting $d$-wave order
parameter. We emphasize that Eq. (\ref{ExpressionChiRes1}) does not
rely on the details of the microscopic model but only on the
commutation relations between the $\pi$, $N$, and $\Delta$ given
by the equation (\ref{pi_n}) (this is somewhat analogous
to the $f$-sum rule \cite{NOZIERES1966}). To relate the order
parameter to what one typically measures in experiments we use BCS
type arguments to connect the order parameter to the quasiparticle
gap (see however \cite{UEMURA1989,EMERY1995}) $\langle 0 |
\Delta | 0 \rangle=C \Delta_0/V_{BCS}$. Here $\Delta_0$ is the
maximal gap for Bogoliubov quasiparticles at the antinodal point,
$V_{BCS}$ is the interaction strength that we expect to be of the
order of the nearest-neighbor exchange coupling $J$, and $C$ is a
dimensionless constant of the order of unity. Therefore, we find
\begin{eqnarray}
\chi^{\rm res}(\Pi,\omega) = C^2\,\,
\frac{|\Delta_0|^2}{J^2(1-n)}\frac{1}{\omega-\omega_\pi+i0}.
\label{ExpressionChiRes2}
\end{eqnarray}
As we go to the underdoped regime, $\Delta_0$ remains constant or
increases slightly, and the factor $1-n$ decreases.
Eq. (\ref{ExpressionChiRes2}) predicts that the
intensity of the resonance should increase;
this increase has been observed in the experiments 
in Ref. \cite{FONG2000}.

It is useful to note that contributions from modes other than the
$\pi$ excitation do not spoil the result in
(\ref{ExpressionChiRes1}). If most of the $\pi$-spectrum is
accommodated in an interval $(\omega_\pi - \nu, \omega_\pi +
\nu')$, one can use the Cauchy-Schwarz inequality to prove a
rigorous and model-independent result\cite{DEMLER1998A} that
\begin{eqnarray}
\frac{1}{\pi} \int_{\omega_\pi - \nu}^{\omega_\pi + \nu'} d \omega
{\rm Im} \chi^{\rm res}(\Pi,\omega) \geq \frac{|\Delta|^2 }{1-n} .
\label{lower-b}
\end{eqnarray}
 The left hand side of this equation represents the
contribution of the $\pi$ mode to the spin excitation spectrum and
the right hand side gives its lower bound. Exact equality holds
when $\pi$ operator is an exact eigenoperator and hence there is
only one energy eigenstate which satisfies $ \langle 0 |
\pi_\alpha | n \rangle \neq 0$.

Thus, a simple picture of the resonant neutron scattering is as
follows: when an incoming neutron is scattered off one of the
electrons in a Cooper pair, it transfers a momentum of $(\pi,\pi)$
to this electron and flips its spin. At the end of the
scattering process the Cooper pair has quantum numbers of the
$\pi$-mode, spin $S=1$ and momentum $\Pi$.  If the energy transfer
matches the energy of the $\pi$ excitation, we have a resonance.
In the next two sections we build upon this simple argument to
establish a more detailed picture of the $\pi$-resonance in  the
two cases - the strong coupling limit described by the $SO(5)$
non-linear $\sigma$ model, and the weak coupling limit where the
Fermi-liquid type analysis can be applied.

\subsection{$\pi$-resonance in the strong coupling:
the SO(5) non-linear $\sigma$ model and the projected SO(5) model}

\label{PiPseudoGoldstone}

In this section we review how the resonant peak observed in the
inelastic neutron scattering experiments appears in the $SO(5)$
non-linear $\sigma$ model, signalling competition between the AF
and SC ground states. We use the Hamiltonian of this model
(see equation (\ref{Hsigma})) to write the operator equations of
motion ($\dot{O}=i[{\cal H},O]$) for the order parameters, $n_a$,
and symmetry generators, $L_{ab}$, with $a,b=\{1,\dots,5\}$.  For
$\mu> \mu_c=\frac{1}{2}\sqrt{g/\chi}$ the system is in the SC
ground state, which we can take to be along the $n_1$ direction.
Linearizing equations of motion around $n_1$ we obtain
\begin{eqnarray}
\chi \partial_t^2 n_5 &=& \rho \nabla_k^2 n_5
\label{EOM_bogoliubov}
\\
\chi \partial_t^2 n_\alpha &=& \rho \nabla_k^2 n_\alpha -[\chi (2
\mu)^2-g] n_\alpha. \label{EOM_pimode}
\end{eqnarray}
The first equation describes the Goldstone mode of the
spontaneously broken charge U(1) symmetry (Bogoliubov-Anderson
mode) and the second equation describes a triplet massive
excitation of the superspin in the direction of the
AF state (see Fig. \ref{sphere}).

In a model with exact $SO(5)$ symmetry superspin ordering reduces
the symmetry from $SO(5)$ to $SO(4)$ and should be accompanied
by the appearance of four Goldstone modes ($SO(5)$ and $SO(4)$ have
ten and six symmetry generators respectively). In the case of
approximate $SO(5)$ symmetry that we discuss here, explicit
symmetry breaking turns some of the Goldstone modes into
pseudo-Goldstone excitations, i.e. they acquire a finite energy.
This is similar to a  chiral symmetry breaking in quantum
chromodynamics, where a small mass of the quarks leads to a finite
mass of pions, which are the Goldstone bosons of the chiral
symmetry breaking \cite{WEINBERG1995}, but it does not change the
fundamental nature of the latter.

The doping dependence of the resonance energy follows immediately
from the equation (\ref{EOM_pimode})
\begin{eqnarray}
\omega_\pi = 2\sqrt{\mu^2-\mu_c^2}.
\label{PiResonanceEnergyNonLinearSigmaModel}
\end{eqnarray}
The resonance energy is zero at the $SO(5)$ symmetric point
$\mu=\mu_c$ and increases with doping. Vanishing of the resonance
energy at $\mu_c$ is a special property of the $SO(5)$ symmetric
point, and for a generic first order transition between the AF and
SC phases (see Fig. \ref{Fig_T_mu}a) the resonance energy
would remain finite at the transition point. When there is an
intermidiate uniform mixed  AF/SC phase (``type 2" transition shown
in Fig. \ref{Fig_T_mu} (c)), the doping dependence of the
resonance energy also obeys
(\ref{PiResonanceEnergyNonLinearSigmaModel}) with $\mu_c$
corresponding to the boundary between the SC and AF/SC phases 
($\mu_{c2}$ in Fig. \ref{Fig_T_mu}). Softening of the $\pi$ mode in this
case demonstrates a continuous transition into a state with
magnetic order \cite{SACHDEV2000,DEMLER2001}.

The dispersion of the $\pi$ resonance mode is model dependent. Hu
and Zhang\cite{HU2001A} studied the dispersion of the $\pi$
resonance mode in the projected $SO(5)$ model using the strong
coupling expansion, and concluded that the $\pi$ mode can have a
downward dispersion away from the $\Pi$ point, reaching a minimum
at some incommensurate wave vector. This model could possibly give
a unified description of the neutron resonance mode and the
incommensurate magnetic fluctuations in the HTSC cuprates.

In Section \ref{sec:so5bosons} we discussed the projected $SO(5)$
model that forbids double occupancy of the Cooper pairs by
introducing chirality into SC rotations. As was pointed out
before, such a projection does not affect small fluctuations around
the SC state (see Fig. \ref{Fig_chiral}) and does not change the
relation (\ref{PiResonanceEnergyNonLinearSigmaModel}).

\subsection{$\pi$-resonance in weak coupling: the Fermi liquid analysis}
\label{PiFermiLiquid}

In this section we consider a weakly interacting electron gas in a
two dimensional square lattice and show that the Fermi liquid
analysis of this system gives rise to the $\pi$-mode that is very
similar to the collective mode we discussed earlier in the
strong coupling limit.  Using perturbative Fermi liquid analysis
to describe strongly interacting electron systems, such as
cuprates, may cause reasonable objections from some readers. We
remind the reader, however, that the goal of this exercise is to complement
strong coupling discussion presented in the earlier sections. The
benefit of the weak coupling discussion is that it provides a
simple picture of the $\pi$-mode as an anti-bound state of two
electrons in the spin triplet state having a center of mass
momentum $\Pi$ and sitting on the neighboring lattice sites.

Our starting point is the $t$-$J$ type model on a two-dimensional
lattice
\begin{eqnarray}
{\cal H} = -t \sum_{\langle ij \rangle \sigma} c^\dagger_{i\sigma}
c_{j\sigma} + U \sum_i n_{i\uparrow} n_{i\downarrow} + J
\sum_{\langle ij \rangle } \vec{S}_i \vec{S}_j.
\label{HubbardtJHamiltonian}
\end{eqnarray}
Note that we do not impose a no-double occupancy constraint but
include the on-site Hubbard repulsion. Within the Hartree-Fock
discussion presented here, the Hubbard $U$ only renormalizes the
band structure, but it does not affect collective excitations of
the order of $J$. Therefore, in the rest of the paper we will
disregard the $U$ term in the Hamiltonian
(\ref{HubbardtJHamiltonian}) and assume that we work with the
renormalized parameters.

To begin, we consider adding two non-interacting electrons into an
empty two dimensional lattice with the condition that the center of
mass of the pair has momentum $q$. For a general $q$ the energy of
such a pair, given by $\epsilon_{q-k}+\epsilon_k$, depends on the
relative momentum of the two electrons.  Therefore, we have a
continuum of particle-particle excitations.  When the center of mass
momentum is $\Pi=(\pi,\pi)$ the whole particle-particle continuum
collapses to a point. This can be verified by taking the tight binding
dispersion $\epsilon_k=-2t (cos k_x + cos k_y)$ and is shown
schematically in Fig.  \ref{continuum}. The collapse of the continuum
makes it easier to create resonant states by adding interaction between
the electrons. For example, the $J$ term in the Hamiltonian
(\ref{HubbardtJHamiltonian}) introduces an energy cost of $J/4$ for
electrons sitting on the nearest neighbor sites when their spins point
in the same direction. Thus, if we make a two electron pair in such a
way that the two electrons form a triplet pair on the nearest neighbor
sites and have a center of mass momentum $(\pi,\pi)$, we get an
anti-bound state separated from the continuum by energy $J/4$.
  The argument above can
be generalized to the case of adding two electrons on top of the
filled Fermi sea. We recall that collective modes correspond to
poles of the vertex functions \cite{ABRIKOSOV1993}.  In the case
of the $\pi$-resonance, we are interested in the particle-particle
vertex, which we describe by the Dyson's equation
\cite{DEMLER1995} after separating the spin triplet component of
the interaction at the center of mass momentum $\Pi$ with the
$d$-wave symmetry of the electron pair
\begin{eqnarray}
{\cal H}_J= \frac{J}{4} \sum_{pp'} d_p d_{k} c^\dagger_{p+\Pi\alpha}
(\sigma_2 \vec{\sigma} )_{\alpha\beta} c^\dagger_{-p\beta}
c_{-k\gamma} (\vec{\sigma} \sigma_2)_{\gamma \delta} c_{k+\Pi\delta}
+ \dots
\end{eqnarray}
From the equation presented in Fig. \ref{dyson} we find the
triplet particle-particle vertex
\begin{eqnarray}
T(p,p',\Pi,\omega)=\frac{\frac{J}{4} d_p d_{p'}}{1-\frac{J}{4}\sum_k
\,d_k^2\,\,
\frac{1-n_k-n_{k+\Pi}}{\omega-\epsilon_k-\epsilon_{k+\Pi}}}
\end{eqnarray}
and observe that it has a pole at energy
\begin{eqnarray}
\omega_\pi = -2 \mu +\frac{J}{4}(1-n).
\label{WeakCouplingResonance}
\end{eqnarray}
The first term in (\ref{WeakCouplingResonance}) originates from
the kinetic energy of the tight binding Hamiltonian $ \epsilon_p +
\epsilon_{p+\Pi} = -2 \mu $, and the second part of
(\ref{WeakCouplingResonance}) describes the nearest-neighbor
exchange interaction of the triplet pair of electrons in the
presence of a filled Fermi sea. The $(1-n)$ factor describes
the blocking of the states below the Fermi energy from the phase space
available for two particle scattering.  In the Hartree-Fock theory
the chemical potential is proportional to doping; hence, we find
that the resonance energy in Eq. (\ref{WeakCouplingResonance}) scales
with $x$. It is useful to point out that including the
near-neighbor density interaction $V\sum_{\langle i j \rangle} n_i
n_j$ in the Hamiltonian (\ref{HubbardtJHamiltonian}) will not
change our discussion as long as the system remains in the
$d$-wave SC state \cite{MEIXNER1997,DEMLER1998A}. Such an interaction
affects equally the $\pi$ mode and Cooper pairs that constitute
the ground state.

One can also ask how to use the perturbative approach to
demonstrate the appearance of the $\pi$-resonance in the
spin-fluctuation spectrum below $T_c$.  In Fig. \ref{chi} we show
that when we compute the spin-spin correlation function in the SC
state, we need to include scattering of spin fluctuations at
momentum $(\pi,\pi)$ into the $\pi$ pair, which corresponds to
mixing the particle-particle ladder of diagrams into the
particle-hole bubble.
This contribution requires two anomalous Green's functions
and is therefore proportional to $|\Delta|^2$. Detailed calculations based on
generalized random phase approximation for the model
(\ref{WeakCouplingResonance}) were presented in \cite{DEMLER1998A};
in Fig. \ref{qplot}, we only show a representative plot of
spin-spin correlation function $\chi(q,\omega)$ computed with an
account of the $\pi$ channel.

In summary, we used a Fermi liquid analysis to establish a
simple picture of the $\pi$-resonance as a triplet pair of
electrons sitting on the nearest neighbor sites with the $d$-wave
function of the pair and with the center of mass momentum
$\Pi$.

\subsection{Resonance precursors in the underdoped regime}

In the underdoped cuprates the resonance does not disappear above
$T_c$ but remains as a broad feature at higher temperatures
\cite{DAI1996,FONG1996,MOOK1998,DAI1998}, with only a cusp in the
temperature dependence of the intensity signalling the onset of
the long range $d$-wave SC order \cite{FONG2000}.  In Ref.
\cite{ZHANG1998,DEMLER1999A}, it was pointed out that the most
likely origin of these resonance precursors is the existence of
strong $d$-wave SC fluctuations in the pseudogap regime of the
underdoped cuprates. Precursor of the $\pi$-resonance in the
spin-spin correlation function can be identified with a process in
which a $\pi$-pair and a preformed Cooper pair propagate in
opposite directions, as shown in Fig. \ref{PiResonancePrecursor}.
Because uncondensed Cooper pairs have a finite energy, we expect
precursors to appear at a slightly higher energy than the resonance
itself and have a width of the order of temperature
\cite{DEMLER1999A}.

\subsection{Implications for experiments and comparison to other theories}

In Section \ref{PiPseudoGoldstone} we discussed the
$\pi$-resonance as a pseudo-Goldstone mode of the $SO(5)$
non-linear $\sigma$-model, and in Section \ref{PiFermiLiquid} we
gave a simple microscopic picture of the $\pi$-mode as a sharp
collective mode in the particle-particle channel with spin $S=1$
and momentum $\Pi=(\pi,\pi)$. From Eq. (\ref{ExpressionChiRes1}),
we see that the $\pi$ resonance intensity due to the contribution
from the particle-particle channel scales with the square of the
SC order parameter, namely
\begin{eqnarray}
I(\Pi) = \int d\omega Im \chi^{\rm res}(\Pi,\omega) \propto
|\langle \Delta (x, B, T) \rangle|^2. \label{intensity}
\end{eqnarray}
Here we have explicitly exhibited the dependence of the SC order
parameter $\Delta(x, B, T)$ on doping $x$, magnetic field $B$ and
temperature $T$. Therefore, this simple scaling relation makes
powerful predictions on the resonance intensity and has been
tested in a number of experiments. Our analysis explains several
puzzling features of the resonance observed in experiments. The
first is the striking contrast between its temperature dependent
intensity and temperature independent energy. Taking the
Bardasis-Schrieffer exciton \cite{BARDASIS1961} that appears as a
bound state below the quasiparticle gap for $s$-wave
superconductors, both energy and intensity of the exciton will be
determined by the SC gap; hence, as the temperature is increased
in the SC state, both the resonance energy and its intensity
decrease. In the case of the $\pi$-mode, on the other hand,
different behavior of the resonance intensity and energy are
expected. The energy is essentially given by the energy of the
$\pi$-mode in the normal state and does not change with
temperature. The resonance intensity is set by the $d$-wave SC
order parameter, as given in Eq. (\ref{intensity}), and 
decreases with increasing temperature and vanishes at $T_c$. Eq.
(\ref{intensity}) also predicted that the suppression of the SC
coherence by a magnetic field should lead to a rapid decrease in
the resonance intensity without changing the resonance energy.
This prediction was confirmed experimentally in a striking
experiment by Dai {\it et.al} \cite{DAI2000}, reproduced here in
Fig. \ref{FigDai2000}. The $SO(5)$ theory predicted that with
decreasing doping the resonance intensity should increase (see Eq.
(\ref{ExpressionChiRes2})) and its energy should decrease (see
Eqs. (\ref{PiResonanceEnergyNonLinearSigmaModel}) and
(\ref{WeakCouplingResonance})) \cite{DEMLER1995,ZHANG1997}, both
of which have been observed in experiments, as we show in Fig.
\ref{FigFong2000}. Note that for small
values of the chemical potential there is a small difference in the
precise $\omega_\pi$ vs $\mu$ relation obtained from the
non-linear $\sigma$ model and the Fermi liquid analysis. We expect the strong
coupling expression (\ref{PiResonanceEnergyNonLinearSigmaModel})
to be more reliable close to the AF/SC transition where $\mu\sim
\mu_c$ and suggest that comparison of the doping dependence of
the resonance energy \cite{FONG2000} and the chemical potential
\cite{INO1997,FUJIMORI1998} should be an important test of the
$SO(5)$ theory.

After the $\pi$ resonance theory\cite{DEMLER1995} was developed,
alternative descriptions of the resonance
\cite{MAZIN1995,LIU1995,BLUMBERG1995,MILLIS1996,BULUT1996,ONUFRIEVA1995,BARZYKIN1995,YIN1997,MORR1998,ASSAAD1998,WENG1998,YOSHIKAWA1999,BRINCKMANN1999,SACHDEV2000}
have also been proposed. These typically discuss the resonance as a magnetic
exciton that is overdamped in the normal state but becomes sharp
in the $d$-wave SC state when a gap opens up for single particle
excitations. In the $d$-wave SC state, the particle-particle
channel and the particle-hole channels are mixed into each other
and there are, strictly speaking, no rigorous distinctions among these
different theories. However, important quantitative predictions
differ in details. Near the $T_c$ transition, the $\pi$ resonance
theory predicts a sharp onset of the magnetic resonance due to
the coupling to the particle-particle channel, whose contribution
to the magnetic scattering can be rigorously established via the
Cauchy-Schwarz inequality, as shown in Eq. (\ref{lower-b}). Some
of these alternative theories expect a gradual broadening of the
resonance rather than a sharp reduction of the intensity as $T_c$
is approached from below. The $\pi$ resonance theory predicts that
the energy of the magnetic resonance mode is independent of the
temperature near $T_c$, while some of the alternative theories
predict that the mode energy should vanish as the SC gap. In
section \ref{sec:p-p}, we shall discuss a rigorous distinction
between the $\pi$-mode in the particle-particle channel and the
magnetic exciton in the particle-hole channel in the {\it normal
state}, and discuss an experimental proposal where this
distinction can be tested.



Several proposals have been made regarding implications of the
resonance peak for various properties of the cuprates (see
\cite{KEE2002} for a critical review).  Scattering of
quasiparticles on the $\pi$-mode was argued to be responsible for
the ``kink'' in the quasiparticle dispersion \cite{JOHNSON2001},
``peak-dip-hump''structure measured in ARPES
\cite{ESCHRIG2000,ABANOV2001}, and the pseudogap seen in optical
conductivity \cite{SCHACHINGER2001}.  SC pairing mediated by the
resonance was suggested in
\cite{ABANOV2001,CARBOTTE1999,ORENSTEIN1999,ZASADZINSKI2003}, 
and relation between
the resonance intensity and the condensate fraction was pointed
out in Ref. \cite{CHAKRAVARTY2000}. We do not discuss these
proposals here, but in the next section we will review an
important role that the resonance plays in thermodynamics of the
SC state. We will argue that the SC condensation energy may be
accounted for by lowering of the spin exchange energy due to the
appearance of the resonance below $T_c$ \cite{DEMLER1998}.

\subsection{Microscopic mechanism and the condensation energy}
The central question in the field of HTSC concerns the microscopic
mechanism of superconductivity. In conventional superconductors,
the pairing interaction is mediated by the phonon interactions
(see Ref.\cite{MAKSIMOV1997} for a review).
Within the weak coupling BCS theory, the vertex corrections are
suppressed by a small parameter, namely the ratio of the electron
mass to the nuclei mass. Thus, the interaction which
mediates the pairing of electrons can be unambiguously determined.
In the case of HTSC, the dominant interaction is the Coulomb
interaction and the AF exchange interaction. In such a strongly
coupled system, the traditional approach based on the Feymann
diagram expansion does not work, and the nature of the pairing
interaction is not easily revealed by studying low order diagrams.
However, the mechanism of superconductivity can still be addressed
by identifying the interaction terms in the Hamiltonian which is
lowered in the SC state. By comparing the magnitude of the energy
saving associated with a particular interaction term with the
actual experimental measurement of the condensation energy, the
mechanism of superconductivity can be unambiguously identified. In
our discussion in the previous section we showed that the
$\pi$-mode contributes to the spin fluctuation spectrum below
$T_c$ and, therefore, enhances AF correlations in the SC state. In
Ref. \cite{DEMLER1998}, it was shown that 
the $\pi$-resonance can be promoted from being
a consequence of superconductivity to being the real driving force
behind the electron pairing. By analyzing the neutron
scattering data, Demler and Zhang demonstrated that lowering of
the AF exchange energy in the supercoducting state due to the
appearance of the $\pi$-resonance can be sufficient to stabilize
superconductivity in the first place.  In this section we provide
the details of this condensation energy argument focusing on the
microscopic $t$-$J$ model and discuss its relevance to the
condensation energy of $YBa_2Cu_3O_{6+x}$ materials. We also
demonstrate that this scenario can be formulated
as an additional contribution to the BCS coupling constant
in weak coupling.

\subsubsection{The $\pi$ Resonance Contribution to the Condensation Energy}
\label{Condensation_Microscopic}

The SC condensation energy is defined as the energy difference
between the SC and the normal states at $T=0$
\cite{SCHRIEFER1964,TINKHAM1995}. In type I superconductors it can
be obtained directly by measuring the critical value of the
magnetic field, $H_c$, at the first order transition between the
normal and SC states. At the transition point, the
energies of the two phases are equal (note that at $T=0$ the free
energy is equal to the energy) and, assuming that the normal state
is not affected by the magnetic field, we obtain the condensation
energy per unit cell
\begin{eqnarray}
E_C = E_N-E_S= \frac{V_0 H^2_c}{8\pi},
\end{eqnarray}
where $V_0=a\times b \times c$ is the volume of the unit cell. For
type II superconductors including the HTSC, such a simple argument
is not available. However, one can use LG theory to relate the
condensation energy to $H_{c1}$ and $H_{c2}$, or alternatively to
the SC coherence length $\xi_0$ and London penetration depth
$\lambda$ \cite{TINKHAM1995}:
\begin{eqnarray}
H^2_c = \frac{\Phi_0}{8\pi \xi_0 \lambda},
\end{eqnarray}
where $\Phi_0=hc/2e$ is the SC flux quantum.  An alternative
approach to measuring the condensation energy is to integrate the
difference between the SC and the normal state specific heat from
$T=0$ to $T_c$, where the normal state specific heat below $T_c$
is defined as extrapolation from temperatures above the transition
point \cite{LORAM1990,LORAM1994}.  To be more precise, let us consider
the condensation energy of the optimally doped $YBa_2Cu_3O_7$. 
Taking the characteristic values $\xi_0 =12 -20 \AA$ and $\lambda =
1300-1500 \AA$, with $a=b=3.85 \AA$ and $c=11.63 \AA$, we find the
condensation energy of $E_C=3.5-12\;K$ per unit cell.
The determination of the $E_C$ of this material using specific heat
measurements by Loram {\it et.al.} \cite{LORAM1990,LORAM1994} gave
$E_C=6\;K$ per unit cell.

Ideally, one would like to start with a microscopic model that has
kinetic energy of electrons and ions, and the Coulomb energy of all
particles, and calculate the condensation energy from first
principles. Although possible in principle, in practice
this approach is very
hard to accomplish  because of large scales involved in
both the kinetic and the Coulomb energies. A method that is easier
to pursue in practice is to start with an effective model defined
on a much smaller energy scale and try to calculate the
condensation energy within this effective model. 
This approach has been undertaken by Scalapino and
White \cite{SCALAPINO1998A} within the
$t$-$J$ model.  In the $t$-$J$ Hamiltonian in
equation (\ref{t-J}), we have two terms: the kinetic energy of
electrons (with the Gutzwiller projection operator) and the
exchange energy of electrons. Analogous to conventional
superconductors, we expect that the transition into the SC state
is driven primarily by lowering the interaction part of the
Hamiltonian , i.e. the exchange term (in conventional
superconductors the relevant interaction is electron-ion Coulomb
interaction). Is it possible then to find the change in the
exchange energy between the normal and SC states? Scalapino
and White made the insightful observation that the value of the
$J$ term in equation (\ref{t-J}) is directly related to the
dynamic spin structure factor $\chi''(q,\omega)$, the quantity
that is being measured directly in neutron scattering experiments.
And the change in the exchange energy, $\Delta E_J = E_J^N -
E_J^S$, can be directly expressed as a frequency and momentum
integral of the difference in dynamic spin structure factors
$\chi''_N(q,\omega)-\chi''_S(q,\omega)$ with a form factor coming
from the interaction being near neighbor
\begin{eqnarray}
\Delta E_J = 3J ( \frac{a}{2\pi} )^2 \int_{-\pi/a}^{\pi/a} d^2 q
\int_0^\infty \frac{d (\hbar \omega)}{\pi}
(\chi''_N(q,\omega)-\chi''_S(q,\omega))(cos (q_x a)+ cos (q_y a)).
\label{EJ}
\end{eqnarray}
This equation applies to the quasi-two-dimensional systems, and
$q=(q_x,q_y)$ is a two dimensional in-plane momentum.
The generalization of  (\ref{EJ})  to the bilayer systems, the case
relevant for $\rm{YBa_2Cu_3O_{6.35}}$, is given in
Ref. \cite{DEMLER1998}.

The quantity $\chi''_N(q,\omega)$ in equation (\ref{EJ}) is not
the normal state spin structure above $T_c$ but rather an
extrapolated normal state quantity at $T=0$. Experimentally, one
has to carefully identify features in $\chi''(q,\omega)$ which
change abruptly at $T_c$. From inelastic neutron scattering
experiments we know that the most drastic change between the SC
and the normal state spin structure factors is the appearance of
the 41 meV resonance. Even for underdoped materials, which have
many more AF fluctuations in the normal state, the main change
between the normal and SC states is the appearance of the
resonance \cite{FONG2000}. It is then reasonable to take formula
(\ref{EJ}) for $\Delta E_J$, calculate the contribution of the
$\pi$-resonance and argue that this will be the dominant
contribution. For optimally doped $YBa_2Cu_3O_{6.35}$ Fong {\it
et.al} \cite{FONG1996} measured the absolute intensity of the
resonance $ \int_0^\infty d (\hbar \omega) \chi''_S(\Pi,\omega)$
to be $0.52$ at $T=10\;K$. This resonance has a Gaussian profile
centered at $\Pi$ with a width $\kappa_{2D}=0.23 \AA^{-1}$, so the
two dimensional integral can be easily estimated, and
\begin{eqnarray}
\Delta E_J = \frac{3}{2} \pi (\frac{a}{2} \kappa_{2D})^2
\frac{1}{2} \; \frac{0.52}{\pi} = 0.016J. \label{dEJ}
\end{eqnarray}
Taking $J=100 meV$ we find that the change in the exchange energy
between the normal and  SC states is approximately $18\;K$ per
unit cell. This remarkable number tells us that
the resonance alone can account for the SC condensation
energy.

Regarding our estimate of $\Delta E_J$ in
equation (\ref{dEJ}), a comment must be made.  The dynamic spin structure factor
$S(q,\omega)$ satisfies the sum-rule \cite{SCALAPINO1998}
\begin{eqnarray}
3 ( \frac{a}{2\pi} )^2 \int _{-\pi/a}^{\pi/a} d^2 q \int_0^\infty
\frac{d (\hbar \omega)}{\pi} \chi'' (q,\omega) = (1-x) S(S+1).
\end{eqnarray}
Therefore, the spectral weight for the resonance needs to come
from other regions in $q$-$\omega$ space. In obtaining (\ref{dEJ})
we made an additional assumption that this weight was spread
uniformly in $q$ in $\chi''(q,\omega)$, and it did not contribute to
(\ref{EJ}), since any uniform component in $\chi''(q,\omega)$ is
cancelled by the $(cos (q_x a)+ cos (q_y a))$ factor in Eq. (
\ref{EJ}). It is also useful to point out that the weight of the
resonance is less than 1$\%$ of the total sum
rule\cite{DEMLER1998,KEE2002}, which, when multiplied by the AF
exchange energy $J$, gives the correct order of magnitude for the
condensation energy.

The condensation energy argument can be generalized to finite
temperatures. In this case the resonant peak intensity at
temperature $T$ should be related to the free energy difference
between the SC and normal states, which in turn is given by
the integral of the specific heat difference above $T$. This
hypothesis has been analyzed by Dai {\it et.al.} \cite{DAI1999},
who showed that the temperature derivative of the resonant peak
intensity follows very closely the specific heat anomaly for
different dopings of $YBa_2Cu_3O_{6+x}$. We show this comparison
in Fig. \ref{FigDai1999}. For optimal doping there is a BCS type
anomaly in the specific heat at $T_c$, which corresponds to the
resonance appearing abruptly in the SC state. For underdoped
samples the specific heat anomaly is broadened, which agrees with
the resonance precursors appearing above $T_c$. This highly
non-trivial experimental test establishes the connection between
the $\pi$ resonance and its contribution to the condensation
energy.


Therefore, we see that the $\pi$ resonance mode naturally accounts
for the condensation energy in the HTSC. The AF exchange
interaction is lowered in the SC state, and this energy saving can
drive the transition from the normal state to the SC state. Within
this scenario, the AF exchange energy is decreased, while the
kinetic energy is increased below the SC transition. On the other
hand, a number of theories argue that the dominant driving
mechanism of HTSC is the saving of the kinetic energy, either
along the $c$ axis, or in the $CuO_2$
plane\cite{Anderson1997,Chakravarty1999,Hirsch2000}. The $c$ axis
kinetic energy saving mechanism has been definitively ruled out by
experiments\cite{Moler1998}. The experimental measurement of the
$ab$ plane kinetic energy has not yielded conclusive
results\cite{Molegraaf2002,Boris2004}. The $\pi$ resonance based
AF exchange energy saving is an experimentally established
mechanism which can account for the condensation energy in the
HTSC cuprates. Recent experiments indicate that phonon mediated
attraction also plays a role in the mechanism of
HTSC\cite{LANZARA2001}. It is possible that various mechanisms
contribute constructively to the condensation energy in the HTSC.
In this case, it is important to quantitatively measure the
relative magnitudes of various contributions and identify the
leading contribution to the condensation energy.

\subsubsection{Microscopic Discussions and Relation to the BCS Pairing}
In the theory of Ref. \cite{DEMLER1998}, the saving of the AF
exchange energy arises from the coupling of the AF order parameter
$\vec{N}$ to the $\vec{\pi}$ operator in the SC state. This
coupling leads to the additional spectral weight, proportional to
$|\Delta|^2$, in the AF spin correlation function, thus lowering
the AF exchange energy. This argument is generally valid, in  both
strong and weak coupling limits. However, it is also useful to
connect this theory to the conventional BCS pairing theory in the
weak coupling limit. In the limit of weakly interacting electron
gas, we can formulate this scenario as a contribution to the BCS
coupling in the $d$-wave channel. In Fig. \ref{c_fig1}, we show a
schematic representation of such a contribution: a Cooper pair
splits into two virtual excitations -- a magnon ($\vec{N}$) and a
$\pi$-particle ($\vec{\pi}$) -- which then recombine into a Cooper
pair. One can easily verify that the quantum numbers are matched
in this process: quantum numbers of the combination of the
$\pi$-mode (charge $2$, momentum $\Pi$, spin $S=1$) and the magnon
(charge $0$, momentum $\Pi$, spin $S=1$) sum to exactly the
quantum numbers of the Cooper pair (charge $0$, momentum $q=0$,
spin $S=0$). This may also be formulated using electron Greens
functions, as shown in Fig. \ref{c_fig2}. We start with a Cooper
pair formed by the electrons $(p\uparrow)$ and $(-p\downarrow)$.
After the latter electron emits a magnon, shown as an upper
particle-hole ladder with total momentum $\Pi$ and spin $S_z=-1$,
we have two electrons with momentum $\Pi$ and spin $S_z=1$. These
are exactly the quantum numbers of the $\pi$-mode that we
describe by the lower particle-particle ladder in Fig.
\ref{c_fig2}.



\section{KEY EXPERIMENTAL PREDICTIONS}
\label{sec:experiment}
\subsection{The antiferromagnetic vortex state}
\label{sec:AFvortex} A fundamental prediction of the $SO(5)$
theory is the smooth rotation from the AF state to the SC state as
the doping density is varied. As shown in sections
\ref{sec:variational} and \ref{sec:diagonal}, this prediction has
been tested numerically within the $t-J$ model, with good
agreement. However, testing this prediction directly in
experiments would be much harder, since the doping level of most
cuprates cannot be controlled well in the regime where the
transition from the AF to SC state is expected to occur.
Therefore, Zhang\cite{ZHANG1997} and Arovas et al\cite{AROVAS1997}
proposed testing this prediction in the vortex state of underdoped
cuprates. Around the center of the vortex core, the phase of the
SC order parameter winds by $\pm 2\pi$, and the amplitude of the
SC order parameter is constrained to vanish at the center for
topological reasons. In conventional BCS superconductors, the metallic
Fermi liquid ground state is realized inside the vortex core. In
the $SO(5)$ theory, the SC order parameter is embedded as
components of a higher dimensional order parameter, namely the
superspin. When the amplitude of the SC order parameter vanishes in
the vortex core, the amplitude of the superspin order parameter
can still remain constant, provided that the superspin vector
slowly rotates from the SC direction into the AF direction as the
vortex core is approached. The superspin configuration near the
vortex core is shown in Fig. \ref{Fig_vortex}. This type of
topological field configuration is known as the meron solution,
meaning half of a Skyrmion\cite{Rajaraman1982}. 
Fig. \ref{Fig_vortex} shows the rotation of the superspin
in the vicinity of a vortex core. 
The AF order, which develops around the center of the
vortex core, can be measured directly in experiments and can
provide a quantitative test of the $SO(5)$ symmetry.


The key idea behind this prediction is more general. {\it When the
SC order is destroyed in the vortex core, the closest competing
order develops in the vortex state}. Aside from the commensurate or
incommensurate magnetic and charge order, a number of novel
correlation states have been proposed, including, for example, the
circulating orbital currents\cite{CHAKRAVARTY2001,LEE2002} and the
fractionalized excitations\cite{SACHDEV1992,SENTHIL2001}.
Therefore, the vortex core state can provide a key test for
various forms of the competing orders that have been
proposed\cite{SACHDEV2002}.

Magnetic field provides a clean tuning parameter
that can be used to investigate quantum transitions between 
the SC and AF phases.  By solving
both the $SO(5)$ non-linear sigma model and the LG model of competing
AF and SC order parameters, Arovas et al\cite{AROVAS1997} predicted
the existence of the AF vortex state in the underdoped cuprates and
further suggested a systematic experimental search for the AF vortex
state in neutron scattering and muon spin rotation experiments. These
authors also predicted that the magnetic field induced AF moment
should increase linearly with the applied magnetic field, or the
number of vortices in the system, when the applied magnetic field is
small compared to the upper critical field $B_{c2}$.  While the
original analysis of Arovas {\it et al} focused on the regime where
the transition between AF and SC is a direct first order transition
(corresponding to Fig. \ref{Fig_T_mu}a of the phase diagram), Demler
{\it et al}\cite{DEMLER2001,ZHANG2002} analyzed the case in which
there are two second order phase transitions with an intervening
uniform AF/SC mixed phase, corresponding to Fig.  \ref{Fig_T_mu}c of
the phase diagram. In this case the AF order extends far beyond the
vortex core region.  Analysis in Refs. \cite{DEMLER2001,ZHANG2002}
demonstrated that the suppression of the SC order in this regime is
dominated by the circulating super-currents and leads to a logarithmic
correction to the linear dependence of the field induced
moment. Recently, a number of experiments have been performed to test
the prediction of AF order in the vortex state. Neutron scattering
under a magnetic field can directly measure the field induced AF
moment. Katano et al\cite{KATANO2000} measured enhanced magnetic
scattering in the $La_{2-x}Sr_xCuO_4$ crystal at $x=12\%$ doping. The
intensity of elastic magnetic peaks around the $(\pi,\pi)$ point
increases at $B=10 T$ by as much as $50\%$. Lake et al\cite{LAKE2001}
observed enhanced dynamic AF spin fluctuations in optimally doped
$La_{2-x}Sr_xCuO_4$ crystal at $x=16\%$ doping in an applied magnetic
field. Without an applied field, the SC state has a spin gap of about
$6meV$. An applied field of $B=7T$ introduces a spectral weight in the
energy range of $3\sim 4 meV$. The mixed AF/SC phase has been also
investigated in both the underdoped $La_{2-x}Sr_xCuO_4$ crystal at
$x=10\%$ doping and in the $La_2CuO_{4+y}$ crystal. In both materials
the applied magnetic field strongly enhances the quasi-static AF
ordering\cite{LAKE2002,KHAYKOVICH2002}.  The field dependence of the
induced AF scattering is approximately linear, as predicted in Ref.
\cite{AROVAS1997}, and it agrees quantitatively with the $B \log
(B/B_{c2})$ form proposed in Ref. \cite{DEMLER2001}, with the correct
value of $B_{c2}$.  Another method to measure the AF order is the
nuclear magnetic resonance (NMR). In the vortex state, the magnetic
field is distributed inhomogeneously over the sample, with the maxima
centered at the vortex cores. Therefore, the NMR frequency correlates
directly with the location of the nucleus in the vortex lattice. Using
NMR on the $^{17}O$ nucleus of $YBa_2Cu_3O_7$ under a magnetic field
as high as $40 T$, Mitrovic et al\cite{MITROVIC2001,Mitrovic2002}
detected a sharp increase of the $1/T_1T$ rate near the vortex core as
the temperature is lowered, indicating enhanced AF ordering (see also
Ref.\cite{CURRO2000}).  Kakuyanagi et al\cite{Kakuyanagi2002}
performed $Tl$ NMR in the $Tl_2Ba_2CuO_{6+\delta}$ sample. $Tl$ NMR
provides a more direct test of the AF ordering, since $^{205}Tl$
nucleus is located directly above the $Cu$ spins. The temperature
dependence of the $1/T_1T$ rate shows that the AF spin correlation is
significantly enhanced inside the vortex core, compared with regions
outside.  The last class of magnetic experiments we discuss is the
muon spin rotation ($\mu$sR) experiments. When muons are stopped
inside a solid, their spin precesses around the local magnetic
field. Since the muon decays predominantly along the direction of its
spin, the spatial decay pattern yields direct information about the
local magnetic field distribution in a solid. Miller et
al\cite{MILLER2002} performed a $\mu$sR experiment in the underdoped
$YBa_2Cu_3O_{6.5}$ system under a magnetic field of $B=4T$. They found
that the local magnetic field distribution has a staggered pattern,
superimposed on a uniform decay away from the vortex core. The
staggered magnetic field detected at the muon site is about $18
Gauss$. All the experiments discussed above were carried out at fields
far below the upper critical field $B_{c2}$, which in hole doped
materials typically exceeds $60T$. In order to establish the nature of
the competing state, one has to perform experiments close to
$B_{c2}$. This was achieved in recent neutron scattering
experiments on the electron doped $Nd_{1.85}Ce_{0.15}CuO_4$ crystal in
magnetic fields up to $14T$, far above the upper critical field,
$B_{c2}$ \cite{Kang2003}. Kang et al found field
induced AF scattering at $(\pi,\pi,0)$ and observed that the AF moment scales
approximately linearly with the applied field up to $B_{c2}$. The AF
moment decreases with the magnetic field in the range between $B_{c2}$
and $14T$. Their experimental data and the theoretical fit are shown
in Fig.\ref{Fig_AF_moment}. The experimental findings of Kang et
al\cite{Kang2003} have been confirmed by Fujita et al\cite{Fujita2003}
in a related, electron doped material $Pr_{1-x}LaCe_xCuO_4$. While
$Nd_{1.85}Ce_{0.15}CuO_4$ material contains the magnetic $Nd$ moment,
the $Pr_{1-x}LaCe_xCuO_4$ material studied by Fujita et al does not
contain such magnetic ions, thus confirming that the field induced AF
moment cannot be due to any spurious effects associated with the $Nd$
moments\cite{Mang2003}. As we shall see below, the wide field range of
the neutron data enables quantitative comparisons with theoretical
models.

Since the original theoretical prediction of the AF vortex state,
tremendous theoretical progress has been made on the subject
of AF vortex
lattices\cite{ALAMA1999,BRUUS1999,OGATA1999,Hu1999A,MORTENSEN2000,HAN2000,ANDERSEN2000A,DEMLER2001,HU2002,ZHANG2002,JUNEAU2002A,FRANZ2001,FRANZ2002A,GHOSAL2002,KIVELSON2002,CHEN2002B,CHEN2002C}.
Based on the variational solution of the
$t-J$ model, Ogata\cite{OGATA1999} concluded that the vortex core 
has AF with
an ordered moment about $10\%$ of the full moment. This calculation
established the microscopic basis of the AF vortex core. The
initial AF vortex solutions were based on the static mean field theory.
In the weak field regime where the vortex cores are  separated far
from each other, the enhanced AF order can be viewed either as
dynamic fluctuations of the AF order parameter due to the finite
size of the vortex core, or as the bulk AF fluctuation pulled
below the spin gap and spatially bound near the vortex cores. This
dynamic picture was developed in
Refs.\cite{BRUUS1999,DEMLER2001,HU2002} and could apply to
experiments by Lake {\it et al} in optimally doped LSCO. Classical
Monte Carlo calculations of the $SO(5)$ model also show the existence
of the AF vortex lattice\cite{Hu1999A}. While the original theory
of the AF vortex state was developed for the commensurate AF
order, it can also be generalized to the case where the AF
ordering wave vector deviates from the $(\pi,\pi)$ point, as in
the case of the LSCO system\cite{ZHANG2002,HU2002}. The
AF ordering inside the vortex core has a profound effect on the
electronic structures of the vortex, since it opens up an
insulating-like energy gap inside the vortex core where the
conventional SC gap vanishes. The conventional theory for $d$-wave
vortices based on Bogoliubov-de Gennes (BdG) mean-field theory
predicts a large and broad peak at the Fermi energy in the local
density of states (LDOS), the so-called zero-energy peak (ZEP), 
and at the
vortex core\cite{WANG1995}. However, scanning tunneling
spectroscopy (STS) spectrum in BSCCO, giving the LDOS
around the vortex core directly, shows only a small-double peak structure
at energies ±7 meV\cite{PAN2000}. A similar situation was
observed in YBCO compounds\cite{MAGGIOAPRILE1995}. The suppression
of the local density of states due to the AF ordering inside the
vortex core could naturally explain this
phenomenon\cite{OGATA1999,ANDERSEN2000A,CHEN2002C}. However, other
forms of order, or the smallness of the core size could also offer
alternative explanations\cite{TSUCHIURA2003}.


While the experimental observation of the AF vortex state confirms
a major prediction of the $SO(5)$ theory, most of these
experiments have not directly tested the symmetry between AF and
SC in the strictest sense. In the following, we shall discuss two
aspects of the AF vortex state which directly pertain to the
$SO(5)$ symmetry. The spatial variation of the AF and  SC order
parameters around vortex core lead to a region of space where both
order parameters coexist. In this region, the $\pi$ order
parameter, whose magnitude can be quantitatively
predicted by the $SO(5)$ orthogonality relation in Eq.
(\ref{so5_orthogonal}), also develops. Ghosal, Kallin and
Berlinsky\cite{GHOSAL2002} have quantitatively verified this
relationship from their numerical solution of the $t-J$ model
around the vortex core. It would be desirable to find a way to
measure the $\pi$ order parameter and test this relation
experimentally.

The detailed experimental data now available up to $B_{c2}$ in
electron doped cuprates allows for a quantitative test of the $SO(5)$
symmetry. As discussed in section \ref{sec:classical}, within models
of competing AF and SC order, a crucial test for the $SO(5)$ symmetry
is the relation $u^2_{12}=u_1 u_2$ for the quartic term in
Eq. (\ref{GL}). Deviation from the $SO(5)$ relation determines the
curvature of the ground state energy versus doping plot, which
can be used to determine the nature of the transition between the AF
and SC states.  Recently, Chen, Wu and Zhang\cite{Chen2003B}
numerically solved the LG model with competing AF and SC order in the
vortex state and found that the deviation from the $SO(5)$ relation
$u^2_{12}=u_1 u_2$ also determines the curvature  of the field 
induced AF moment versus the magnetic field plot for magnetic fields up to
$B_{c2}$. The neutron scattering data obtained in the NCCO
superconductors\cite{Kang2003} can be fitted by $u^2_{12}/u_1
u_2=0.95$, showing that this system only has a $5\%$ deviation from
the $SO(5)$ symmetry. When the magnetic field exceeds $B_{c2}$, it
causes canting of the spin moments, thereby reducing the AF moment
while increasing the ferromagnetic moment. Thus the $SO(5)$ theory
quantitatively explains the experimental data in the entire magnetic
field range below $14T$. The experimental results of Fujita et
al\cite{Fujita2003} in the $Pr_{1-x}LaCe_xCuO_4$ material are
quantitatively similar.  We note that the mean-field analysis of the
GL free energy does not include quantum fluctuations of the AF order
(the first term in equation (\ref{Hsigma})). The latter should be
important when the AF moments are strongly localized inside the vortex
cores. We expect that proximity effect type coupling between
neighboring AF vortices should be sufficient to suppress such
fluctuations.

In the above discussions we focused on the AF moments of static votices in the
SC state. The $SO(5)$ model has also been extended  to
study thermally activated phase slips in one-dimensional wires
\cite{SHEEHY1998}. One can also construct a dual effect to
the AF vortices: Goldbart and Sheehy proposed AF hedgehogs with SC cores
in Ref. \cite{GOLDBART1998}.

\subsection{The pair density wave state}
\label{sec:PDW} In the quantum disordered phase of the $SO(5)$
model, the hole pair bosons become localized, forming a pair density
wave. Since the superfluid density is low and pairing is strong in the
underdoped regime of HTSC cuprates, the pair density wave state
competes with the $d$ wave SC state. In the global phase diagram shown
in Fig. \ref{Fig_global}, aside from the half-filled AF insulator, there
are several possible pair density waver states surrounded by the SC
phases. In contrast to the supeconducting state, which
can be realized for any charge density, each pair
density wave state has a preferred charge density, the dominant one
being at doping level $x=1/8$. Since the projected $SO(5)$ model is
formulated on the plaquettes of the original lattice, the pair density
wave naturally forms a checkerboard pattern, as depicted in Fig.
\ref{PDW}. This state has a rotationally symmetric charge
periodicity of $4a\times 4a$ near doping level $x=1/8$.
However, connecting
period of charge modulation to hole density  in realistic systems
is not always
straightforward. In most cases we find states that have both
superconductivity and periodic density modulation.  Hence, they may be best
described as supersolids.  Supersolid phases are compressible and can
accomodate extra charge without changing the period. Expressed differently,
the excess charge can always be taken by the superfluid part of
the Cooper pair density without affecting the
localized part.
The pair density wave state differs
from the stripe state\cite{EMERY1999,ZAANEN1989}, since
it does not
break the symmetry of $\pi/2$ lattice rotations.
It is also distinct from the Wigner crystal 
of individual holes proposed in Ref.\cite{FU2004},
which should have a charge periodicity of
$\sqrt{8}a\times \sqrt{8}a$ at the same doping level. The pair
density wave state was first proposed by Chen et
al\cite{CHEN2002B} in the context of the $SO(5)$ theory of the
vortex state. It also arises naturally from the plaquette boson
approach of Altman and Auerbach\cite{ALTMAN2002}.
Podolsky et al. \cite{PODOLSKY2003} discussed how unconventional
states with translational symmetry breaking, including
the pair density wave state, can be detected in STM
experiments. Relevance of this state to tunneling experiments has
also been considered in \cite{ANDERSEN2003,VOJTA2002,Chen2004}.


As we see in the global phase diagram shown in Fig.
\ref{Fig_global}, the pair density wave state can be stabilized
near doping of $x=1/8$, when the superfluid density (or the
kinetic energy of the hole pairs) is small compared to interaction
energy. This situation can be realized in the vortex core, near
the impurities, in the underdoped cuprates or in the pseudo-gap
phase. The STM experiments\cite{HOFFMAN2002} measuring the local
density of states near the vortex core demonstrated a $4a\times 4a$
checkerboard pattern, consistent with the hole pair checkerboard
state\cite{CHEN2002B} shown in Fig. \ref{PDW}. The
vortex core can be either positively or negatively charged,
depending on whether the bulk density is greater or smaller than
that of the nearby pair density wave state\cite{Wu2002}. For
example, if the chemical potential is such that the bulk SC state
is on the left (right) side of the $\delta=1/8$ insulator, we
expect the vortex core to have more (less) hole density. The STM
experiment of Howald et al \cite{HOWALD2002} sees a similar real
space modulation without the applied magnetic field, possibly
induced by impurities\cite{McElroy2003} . 
More recently, Vershinin et al
\cite{Vershinin2004} discovered a real space modulation of the
density-of-states in the pseudo-gap phase above $T_c$.
Enhancement of the translational symmetry breaking
in the pseudogap regime of the cuprates has been 
proposed theoretically in Ref. \cite{SACHDEV2003}. The microscopic
picture of this phenomenon has been studied in Ref. \cite{Chen2004}
using an extension of the formalism in Ref. \cite{PODOLSKY2003} 
for the pseudogap regime. Analysis of Ref. \cite{Chen2004} shows
that the experimentally observed
modulation is inconsistent with an ordinary 
site centered charge density wave and 
the corresponding modulation of the Hartree-Fock potential.
However, the pair density
wave state provides good agreement with the experimental data.

\subsection{Uniform mixed phase of antiferromagnetism and superconductivity}
\label{sec:experiment_coexistence} The phase diagram obtained from
the classical competition between the AF and SC states is
shown in Fig. \ref{Fig_T_mu}. We classified the phase transition
broadly into three different types. The ``type 1" transition
involves a direct first order phase transition between the AF and
the SC phases, and the ``type 2" transition involves two second order
phase transitions, with an intermediate phase which is a uniform
mixture of AF and SC. The marginal ``type 1.5" transition
describes the special $SO(5)$ symmetric case where the chemical
potential remains constant in the entire uniform mix phase.
Therefore, both ``type 2" and ``type 1.5" transitions predict a
uniform mixed phase of AF and SC.

Evidence for the AF/SC mixed phase exists in the excess oxygen doped
$La_2CuO_{4+y}$ material. Neutron scattering measurement detects
the onset of the AF or spin-density-wave orders at the same
temperature as the SC $T_c$\cite{Lee1999}. This remarkable
coincidence is the hallmark of a multi-critical point, which we
shall return to later. Because the $La_2CuO_{4+y}$ system has an ordering
wave vector similar to that of the $La_{2-x}Sr_xCuO_4$ system,
it should also be classified as the ``class B3" trace
in the global phase diagram of Fig. \ref{Fig_global}, passing
through the ``1/8" Mott lobe. However, in this case, the Mott
phase boundary likely belongs to ``type 1.5 or 2," where the AF
and SC order can coexist.

For $YBa_2Cu_3O_{6+x}$ materials, static magnetic ordering extending
to $x \approx 0.5$ has been observed recently using muon spin
rotation/relaxation measurements in \cite{Miller2003}.  Preliminary
neutron scattering experiments in \cite{SIDIS2001} and \cite{MOOK2002}
also reported magnetic ordering with a wavevector $(\pi,\pi)$. Thus,
in this case we have AF coexisting with SC without any additional
charge order.  However, it is unclear if two phases coexist uniformly
in these materials. Assuming that future experiments verify the existence of
a homogenious phase with AF and SC orders, we conclude that the
phase diagram for $YBa_2Cu_3O_{6+x}$ may be understood as moving along
the B1 line line in Figure \ref{Fig_global}, when the system avoids
all the PDW lobes but only has AF and SC orders either separately or
in a uniform mixed phase.

Evidence for the mixed phase of superconductivity and
antiferromagnetism has also been obtained recently in the
five-layered HTSC cuprate $HgBa_2Ca_4Cu_5O_y$. In this system, the
three inner layers are predominantly antiferromagnetic, while the
two outer layers are predominantly superconducting. In a Cu-NMR
study, Kotegawa et al\cite{Kotegawa2004} obtained firm evidence
that the AF inner layers induce a small magnetic moment in the
outer layers, establishing the case of an AF/SC uniform mixed phase in this
system. However, this type of AF/SC proximity effect was not
observed in the artificially grown layer
structures\cite{BOZOVIC2003}.

Above discussions show that there is evidence of a uniform
mixed phase of AF and SC in the HTSC cuprates. On the other hand,
microscopic probes such as STM\cite{PAN2001}
reveal electronic inhomogeneities characteristic of the ``type 1"
direct first order transition between AF and SC. Therefore,
depending on material details, some HTSC compounds show AF/SC
mixed phase, characteristic of the ``type 2" behavior, while
others show microscopic separation between these two phases, 
a characteristic more consistent with the ``type 1" behavior. 
It is quite remarkable that
such different physical effects can be obtained in  materials
that are so similar. A reasonable explanation is
that the system is actually very close to the $SO(5)$ symmetric
point exhibiting ``type 1.5" behavior. Only in this case can a slight
variation tip the balance towards either the ``type 1" or
``type 2" behaviors.

A genuine uniform mixed phase of AF and SC has been
observed in several heavy-fermion systems in some regions of the
pressure (P) versus temperature (T) phase
diagram~\cite{KITAOKA2001,KITAOKA2002}. Recently, such coexistence
was observed through NMR and NQR spectrum measurements in
$CeCu_2(Si_{1-x}Ge_x)_2$ with a small concentration $x=0.01$ of
Ge. In  $CeCu_2Si_2$, SC coexists with slowly fluctuating magnetic
waves. However, for AF $CeCu_2Ge_2$, which has the same
lattice and electronic structure as $CeCu_2Si_2$, it was found
that a SC phase can be reached at a critical pressure $P_c\sim 7.6
GPa$. Since $CeCu_2Si_2$ behaves at $P=0$ like $CeCu_2Ge_2$ at
$P_c$, it is argued that
 SC in $CeCu_2Si_2$ occurs
 close to an AF phase at $P=0$ corresponding to a critical lattice
 density $D=D_c$. This appears to be
 the reason for the strong AF
fluctuations at $P=0$. A small concentration of Ge expands the
unit-cell volume reducing $D$
 below $D_c$ and
is thus sufficient to pin the magnetic fluctuations and to
 produce AF long-range order within the SC phase.
Noting that $D = D_{Si}[1 (V_{Ge}- V_{Si})x/V_{Ge}]$ for Ge doping
and that $D$ increases with pressure, one can draw a combined
phase diagram as a function of lattice density
$D$~\cite{KITAOKA2002}.


In Ref.~\cite{KITAOKA2002}, it was shown that the phase diagram of
Fig.~\ref{kitaoka} could be understood in terms of an $SO(5)$
superspin picture. This could suggest that SC in $CeCu_2Si_2$
 could be mediated by the same magnetic interactions leading to the AF
 state in
 $CeCu_2(Si_{1-x}Ge_x)_2$.

\subsection{Global phase diagram and multi-critical points}
\label{sec:experiment_phase_diagram} 
\begin{figure}[hp]
\begin{center}
      \includegraphics[scale=1.0]{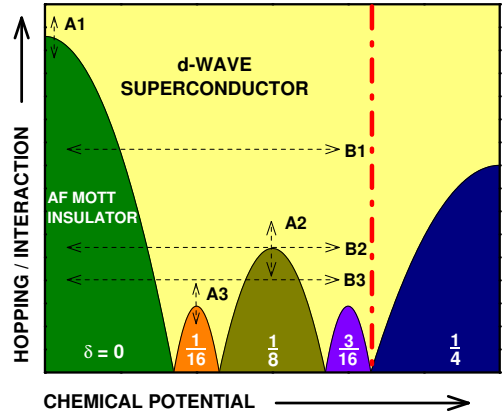}
\caption{A typical global phase diagram of the extended $SO(5)$
model in the parameter space of chemical potential and the ratio
of boson hopping energy over interaction energy (see
Ref. \cite{CHEN2003A} for details). This phase
diagram shows self-similarity among the insulating states at
half-filling and other rational filling fractions. There are two
types of superfluid-insulator transition. The quantum phase
transition of ``class A" can be approached by varying the hopping
energy, for example, by applying a pressure and magnetic field at
constant doping. The quantum phase transition of ``class B" can be
realized by changing the chemical potential or doping. This
theoretical phase diagram can be compared with the global phase
diagram of the HTSC cuprates. Different families of cuprates
correspond to different traces of ``class B." For example, we
believe $YBCO$ is $B1$-like, $BSCO$ may be close to $B2$-like and
$LSCO$ is $B3$-like. The vertical dash-dot line denotes a boundary
in the overdoped region beyond which our pure bosonic model
becomes less accurate.  
All the phase boundaries in this figure can be classified
into direct first order (type 1), two second order
(type 2), or a marginal case with enhanced symmetry
(type 1.5). Type 2 transitions between CDW lobes and
the superconducting state lead to intermediate supersolid
phases.
} \label{Fig_global}
\end{center}
\end{figure}

\begin{figure}[hp]
\begin{tabular}{cc}
    \includegraphics[scale=1,clip]{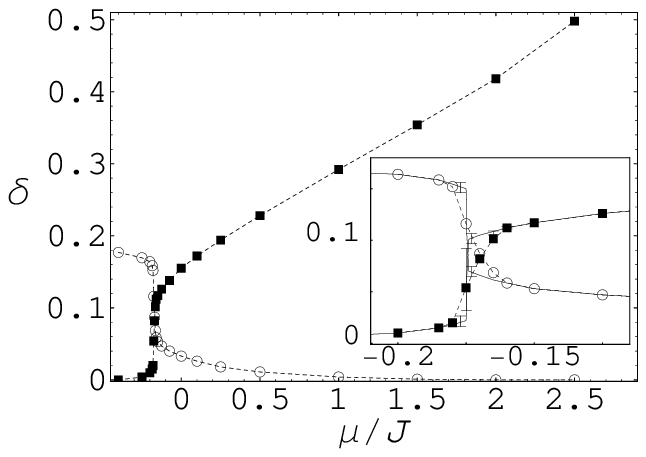}&
    \includegraphics[scale=1,clip]{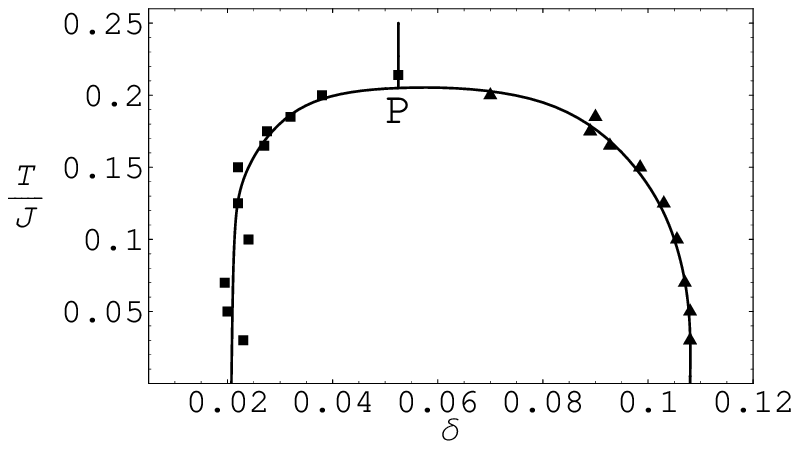}\\
(a)&(b)
\end{tabular}
\caption{(a) Hole concentration
  $\delta=\rho/2=\frac{1}{2}\langle t_h^{\dagger} t_h^{} \rangle$ (filled
  squares) and magnon density $\sum_{\alpha} \frac{1}{2}\langle
  t_\alpha^{\dagger} t_\alpha^{} \rangle$ (circles) as a function of the
  chemical potential $\mu$ at $T/J=0.03$. The small inlay shows a detailed view to the
  $\mu$ region in which the hole-pair density jumps to a finite value. (b) Hole
  densities of the coexisting phases on the first order transition line
  from (almost) zero to finite hole density at $\mu\!=\!\mu_c$ as a function
  of temperature.} \label{Fig_2D_pso5}
\end{figure}

\begin{figure}[hp]
\begin{tabular}{cc}
    \includegraphics[scale=1,clip]{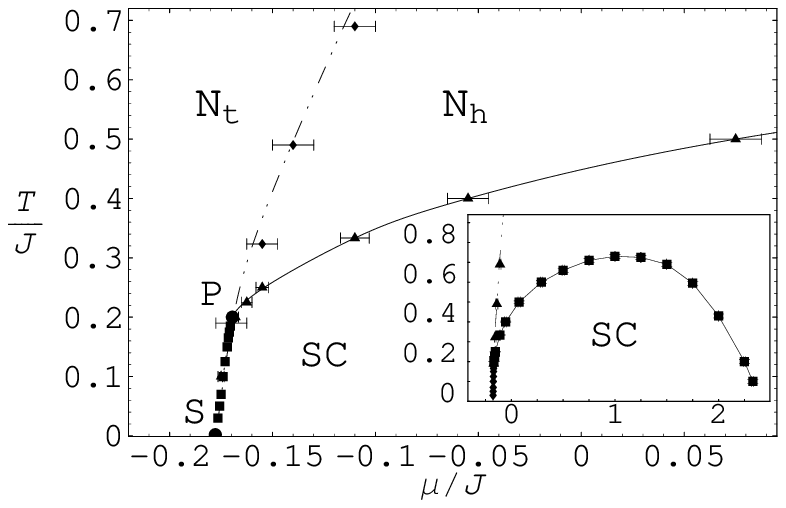}&
    \includegraphics[scale=1,clip]{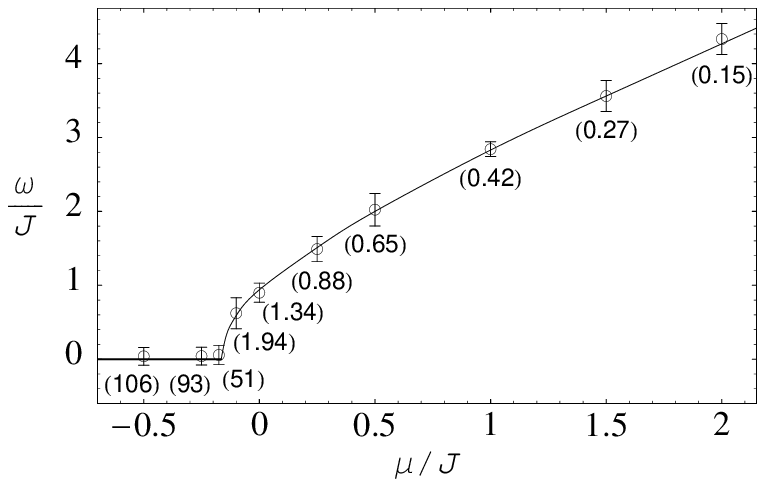}\\
(a)&(b)
\end{tabular}
\caption{(a) Phase diagram of the projected $SO(5)$ model (see Eq.
(\ref{pso5}) with $J_c=J_S$) in D=2: The squares between S and the
tricritical
  point P trace the first-order line of phase separation.  The solid line from
  P to the right edge of the plot traces the Kosterlitz-Thouless transition
  between the SC and the normal state. The dashed line separating $N_t$
  (=triplet dominated region) and $N_h$ (=hole pair dominated region)
  describes the line of equal AF and SC correlation lengths.  The small inlay
  shows the same phase diagram on a larger $\mu$ scale, covering the whole
  KT phase. The tricritical point P appears as a result
 of the Mermin-Wagner theorem, which does not allow spin ordering
 in D=2 at finite temperature.
(b) Energy of a single magnon excitations in the projected $SO(5)$
model as a function of the chemical potential. This corresponds to
the resonance energy of the $(\pi,\pi)$ peak of the spin
correlations in the fermionic model (magnons are defined to carry
the momentum of the AF order). The numbers in parentheses indicate
the peak weights, i.e. the area under the peak.  ($20\!\times\!20$
lattice at temperature $T/J\!=\!0.1$). (from
Ref.~\cite{JOSTINGMEIER2003})
  } \label{Fig_tricritical}
\end{figure}

\begin{figure}[hp]
  \includegraphics[scale=1.0]{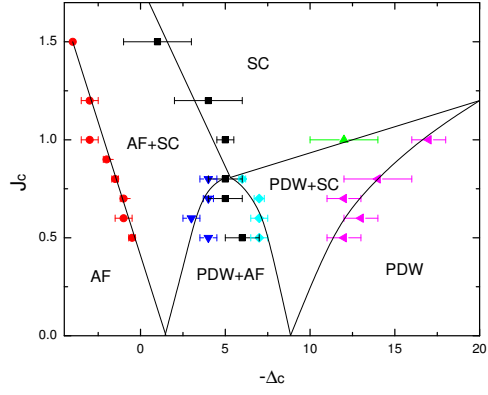}
\caption{The phase diagram of the extended $SO(5)$ model obtained
by the QMC simulation. The parameters used in simulation are
$\Delta_s=4.8$, $V_c=4.1010$, $V_c'=3.6329$ and $J_\pi=V_\pi=0$.
The lines are guides to the eye only. The overall topology of the
phase diagram agrees well with the global phase diagram presented
in Fig. \ref{Fig_global}. } \label{FIG-Phase-QMC}
\end{figure}

\begin{figure}[hp]
  \hspace{2mm}
  \includegraphics[scale=0.6]{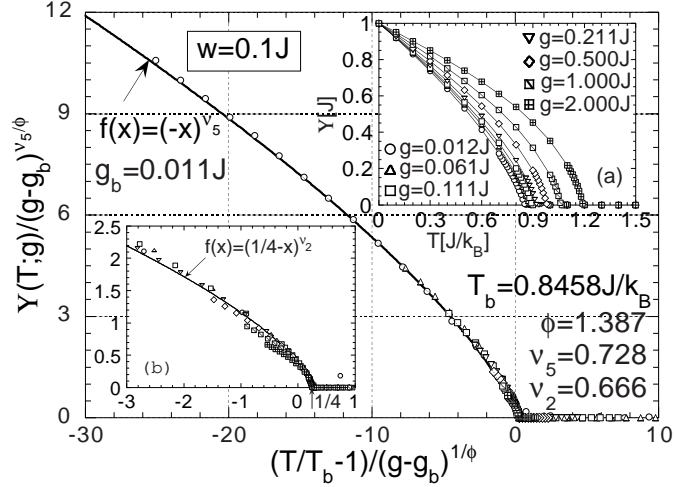} \vspace{0.8mm}
  \narrowtext
\caption{Scaling of the superfluid density near the $SO(5)$
bi-critical point obtained by the classical Monte Carlo
simulations. The critical behavior of the superfluid density for
various $g$ fit into a single scaling curve, from which $SO(5)$
scaling exponents were obtained. Reproduced from Ref.
\cite{HU2001}.} \label{Fig_Hu}
\end{figure}

\begin{figure}[hp]
\begin{tabular}{cc}
    \includegraphics[scale=0.7]{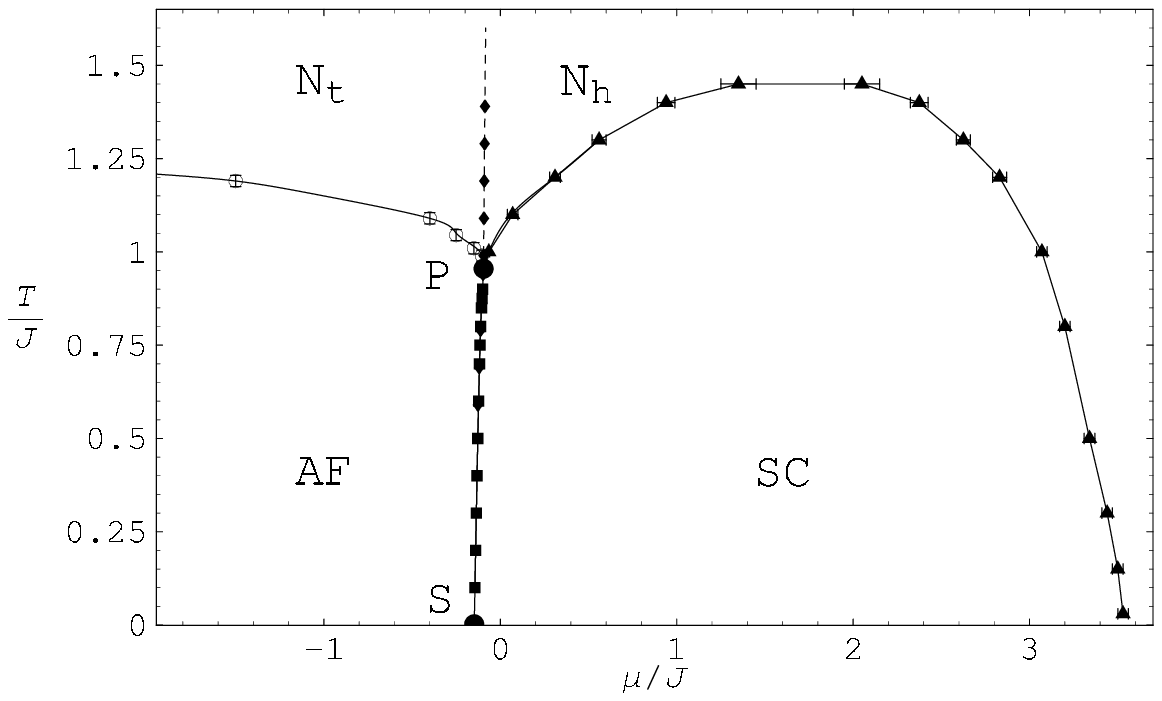}&
    \includegraphics[scale=0.7]{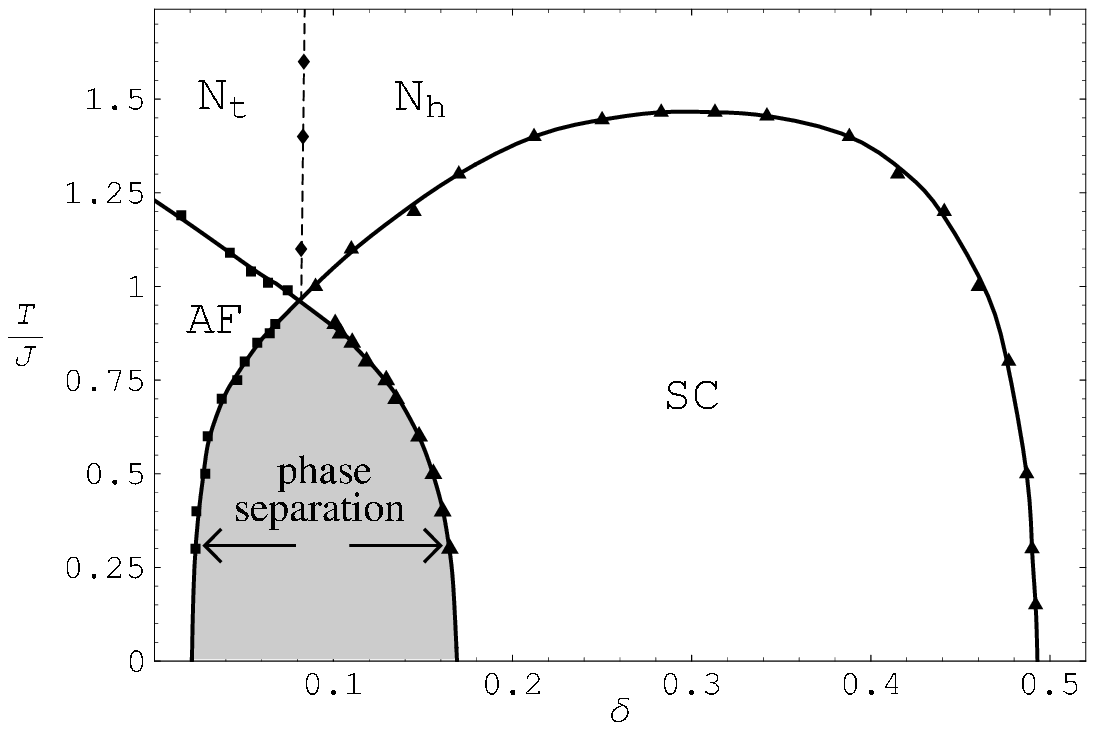}\\
(a)&(b)
\end{tabular}
\caption{(a) The $T(\mu)$ phase diagram of the three-dimensional
projected $SO(5)$ model with $J_s\!=\!J_c/2$ and
$\Delta_s\!=\!\Delta_c\!=\!J$. N$_h$ is the hole-pair dominated
part, N$_t$ the triplet dominated part of the high-temperature
phase without long-range order. The separation line between N$_h$
and N$_t$ is the line of equal spatial correlation decay of
hole-pairs and bosons. (b) The $T(\delta)$ phase diagram of the 3D
projected $SO(5)$ model as a function of hole doping
$\delta=n_h/2$. The first order transition line from $S$ to $P$ in
the $T(\mu)$ diagram becomes a ``forbidden region'' due to phase
separation. These two phase diagrams are consistent with those
presented in Fig. \ref{Fig_T_mu}a-b based on general arguments.}
\label{Fig_3D_pso5}
\end{figure}

The $SO(5)$ theory makes the
key prediction of the existence of a multi-critical point where
$T_N$ and $T_c$ intersect (see Fig. \ref{Fig_T_mu}) and the
general topology of the global phase diagram in the space of
quantum parameters (see Fig. \ref{Fig_global}). The goal of this
section is to establish the connection between the theoretical quantum
phase diagram proposed in Section \ref{sec:quantum} and the
experimental phase diagrams of various families of cuprates. The
underlying assumption for making such a connection is that most of
material specific properties can be absorbed into parameters of
the effective Hamiltonian given in Eq. (\ref{pso5}) and in Eq.
(\ref{extended}).

One of the most studied phase diagrams of the HTSC is for
$La_{2-x}Sr_xCuO_4$. The stripe order's presence in these
materials has been well documented by neutron scattering experiments
\cite{YAMADA1998,WAKIMOTO2000,WAKIMOTO2001}. For less than 5\%
doping the system is in the insulating regime with diagonal
stripes and for higher dopings the system is superconducting with
collinear stripes (see Fig. \ref{LSCOPhaseDiagram}).
It is natural to relate this family of cuprates to the B3 trajectory
on the $J/V$-$\mu$ phase diagram shown in Fig.  \ref{Fig_global}: with
increasing $\mu$ the system goes through a hierarchy of states at
fractional filling factors that correspond to insulating pair density
wave states. Near these magic filling factors, the SC $T_c$ drops
dramatically, while magnetic ordering increases substantially. This is
indeed the behavior observed in Fig. \ref{LSCOPhaseDiagram}. As
discussed in Section \ref{sec:quantum}, the two possible patterns of
charge ordering are checkerboard and stripes. In the case of
$La_{2-x}Sr_xCuO_4$, stripe ordering may be stabilized by tilting the
$Cu O_6$ octahedron toward [100] tetragonal direction (parallel to the
Cu-O bonds). The phase diagram in Fig. \ref{Fig_global} predicts that
the ordering wave-vectors take discrete values that correspond to
different Mott insulating PDW lobes.  
For long range interactions PDW phases are
very densely packed, so experimentally we may observe an almost
continuous dependence of incommensuration on doping, 
such as the one discussed
in Refs. \cite{YAMADA1998} and \cite{WAKIMOTO2000}. However, different
states in the hierarchy are not equivalent. For example, at $1/8$
doping we have a very strong insulating phase which corresponds to
insulating stripes or a simple checkerboard pattern of Cooper pairs
(see Fig.  \ref{Fig_stripe_checker}). This may explain a famous ``1/8
anomaly'' in the $T_c$ vs doping relation for the $La_{2-x}Sr_xCuO_4$
family of cuprates. Another strong PDW phase is for $1/16$ doping,
which may explain why supreconductivity disappears close to this
filling (see Fig.  \ref{LSCOPhaseDiagram}).  A ``staircase'' of ordering
wavevectors for underdoped cuprates has also been discussed in the
context of doping the spin-Peierls insulating phase in
Refs. \cite{VOJTA1999,SACHDEV2002B}.

By adding another external parameter we can tune our system
continuously between B1 and B3 trajectories. This was done in
recent high pressure experiments on
$La_{1.48}Nd_{0.4}Sr_{0.12}CuO_4$
\cite{LOCQUET1998,SATO2000,ARUMUGAM2002,Takeshita2003}, where the
pressure of the order of 0.1 GPa was sufficient to suppress stripe
ordering at 1/8 doping and stabilize the high temperature SC
phase. Such pressure experiments correspond to moving up along the
A2 path in Fig. \ref{Fig_global}. Applying pressure along this
path can directly induce a superconductor to insulator transition.

In contrast to the $LSCO$ family of HTSC cuprates, when one varies the
carrier density in the $YBCO$ or $BSCO$ cuprates, there is no evidence
for the static charge order.  In these materials, charge ordered PDW
states can only be realized around vortex cores\cite{HOFFMAN2002},
when the effective Cooper pair kinetic energy is reduced, or near
impurities\cite{HOWALD2002,McElroy2003,Vershinin2004}.  Therefore, we
identify these materials with the B1 trajectories in the global phase
diagram of Fig. \ref{Fig_global}.  In this case, the AF/SC boundary
can be either be ``type 1" or ``type 2". Given the evidence discussed
in section \ref{sec:experiment_coexistence}, these systems seem to be
close to the ``type 1.5" marginal case in between these two types of
phase transitions, which means that they should have  the approximate $SO(5)$
symmetry.


Within the class of materials exhibiting the ``$B1"$ type of
trajectory in the global phase diagram, the $SO(5)$ theory makes a
distinct prediction of the finite temperature multi-critical point
where $T_c$ and $T_N$ intersect. An interesting issue discussed in
 Secs.~\ref{sec:numerics} and \ref{sec:so5global} is
the possibility of analyzing the critical properties of systems
(such as many HTSC cuprates) showing a direct transition between
an AF and a SC phase. In particular, measuring the critical
exponent associated with various physical quantities near the
bicritical AF-SC point can give information about the dimension of
the symmetry group at the transition~\cite{HU2000}. Unfortunately,
in the HTSC cuprates, sample qualities are not high enough to
enable a reliable measurement of the critical behavior near the
multi-critical points discussed above. On the other hand,
encouraging experimental evidence for an $SO(5)$ bicritical point
does exist in a class of 2D organic superconductors called $BEDT$
salt. These material share most common physical properties with
the cuprates, and the AF to SC transition can be induced by
pressure. In particular, recent experiments on $k-(BEDT-TTF)_2X$
~\cite{Kanoda1997} revealed an interesting phase
diagram in which $T_c$ and $T_N$ intersect each other at a
bicritical point.
 Kanoda~\cite{Kanoda1997} measured the NMR relaxation rate
$1/T_1$ both in the AF and in the SC region near the bicritical
point. Below a characteristic temperature $T^*$, $1/T_1$ diverges
toward the AF transition temperature, while it exhibits a
spin-gap-like behavior on the SC side. Murakami and
Nagaosa\cite{MURAKAMI2000} analyzed this experimental data in
terms of a generalized LG model including both AF and SC
fluctuations near the bicritical point. Their study
concentrated on the dynamic critical phenomena, in particular
the relaxation rate $1/T_1$ around the bicritical point. A
detailed analysis of the data allowed the extraction of the corresponding
critical exponent $x$. Before discussing the NMR line width, we
would like to caution the readers that there is also a first order
metal-insulator transition in addition to the AF/SC transition
discussed here\cite{LEFEBVRE2000}. The presence of the critical
endpoint of the metal insulator transition line may lead to some
additional complications in the analysis.

On the AF side
of the phase diagram,
the NMR line width is proportional to $(T-T_N)^{-x}$ 
when approaching $T_N$ from the normal state. 
For systems far away from the
bicritical point, the dynamical critical behavior is governed by
the $SO(3)$ Heisenberg model, whose exponent $x=x_3\approx 0.315$.
On the other hand, when the $SO(5)$ bicritical point governs the
critical dynamics, the exponent $x$ should change to the $SO(5)$
one $x=x_5\approx 0.584$, as obtained from the
$\epsilon$-expansion. In Fig.~\ref{exp}, we present a
  log-log plot of $1/T_1$ vs $(T-T_c)/T$  
(from Ref.~\cite{MURAKAMI2000}, data from Ref.~\cite{KAWAMOTO1995})
for (A) $\kappa$-(BEDT-TTF)$_{2}$Cu[N(CN)$_{2}$]Cl (solid
squares), and (B) deuterated
$\kappa$-(BEDT-TTF)$_{2}$Cu[N(CN)$_{2}$]Br (open squares). 
System (A) is
located in the AF region away from the bicritical
point and system (B) is nearly at the bicritical point. As one can see
from the figure,
 the critical exponent, $x$,
is $0.30\pm 0.04$ for system (A) and $ 0.56\pm 0.04$ for
system (B).
These values of $x$ are in reasonably good agreement with the
theoretical ones, and, in particular, support the fact that the
AF/SC bicritical point is governed by the $SO(5)$ symmetric fixed
point. This is the first experiment which directly measures the
dimension of the symmetry group close to the AF/SC bicritical point
and determines $n$ to be close to $5$. More extensive study near
the critical region is certainly desired.

A central issue of the HTSC cuprates concerns the phase boundary
between the AF and SC phases. It is also in this region that the
$SO(5)$ theory makes the most direct and distinct predictions.
Current experiments discussed above seem consistent with the
zero temperature and finite temperature phase diagrams
presented in Fig. \ref{Fig_global} and Fig. \ref{Fig_T_mu}.
However, detailed quantitative comparison is still lacking. As the
material properties of the HTSC cuprates improve,
direct quantitative tests of the $SO(5)$ theory, such as those
performed in the organic superconductors, may become possible.

\subsection{The particle-particle resonance mode in the normal state}
\label{sec:p-p} In this paper we discussed the scenario in which
the resonance peak in the inelastic neutron scattering experiments
\cite{ROSSATMIGNOD1991A,MOOK1993,FONG1995,FONG1999A,HE2001}
originates from the triplet $\pi$ mode in the particle-particle
channel. This mode does not disappear above $T_c$, but it
ceases to contribute to the spin fluctuation spectrum, since the
particle-particle and particle-hole channels are decoupled from
each other in the normal state. An important question to ask is
whether one can  couple to the $\pi$-channel directly and
establish the existence of the resonance already in the normal
state. This cannot be done using conventional electro-magnetic
probes, which all couple to the particle-hole channels only, but
it is possible using tunnelling experiments. Before we discuss the
specific proposal of Bazaliy {\it et.al.}\cite{BAZALIY1997} for
detecting the $\pi$ excitations, it is useful to remind the
readers about earlier work on measuring pairing fluctuations in
conventional superconductors above their transition temperature
\cite{ANDERSON1970}.  As originally proposed by Scalapino
\cite{SCALAPINO1970}, the latter can be measured in a sandwich
system of two superconductor SC$_1$ and SC$_2$ with different
transition temperatures in the regime $T_{c2}<T<T_{c1}$. Resonant
coupling between Cooper pairs from the superconductor SC$_1$ and
the fluctuating pairing amplitude in SC$_2$ leads to the peaks in
the IV characteristics at voltages that correspond to half the
energy of the ``preformed'' Cooper pairs in SC$_2$. The generalization
of these tunnelling experiments for detecting the $\pi$ mode in
the normal state of the cuprate has been suggested in
\cite{BAZALIY1997} and is shown in Fig. \ref{pifig1}.
In place of the SC$_2$ region we now have some cuprate material
that shows a resonance in the SC state, e.g. an underdoped YBCO,
(electrode C in Fig \ref{pifig1}), and in place of the SC$_1$
materials we have a different cuprate superconductor (electrode A
in Fig \ref{pifig1}) with a higher transition temperature than
material C. The system should be in the temperature regime $T_c^C
< T < T_c^A$. The main difference with the set-up suggested in
Ref. \cite{SCALAPINO1970} is the presence of a thin layer of AF
insulator between the A and C electrodes. The reason for this
modification is straightforward: we need to probe the $\pi$
channel in the C material that corresponds to the
particle-particle mode with spin $S=1$ and momentum
$\Pi=(\pi,\pi)$, whereas the SC electrode  A provides Cooper pairs
with $S=0$ and momentum $q=0$. If the two materials are connected
as shown in Fig. \ref{pifig1}, a Cooper pair travelling across an
AF layer B can emit a magnon, which converts this Cooper pair into
a $\pi$-pair and allows resonant coupling between superconductor A
and the $\pi$-channel of the ``normal'' electrode C.  One
expects to find a resonance in the IV characteristics of the
junction with a peak in the tunnelling current at a voltage which
is exactly half the energy of the $\pi$-resonance in the C
electrode (note that this peak only appears  when electrons are
injected from A to C, so it appears on one side of the IV curve).
A simple qualitative picture described above can be made more
precise by considering a tunnelling Hamiltonian between materials
A and C
\begin{eqnarray}
H_{T}=\sum_{pk\sigma} T^d_{pk}  a^{\dagger}_{p\sigma} c_{k\sigma}
e^{i V t}
 + T^f_{pk} a^{\dagger}_{p+Q\sigma} c_{k-\sigma} e^{i V t} + h.c.
\label{explicit}
\end{eqnarray}
Here $V$ is the applied voltage, the $a_{p\alpha}$ and
$c_{k\alpha}$ operators refer to the electronic operators in A and
C with momenta $p$ and $k$. The ratio of the spin flip matrix
element $T^f_{pk}$ to the direct matrix element $T^d_{pk}$ is on
the order of $\Delta_{SDW}/U$, where $\Delta_{SDW}$ is the
spin-density-wave gap of the AF insulating material B. The diagram
responsible for the resonant contribution to the tunnelling
current is shown in Fig. \ref{pifig2}. The triplet vertex $\Gamma$
takes into account interactions needed to create a sharp
$\pi$-resonance  in the A electrode.
The magnitude of the peak in the tunnelling current was estimated
in Ref. \cite{BAZALIY1997} to be 10 $\mu$A$\mu$V for
a system of area $10^{-4}$ cm$^2$. As argued in section
\ref{sec:pi}, it is not easy to distinguish the particle-hole and
the particle-particle origin of the $\pi$ resonance below $T_c$
since these two channels are mixed. Direct experimental detection
of the triplet particle-particle mode in the normal state would
give unambiguous evidence of the particle-particle nature of the
$\pi$ resonance mode.

\subsection{Josephson effect in the SC/AF/SC junction}
When discussing the relationship between $d$-wave SC and AF in the HTSC
cuprates, one often finds signatures of the nearby magnetic phase
in experiments performed on the SC materials. An important
question to ask is whether the AF insulating phase shows any
signatures of the nearby SC state.  An intriguing set of
experiments that possibly provides such a demonstration is the long
range proximity effect observed in insulating samples of
$YBa_2Cu_3O_{6+x}$ based materials coupled in the $a$-$b$ plane
directions\cite{BARNER1991,HASHIMOTO1992,SUZUKI1994,DECCA2000}.
The AF/SC proximity effect was also observed by Kotegawa et
al\cite{Kotegawa2004}, however, it seems to be absent in the case
of artificially grown $c$-axis coupled layers\cite{BOZOVIC2003}.
The appearance of the long range proximity effect is very natural
from the point of view of the $SO(5)$ theory, in which low energy
degrees of freedom correspond to the order parameter rotation
between the AF and SC configurations. A theory of the long range
proximity effect within the $SO(5)$ non-linear sigma model has
been developed in Ref.\cite{DEMLER1998B}. Let us consider the SC/AF/SC
junction shown in Fig. \ref{fr_geometry}.
If we set $ Re \Delta = \cos \theta \cos \phi$, $ Im \Delta = \cos
\theta \sin \phi$, and $ N_3= \sin \theta$, then according to our
discussion in Section \ref{sec:so5sigma} (see Eqs.
(\ref{Vn})-(\ref{Veff})), the junction can be described by the
effective Lagrangian density
\begin{eqnarray}
{\cal L} (\theta, \phi) = \frac{\rho}{2} \{ (\partial_i \theta )^2
+ \cos^2 \theta~ ( \partial_i \phi )^2 \} - g~ \sin^2 \theta.
\label{angle-functional}
\end{eqnarray}
The anisotropy term is given by $g_A>0$ inside the A region, so
that the AF phase would be established in the bulk. In the SC S
regions on both sides of the junction we have $g_S < 0$, and we
should impose boundary conditions $\theta \rightarrow 0$ as $x
\rightarrow \pm \infty$. As discussed in Ref.\cite{DEMLER1998B}, a
simplified case corresponds to taking a ``strong'' superconductor
limit for which $\theta (x=0,d)=0$. The current phase relation can
now be obtained by writing the Euler-Lagrange equations for the
functional (\ref{angle-functional}) at a fixed current.  The
maximal value of $\theta$ reached at $x=d/2$, $\theta_0$, is
determined by the equation
\begin{eqnarray}
\frac{d}{2\xi_A} &=& \frac{ \cos \theta_0 }{ \sqrt{ \omega_s^2 +
\cos^2 \theta_0 }} K (  k ) \nonumber\\ k^2 &=& \frac{ \sin^2
\theta_0 \cos^2
  \theta_0 }{ \omega_s^2 + \cos^2 \theta_0},
\label{k-eq}
\end{eqnarray}
where $K(k)$ is the the complete elliptic integral of the first
kind, dimensionless current $\omega_s  = I \xi_A$, with $I$ being
the actual current through the junction and the characteristic
length
\begin{eqnarray}
\xi_A = \sqrt{ \rho / 2 g_A}. \label{xi-A}
\end{eqnarray}
On the other hand, the equation for the phase difference across the
junction, $\Delta \Phi$, is given by
\begin{eqnarray}
\Delta \Phi = 2 \omega_s \frac{ \cos \theta_0 }{ \sqrt{ \omega_s^2
+ \cos^2
    \theta_0 }} \Pi_1( -\sin^2 \theta_0, k ).
\label{Delta_Phi_Eq}
\end{eqnarray}
Here $ \Pi_1( n, k )$ is a complete elliptic integral of the third
kind. Immediately, one can see that equation (\ref{Delta_Phi_Eq})
describes two different kinds of behavior for $d$ larger or
smaller than $d_{c0}= \pi \xi_A$. When $d > d_{c0}$ we have a
conventional proximity effect with $I(\Delta \Phi) = I_0(d) sin
\Delta \Phi$ and $I_0(d) \propto exp(-d/\xi_A)$. We observe,
however, that the SC correlation length, $\xi_A$ may be very long
if the system is close to the $SO(5)$ symmetric point ($g_A
\rightarrow 0$ in equation (\ref{xi-A})), which corresponds to the
long range proximity effect. When $d < d_{c0}$ we get more
intriguing behavior in (\ref{Delta_Phi_Eq}), where for small
currents the A region is uniformly superconducting, i.e.
$\theta_0=0$ (proximity to a strong superconductor completely
suppresses the AF order inside the A region), but when the current
exceeds some critical value, the system goes into a state
that has both $d$-wave SC and AF orders, i.e. $0<\theta_0<\pi/2$.
The resulting nontrival $I(\Delta \Phi)$ are shown in in Fig.
\ref{I-fig}. We note that the analysis presented above does not
take into account the long range part of the Coulomb interaction
between electrons. This may become important for systems with
sufficiently wide AF layers and lead to suppression of the
proximity induced SC order in the AF layer.

Several consequences of the non-sinusoidal behavior of the
current-phase relation of the SAS junctions have been explored in
Ref. \cite{DENHERTOG1999}, including current-voltage characteristics in
the presense of thermal fluctuations, Shapiro steps, and the
Fraunhofer pattern. Decca {\it et.al.} \cite{DECCA2000} used
near-field scanning tunnelling microscopy to photo-generate
Josephson junctions in underdoped thin films of
$YBa_2Cu_3O_{6+x}$. They have verified a long range proximity
effect through insulating layers but observed a conventional
Fraunhofer pattern rather than the one predicted in
\cite{DENHERTOG1999}. The geometry of their samples, however, is
different from the system studied in
\cite{DEMLER1998B,DENHERTOG1999}: the intermediate AF layer in
their case is connected to large AF regions on both sides of the
junctions, which suppresses rotation of the superspin into the SC
direction.

In a related context, Auerbach and Altman\cite{AUERBACH2000}
applied the project $SO(5)$ theory to predict multiple Andreev
resonance peaks in the SC/AF/SC junctions.

\section{CONCLUSIONS}
In a large class of materials including the HTSC cuprates, the
organic superconductors and the heavy fermion compounds, the AF
and SC phases occur in close proximity to each other. The
$SO(5)$ theory is developed based on the assumption that these two
phases share a common microscopic origin and should be treated on
equal footing. The $SO(5)$ theory gives a coherent description of
the rich global phase diagram of the HTSC cuprates and its low
energy dynamics through a simple symmetry principle and a unified
effective model based on a single quantum Hamiltonian. A number of
theoretical predictions, including intensity dependence of the
neutron resonance mode, the AF vortex state, the pair-density-wave
state and the mixed phase of AF and SC have been verified
experimentally. The theory also sheds light on the microscopic
mechanism of superconductivity and quantitatively correlates the
AF exchange energy with the condensation energy of
superconductivity. However, the theory is still incomplete in many
ways and lacks full quantitative predictive power. While
the role of fermions is well understood within the exact $SO(5)$
models, their roles in the projected $SO(5)$ models are still
not fully worked out. As a result, the theory has not made many
predictions concerning the transport properties of these
materials.

Throughout the history of our quest for the basic laws of nature,
symmetry principles have always been the faithful guiding light
which time and again led us out of darkness. The enigma of HTSC
poses an unprecedented challenge in condensed matter physics.
Reflecting upon the historical developments of physical theories,
it seems worthwhile to carry out the symmetry approach to the HTSC
problem to its full logical conclusion. The basic idea of unifying
seemingly different phases by a common symmetry principle may also
prove to be useful for other strongly correlated systems.

\section{NOTATIONS AND CONVENTIONS}
\subsection{Index convention}
$\tau^\alpha$ denote Pauli matrices.

$\alpha, \beta=x,y,z$ denote $SO(3)$ vector spin indices.

$\sigma, \sigma'=1,2$ denote $SO(3)$ spinor indices.

$a, b, c=1,2,3,4,5$ denote $SO(5)$ superspin vector indices.

$\mu,\nu=1,2,3,4$ denote $SO(5)$ spinor indices.

$i,j=1,5$ denote $U(1)$ vector indices for superconductivity.

$x, x'$ denote site indices.

\subsection{Dirac $\Gamma$ matrices}
The general method introduced by Rabello et.~al\cite{RABELLO1998}
to construct $SO(5)$ symmetric models uses the five Dirac $\Gamma$
matrices $\Gamma_a$ $(a=1,..,5)$ which satisfy the Clifford
algebra,
\begin{eqnarray}
\{\Gamma^a, \Gamma^b \} = 2 \delta^{ab}.
\end{eqnarray}
Rabello et.~al introduced the following explicit representation
which is naturally adapted for discussing the unification of AF
and $d$-wave SC order parameters,
\begin{eqnarray}
 \Gamma^1\!=\! \left( \begin{array}{cc}
               0          & -i\tau_y  \\
               i\tau_y  & 0            \end{array} \right)
 \ \ \
 \Gamma^{(2,3,4)} \!=\! \left( \begin{array}{cc}
             \vec \tau  & 0  \\
                  0       &  ^t\vec \tau  \end{array} \right)
\ \ \
 \Gamma^5 \!=\! \left( \begin{array}{cc}
                0         & \tau_y  \\
                \tau_y  & 0            \end{array} \right).
\end{eqnarray}
Here $\vec \tau=(\tau_x,\tau_y,\tau_z)$ are the usual
Pauli matrices and $^t\vec \tau$ denotes their transposition.
These five $\Gamma_a$ matrices form the $5$ dimensional vector
irreps of $SO(5)$. Their commutators,
\begin{eqnarray}
\Gamma^{ab} = - \frac{i}{2} \left[ \Gamma^a, \Gamma^b \right],
 \end{eqnarray}
define the $10$ dimensional antisymmetric tensor irreps of
$SO(5)$. In the above representation, the $10$ $\Gamma^{ab}$'s are given
explicitly by
\begin{eqnarray}
 \Gamma^{15} &=&  \left( \begin{array}{cc}
                      -1  & 0  \\
                       0  & 1           \end{array} \right)
\nonumber \\
 \Gamma^{(i+1)(j+1)} &=&  \varepsilon_{ijk}\left( \begin{array}{cc}
                \tau_k  & 0  \\
                     0    &  -{}^t\tau_k  \end{array} \right)
 {} \hskip 7mm (i,j=1,2,3)
\nonumber \\
 \Gamma^{(2,3,4) 1} &=&  \left( \begin{array}{cc}
                0         & - \vec \tau \tau_y  \\
            - \tau_y \vec \tau & 0    \end{array} \right)
                    =  \tau_y \left( \begin{array}{cc}
                0         & {}^t\vec \tau  \\
            - \vec \tau & 0    \end{array} \right)
\nonumber \\
 \Gamma^{(2,3,4) 5} &=&  \left( \begin{array}{cc}
                0         & -i \vec \tau \tau_y  \\
             i\tau_y \vec \tau & 0    \end{array} \right)
                    =  i\tau_y \left( \begin{array}{cc}
                0         & {}^t \vec \tau   \\
              \vec \tau &  0    \end{array} \right).
\nonumber
\end{eqnarray}
These $\Gamma$ matrices satisfy the following commutation relations:
\begin{eqnarray}
 \left[ \Gamma^{ab}, \Gamma^c \right] &=& 2i(\delta_{ac}\Gamma^b -
                                  \delta_{bc}\Gamma^a)    \\
 \left[ \Gamma^{ab}, \Gamma^{cd} \right] &=&
          2i(\delta_{ac}\Gamma^{bd} + \delta_{bd}\Gamma^{ac}
           - \delta_{ad}\Gamma^{bc} - \delta_{bc}\Gamma^{ad}).
\end{eqnarray}
An important property of the $SO(5)$ Lie algebra is the
pseudo-reality of its spinor representation. This means that
there exists a matrix $R$ with the following properties:
\begin{eqnarray}
  R^2 = -1, \ \ \ R^{\dagger} =  R^{-1} = {}^tR = -R,   \\
  R\,\Gamma^aR = -{}^t\Gamma^a, \ \ \ R\,\Gamma^{ab}R = {}^t\Gamma^{ab}.
\end{eqnarray}
The relations
$ R\,\Gamma^{ab}R^{-1} = -(\Gamma^{ab})^*$ indicate that the spinor
representation is real, and the antisymmetric nature of the matrix $R$
indicates that it is pseudo-real. The $R$ matrix plays a role similar
to that of $\epsilon_{\alpha\beta}$ in $SO(3)$. In our representation,
the $R$ matrix takes the form
\begin{eqnarray}
R = \left( \begin{array}{cc}
                     0  &  1  \\
                    -1  &  0  \end{array} \right).
\end{eqnarray}

\section*{ACKNOWLEDGMENTS}

We would first like to thank E. Arrigoni, H.D. Chen, C. Dahnken,
J. Schafer, and C.J. Wu for their kind help with the manuscript
preparation.  We have benefited from long-term collaborations,
discussions, and exchange of ideas and results with a large number of
colleagues.  We particularly mention G. Aeppli, E. Altman,
P. Anderson, D. Arovas, G. Arnold, E.  Arrigoni, A. Auerbach,
S. Balatsky, G. Baskaran, Y. Bazaliy
% ...
G. Zimanyi.

ED is supported by the NSF grant DMR-0132874 and by the Sloan
foundation. WH is supported by the DFG under Grant No.~Ha
1537/16-2. SCZ is supported by the NSF under grant numbers
DMR-9814289 and the US Department of Energy, Office of Basic
Energy Sciences under contract DE-AC03-76SF00515.

\bibliographystyle{apsrmp}
\bibliography{rmp_V3,extra_V3}



\begin{figure}[hp]
  \hspace{2mm}
  \includegraphics[scale=1]{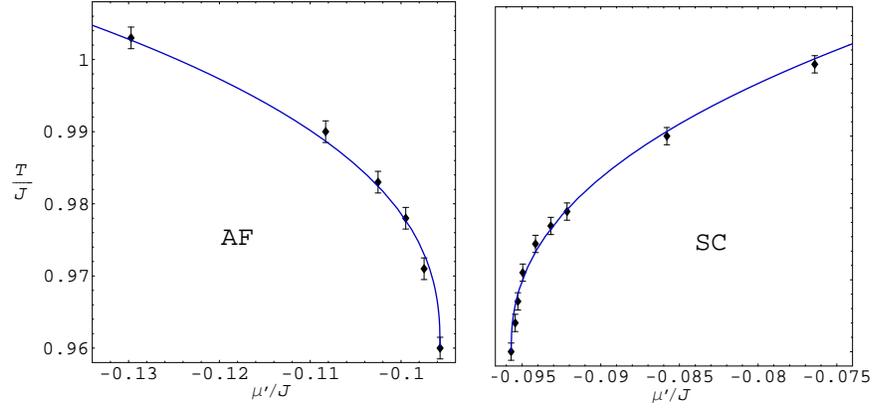} \vspace{0.8mm}
  \narrowtext
\caption{Scaling of $T_N$ and $T_c$ near the $SO(5)$ bi-critical
point. Both $T_N$ and $T_c$ merge into the bi-critical point
tangentially, with the crossover exponent of $\phi = 1.43 \pm
0.05$.} \label{Fig_Tn_Tc}
\end{figure}


\begin{figure}[hp]
\begin{tabular}{cc}
    \includegraphics[scale=0.3]{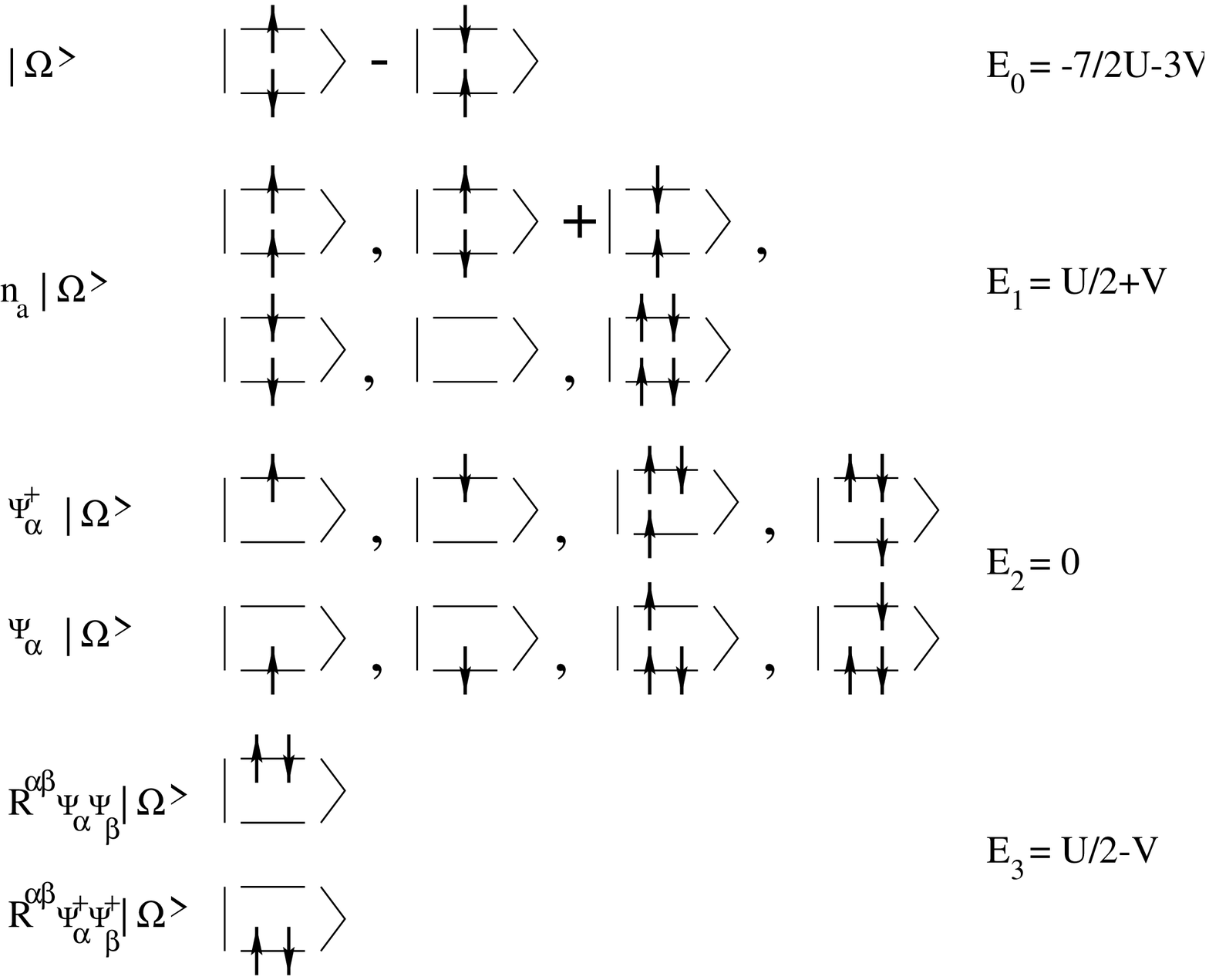}&
    \includegraphics[scale=0.4]{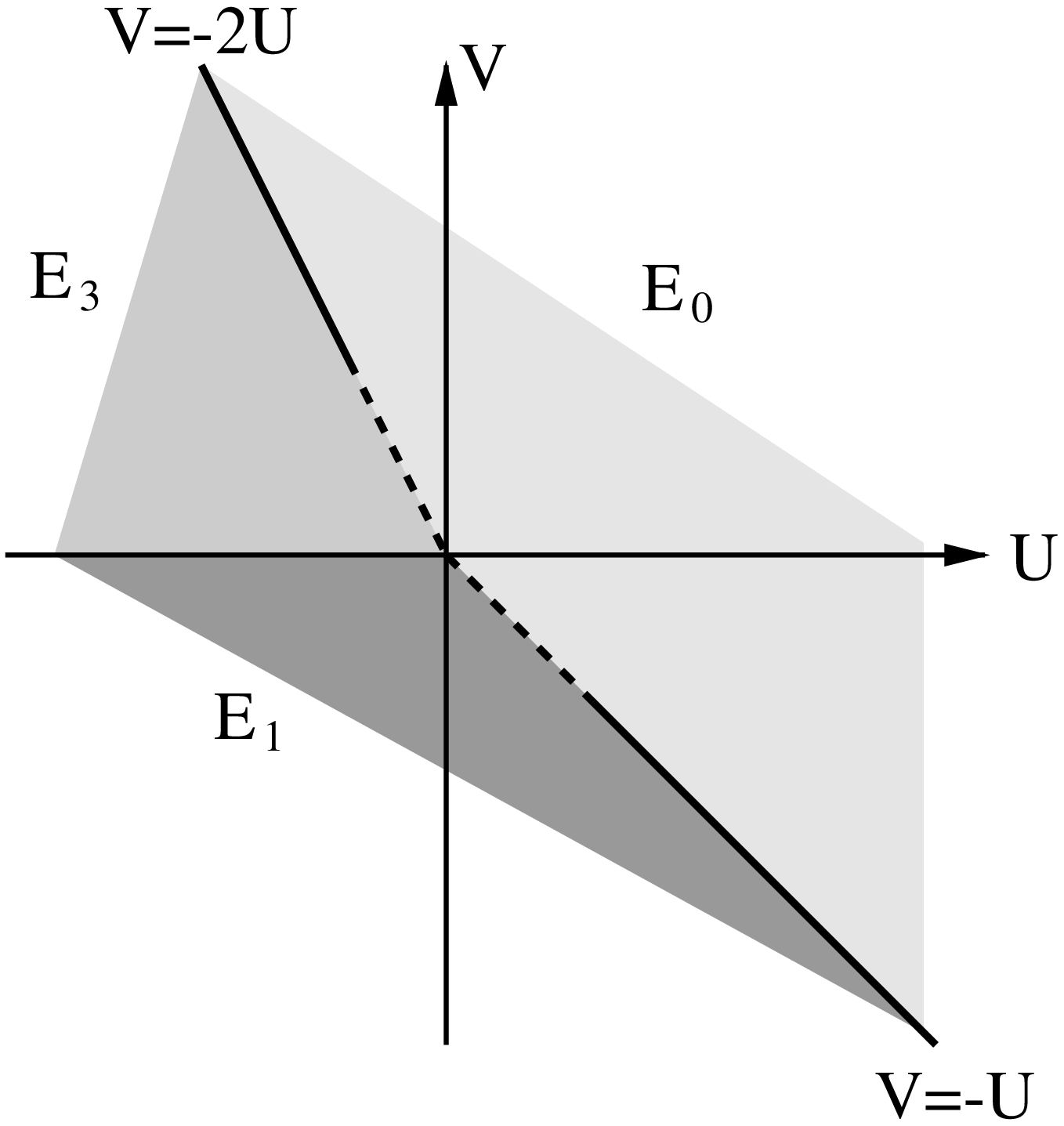}\\
(a)&(b)
\end{tabular}
\caption{(a) Under the condition specified by Eq.
(\ref{so5_condition}), the 16 states on a rung are classified into 6
groups, each transforming irreducibly under the $SO(5)$ group.
$|\Omega\rangle$ is an $SO(5)$ singlet state; $n_a |\Omega\rangle$
desribes five states that transform as $SO(5)$ vectors;
$\Psi_\alpha^\dagger |\Omega\rangle$ are four states that form an
$SO(5)$ spinor; four states $\Psi_\alpha |\Omega\rangle$ also
correspond to a spinor; $R_{\alpha\beta}\Psi_\alpha \Psi_\beta
|\Omega\rangle$ and $R_{\alpha\beta}\Psi^\dagger_\alpha
\Psi^\dagger_\beta |\Omega\rangle$ are two $SO(5)$ singlet states.
The figure also gives energies of all multiplets for the $SO(5)$
symmetric ladder model described by equations (\ref{rung-H}) and
(\ref{so5_condition}). (b) Strong coupling phase diagram of the
$SO(5)$ symmetric ladder model in the $(U,V)$ space. The $E_0$, $E_1$
and $E_3$ phases are regions in parameter space where the respective
states have the lowest energy.} \label{Fig_ladder}
\end{figure}


\begin{figure}[hp]
\begin{center}
        \includegraphics[scale=0.6]{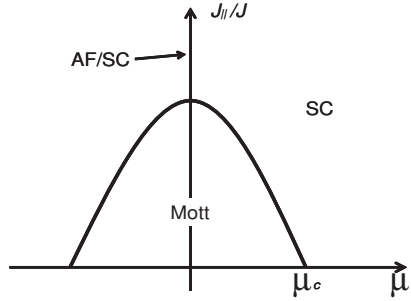}
 \caption{Phase diagram of the bi-layer $SO(5)$ model plotted as
        $J_\parallel/J$ versus $\mu$. The entire phase transition line
        from the Mott phase into any of the ordered phases is a second
        order quantum phase transition. The Mott insulating state has
        5 massive collective modes. The $SO(5)$ symmetric AF/SC
        uniform mixed state at half-filling has 4 gapless collective
        modes. The SC state has a spin triplet $\pi$ resonance mode
        and one massless charge Goldstone mode.}
\label{Fig_bilayer}
\end{center}
\end{figure}


\begin{figure}[hp]
\begin{center}        
\includegraphics[scale=1.0]{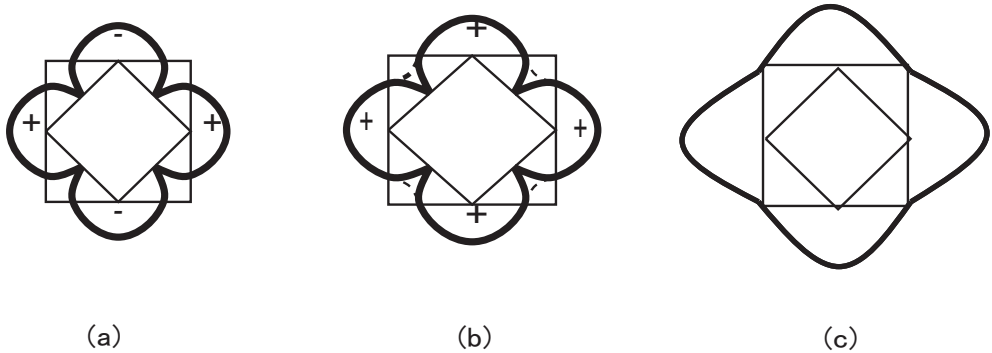}
\caption{Evolution of the quasi-particle states when doping is
reduced. (a) pure $d$ wave SC gap with nodal quasi-particles. (b)
the pure $d$ wave SC gap is rotated into an AF gap of the form
$|\cos p_x - \cos p_y|$. (c) A large uniform component of the
AF/Mott insulating gap is developed on top of the $|\cos p_x -
\cos p_y|$ gap when doping is reduced close to zero. }
\label{Fig_d_wave}
\end{center}
\end{figure}

\begin{figure}[hp]
\includegraphics[scale=0.6]{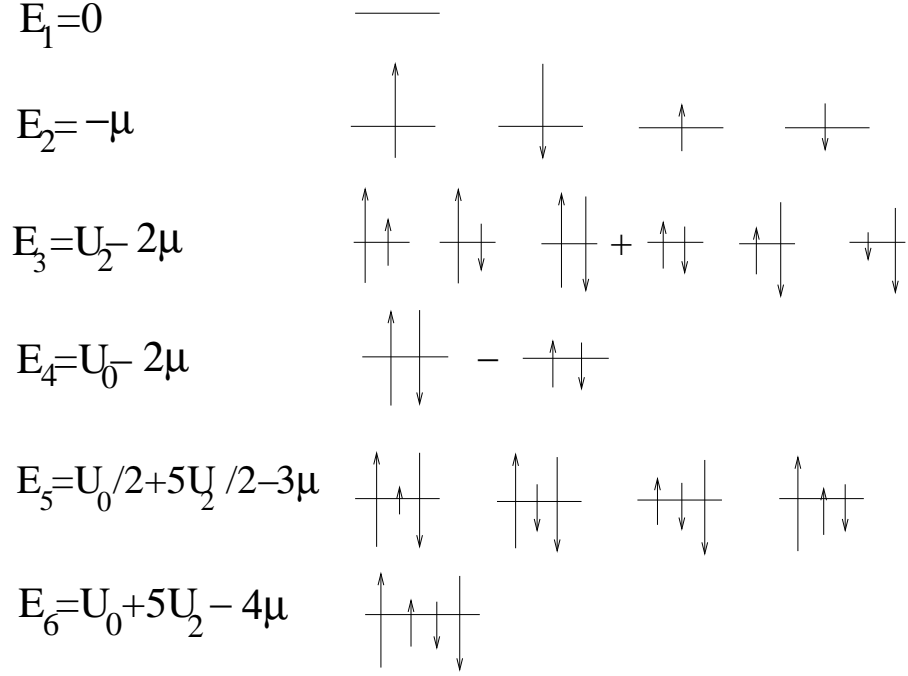}
\caption{Eigenstates the spin 3/2 problem on a single site. The
longer (shorter) arrows denote |Sz| = 3/2 (1/2) and the up (down)
direction denote the '+' (-) sign. The E1,4,6 (singlet), E2,5
(quartet), and E3 (quintet) sets can also be classified as $SO(5)$
singlet, spinor, and vector representations.} 
\label{Fig_spin32}
\end{figure}


\begin{figure}[hp]
\begin{center}
       \includegraphics[scale=0.4]{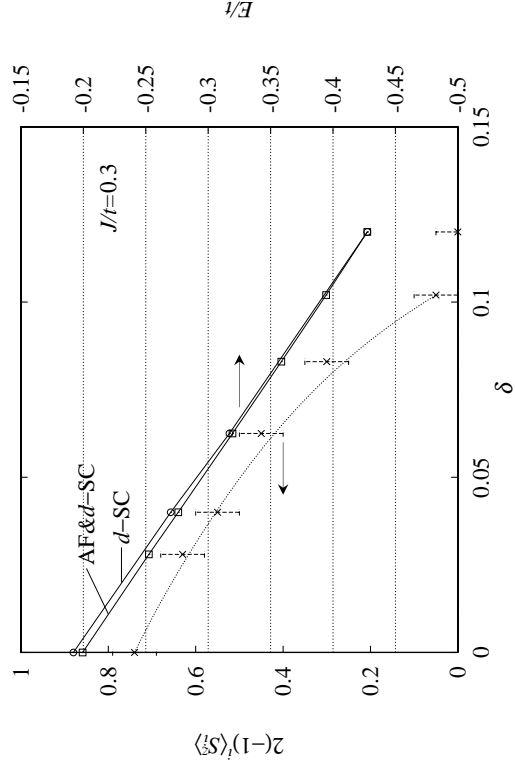}
       \caption{Doping dependence of the ground-state energy 
(two upper curves) and
              staggered magnetization 
(lower curve) for the $t-J$ model with $J/t=0.3$.
             The state with {\it uniform} 
AF and $d$-wave SC order has lower energy compared with
             the pure $d$-wave SC state for $0<\delta<10\%$, furthermore, 
              the energy of the uniform
              AF/SC mixed state depends {\it linearly} 
on $\delta$, fitting into the $SO(5)$
              symmetric ``type 1.5" transition classified in 
Fig. (\ref{Fig_E_delta}).
              Reproduced from Ref. \cite{HIMEDA1999}.}
\label{figHimeda2}
\end{center}
\end{figure}

\begin{figure}[hp]
\begin{center}
\includegraphics[scale=1.0]{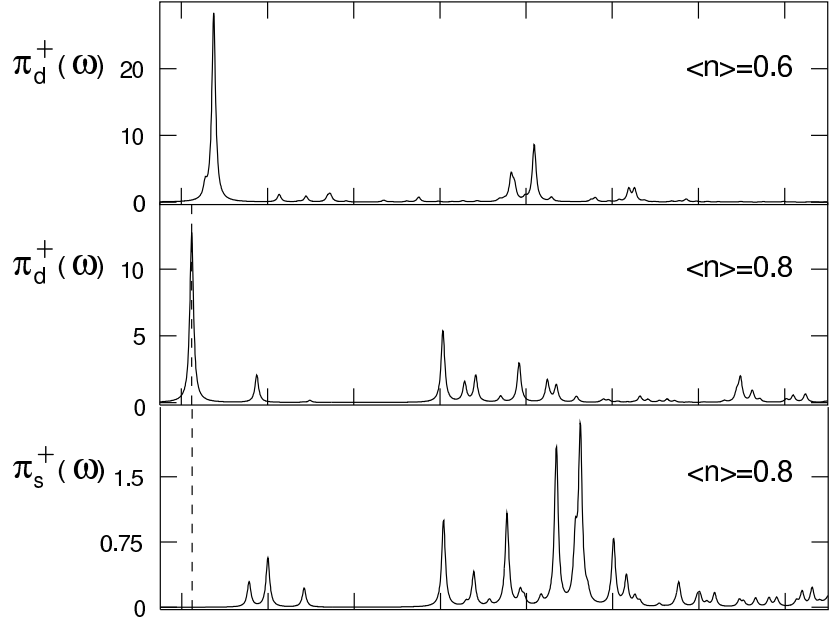}
\caption{ Exact diagonalization results for the dynamical
correlation function of the $\pi$-operator on a 10-site Hubbard
system with $U=8 t$ reproduced from reference \cite{MEIXNER1997}.
A single $\delta$-function-like peak with pronounced weight is
visible near $\omega = 0$ for the $\pi$-operator, proving the
eigenoperator relation (\ref{approximate_eigen}) in the low-energy
regime. This ``precession frequency" $\omega_\pi$ decreases with
decreasing doping. An alternatively constructed ``$s$-wave
$\pi$-operator", with $g(p)$ in Eq. (\ref{pi_k}) given by
$g(p)=cos p_x+cos p_y$, shown in the bottom graph exhibits only
incoherent behavior and hardly any weight (note the difference in
the $y$-scale). Here $\langle n \rangle$ denotes average electron
density, with $\langle n \rangle=1$ being at half-filling.}
\label{figMZ322meixneru8}
\end{center}
\end{figure}


\begin{figure}[hp]
\begin{center}
\includegraphics[scale=0.6]{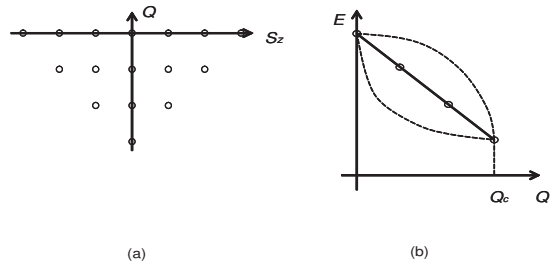}
\caption{After expanding out the coherent state (\ref{coherent}),
we obtain the magnon and hole pair states at level $\nu$, which is
the total number of magnons and hole pairs. These states are
classified by their $(S_z,Q)$ quantum numbers in (a). The energy
is independent of the $S_z$ quantum number because of the $SO(3)$
spin rotation symmetry. The energy can depend on $Q$ with three
generic possibilities, as depicted in (b). (Compare with Fig.
\ref{Fig_E_delta}). If the energy depends linearly on $Q$, there
is no free energy cost to rotate magnons and hole pairs into each
other, and the potential energy is $SO(5)$ symmetric. This
multiplet structure was tested in the $t-J$ model and shown in
Fig. \ref{figFKPirreps}.} \label{Fig_multiplet}
\end{center}
\end{figure}


\begin{figure}[hp]
\vspace{-4mm}
\includegraphics[scale=0.8]{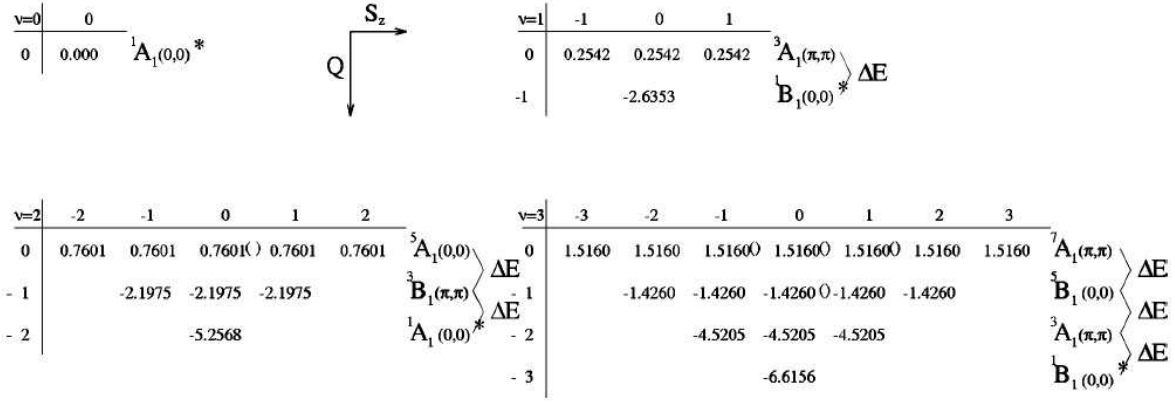}
\narrowtext \vspace{1mm} \caption{The low energy states within
each total spin and charge sector $(S_z,Q)$ of the $18$-site
cluster $t-J$ model with $J/t=0.5$. The states are grouped into
different multiplets and are labelled by the spin, charge, point
group symmetry, and total momentum. $A_1$ denotes the totally
symmetric, $B_1$ the $d_{x^2-y^2}$-like representation of the
$C_{4v}$ symmetry group. The quantum numbers of these states match
that of the magnon and hole pair states shown in Fig.
\ref{Fig_multiplet}. Furthermore, the energy depends approximately
linearly on $Q$, demonstrating the $SO(5)$ symmetry of the
interaction potential among the magnons and hole pairs.}
\label{figFKPirreps}
\end{figure}


\begin{figure}[hp]
\begin{center}
        \includegraphics[scale=0.5]{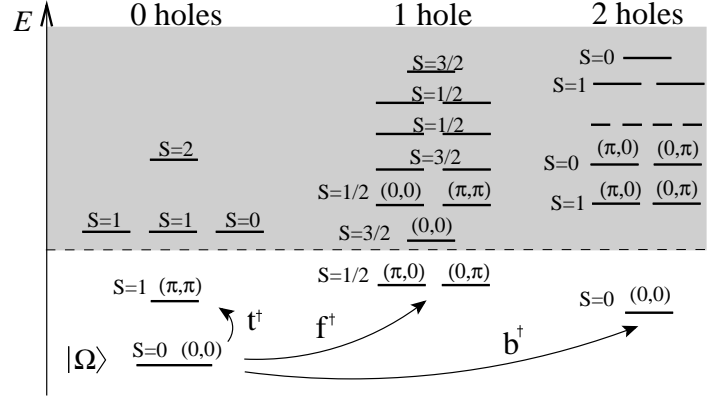}
          \caption{
            Low energy spectrum of the Hubbard model on a plaquette.
Eigenstates by total spin $S$ and plaquette momentum
$q_x,q_y=0,\pi$. Truncated high energy states are shaded. The
vacuum is defined as $\ket\W$, and quantized operators connect the
vacuum to the lowest eigenstates as shown. (In this figure,
$t^\dagger$ denotes the magnon creation operator
$t_\alpha^\dagger$, and $b^\dagger$ denotes the hole pair creation
operator $t_h^\dagger$.) \label{fig5espectrum}}
\end{center}
\end{figure}

\begin{figure}[hp]
  \includegraphics{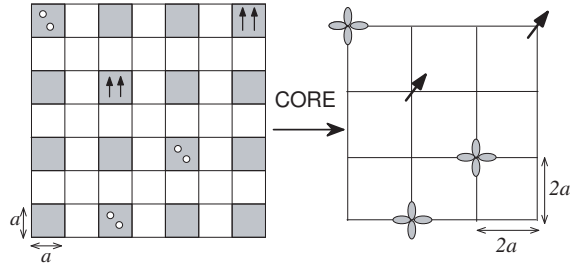}
  \caption{Illustration of the basic idea of the CORE method.
  To implement the CORE method, first decompose the original lattice
  in plaquettes, and then truncate the spectrum of a given plaquette to five lowest states,
  {\it i.e}, singlet, hole-pair and three magnon states. An effective Hamiltonian for
  these bosons can then be calculated using the CORE method.
  Left: local bosons in the original lattice. Gray rectangle denotes
  the singlet RVB vacua, circles denote holes and the set of two
  parallel vertical arrows denote the magnon.
  Right: local bosons on the lattice of plaquette. Leaf-like
  pattern denotes a local $d$-wave hole-pair on a plaquette.
  Canted arrow denotes local magnon on a plaquette.
  The singlet RVB vacuum is denoted by an empty site.}\label{figALplaqlatt}
\end{figure}

\begin{figure}[hp]
\begin{center}
        \includegraphics[scale=0.4]{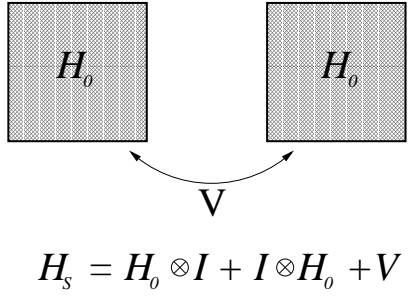}
\caption{ This figure illustrates the construction of the
``superblock'' and its Hamiltonian $H_s$ out of two neighboring
blocks, with intrablock Hamiltonian $H_0$ and interblock coupling
V (in the block basis: $\left( H_0 \right) _{n,n^\prime} =\langle
\alpha_n^\circ \mid H\mid \alpha_n^\circ \rangle =
\epsilon_n^\circ \delta_{n,n^\prime}$ and $\left( V_0 \right)
_{nm,n^\prime m^\prime} =\langle \alpha_n^\circ \alpha_m^\circ
\mid V\mid \alpha_{n^\prime}^\circ \alpha_{m^\prime}^\circ
\rangle$). } \label{fig5eMJ}
\end{center}
\end{figure}

\begin{figure}[hp]
\begin{center}
        \includegraphics[scale=0.4]{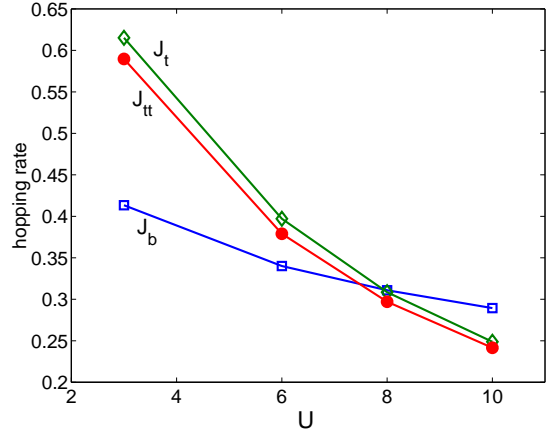}
\caption{ Boson hopping energies versus Hubbard $U$. The
intersection region near $U=8$ is close to the projected $SO(5)$
symmetry point. All energies are in units of $t$. }
\label{fig5epars}
\end{center}
\end{figure}

\begin{figure}[hp]
\epsfxsize=2.25in
\centerline{\epsffile{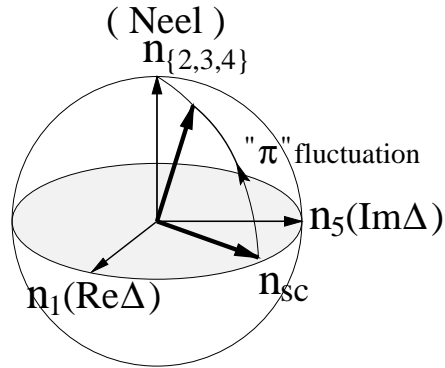}}\caption{ The order
parameter space of the $SO(5)$ theory.  $\pi$ operator performs
a rotation between the AF and the $d$-wave SC states. This small
fluctuation is the new Goldstone mode of the $SO(5)$ theory. }
\label{sphere}
\end{figure}

\begin{figure}[hp]
\epsfxsize=3in
\centerline{\epsffile{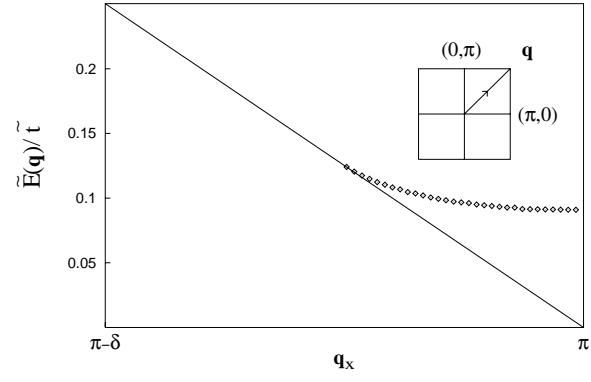}}\caption{ The two
particle continuum and the $\pi$ excitation for the tight-biding
model. Note, that the continuum of two particle states collapses
to a point when the center of mass momentum is $\Pi=(\pi,\pi)$.
The $\pi$-mode emerges as an anti-bound state above the continuum.
} \label{continuum}
\end{figure}
\begin{figure}[hp]
\epsfysize=1in \centerline{\epsffile{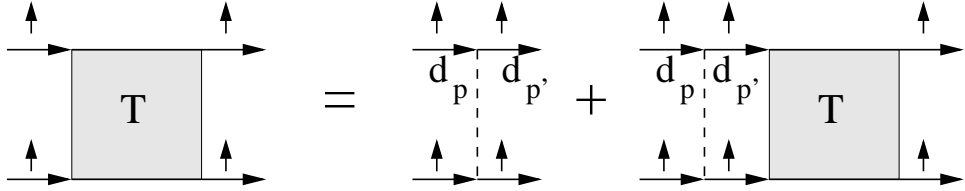}}\caption{
Dyson's equation for the $\pi$-resonance. Function $d(p)$ is
defined in equation (\ref{AFdSC}). } \label{dyson}
\end{figure}
\begin{figure}
\epsfysize=1.5in \centerline{\epsffile{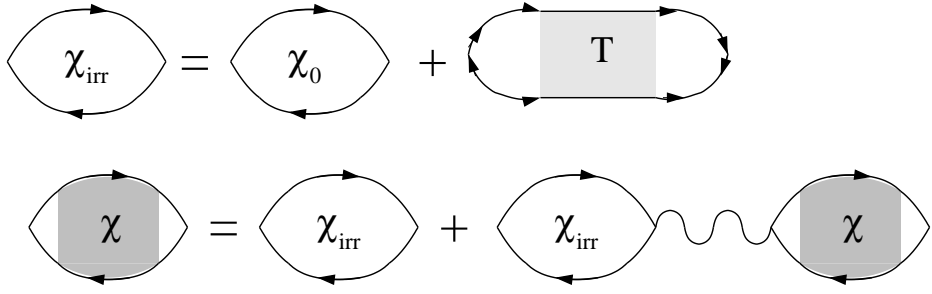}}\caption{
$\pi$ resonance contribution to the spin susceptibility in the SC
state.} \label{chi}
\end{figure}
\begin{figure}[hp]
\epsfxsize=3in \centerline{\epsffile{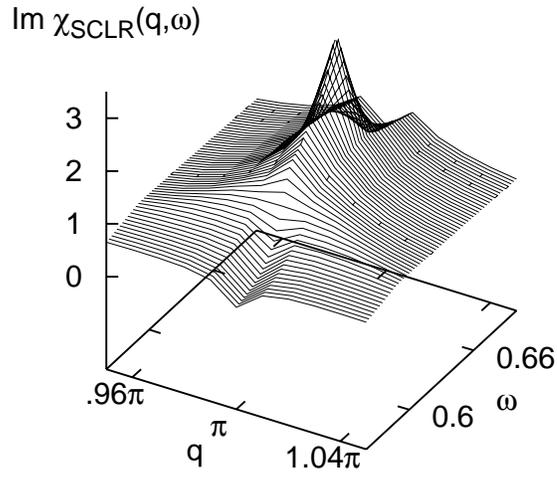}}\caption{
Spin susceptibility in the SC state for the model
(\ref{HubbardtJHamiltonian}) \cite{DEMLER1998A}. The wavevector is
along the 0 to $(\pi,\pi)$ direction. Susceptibility was computed
using the self-consistent linear response formalism in Fig.
\ref{chi}. The peak at $(\pi,\pi)$ comes from the $\pi$-resonance.
} \label{qplot}
\end{figure}

\begin{figure}[hp]
\epsfysize=1.5in
\centerline{\epsffile{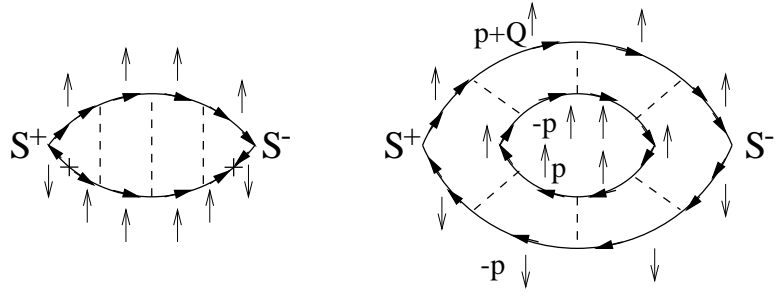}}\caption{ Feynmann
diagram for the $\pi$-resonance below $T_c$ contrasted with the
diagram above $T_c$. The cross denotes the anomalous scattering in
the SC state which converts a particle into a hole, and vice
versa.} \label{PiResonancePrecursor}
\end{figure}

\begin{figure}[hp]
\epsfysize=1.5in
\centerline{\epsffile{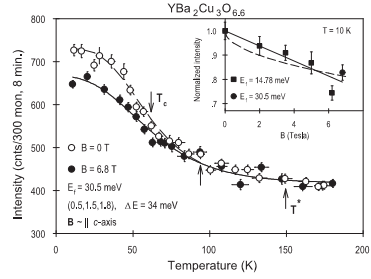}}\caption{
Suppression of the resonance intensity by the magnetic field.
Reproduced from Ref. \cite{DAI2000}. } \label{FigDai2000}
\end{figure}


\begin{figure}
\epsfysize=3in
\centerline{\epsffile{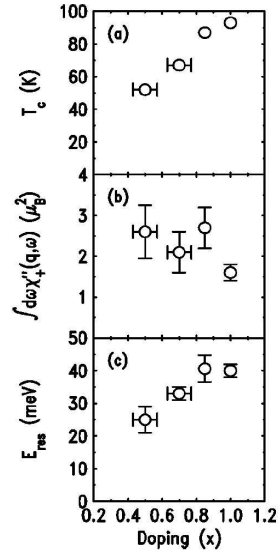}}\caption{ Doping
dependence of the resonance energy and intensity measured in
neutron scattering experiments. Reproduced from Ref.
\cite{FONG2000}. } \label{FigFong2000}
\end{figure}

\begin{figure}
\epsfysize=3in
\centerline{\epsffile{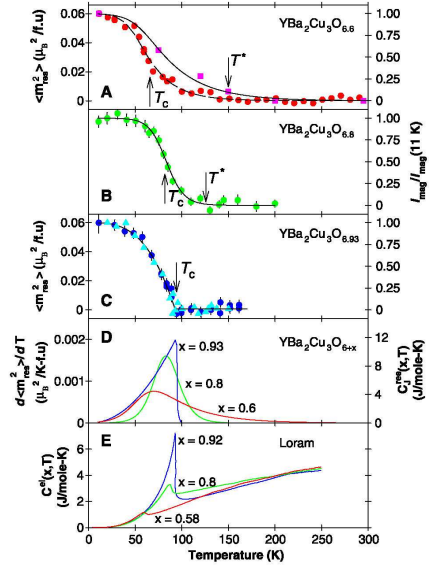}}\caption{
Temperature dependence of the resonance intensity compared to the
specific heat. Reproduced from Ref. \cite{DAI1999}. }
\label{FigDai1999}
\end{figure}

\begin{figure}
\epsfxsize=1.5in
\centerline{\epsffile{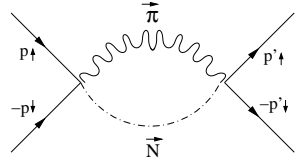}}\caption{ SC pairing
between electrons mediated by exciting a virtual magnon-$\pi$-mode
pair.} \label{c_fig1}
\end{figure}
\begin{figure}
\epsfysize=1.5in \centerline{\epsffile{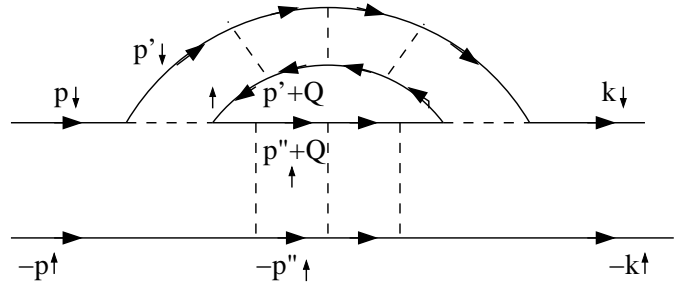}}
\caption{ Diagrammatic representation of the SC pairing mediated
by exciting a virtual magnon-$\pi$-mode pair. Solid lines describe
electron propogators and dashed lines describe interactions. The
upper particle-hole ladder corresponds to the magnon and the lower
particle-particle ladder corresponds to the $\pi$-mode. }
\label{c_fig2}
\end{figure}

\begin{figure}[h]
\begin{center}
      \includegraphics[scale=0.4]{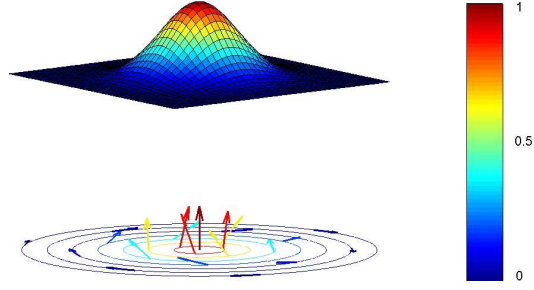}
\caption{ SC vortex with AF core. Far from the center of
the vortex core, the superspin vector lies in the SC plane and
winds around the vortex core by $2\pi$. The superspin vector lifts
up to the AF direction as it approaches the center of the vortex
core. The arrows represent the direction of the superspin and the
color scale represents the magnitude of the AF order parameter.}
\label{Fig_vortex}
\end{center}
\end{figure}

\begin{figure}[h]
\begin{tabular}{ccc}
    \includegraphics{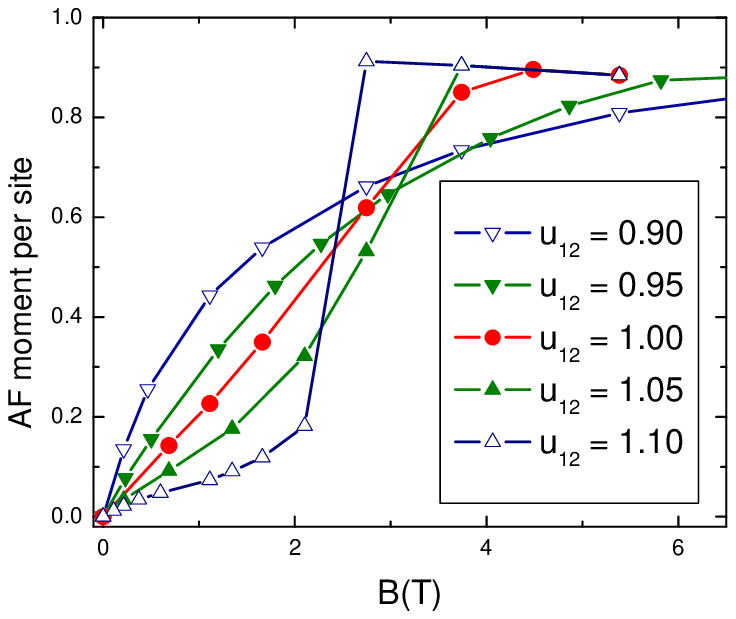}&
    \includegraphics{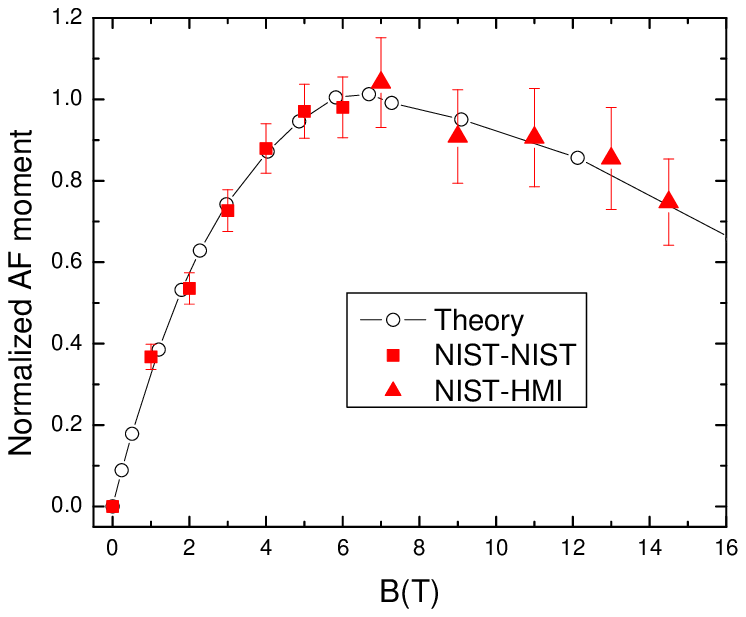}\\
(a)&(b)
\end{tabular}
\caption{The plot of field dependence of the AF moment for different
parameters of the LG theory (defined in Ref. \cite{Chen2003B}).
The parameters are $\rho_1=\rho_2= a^2$, $r_1=-1$, $r_2=-0.85$,
$u_1=u_2=1$ and $\chi=42.4$. Here the parameters are chosen such
that the maximum SC order is $1$ and the SC coherence length at
zero field equals the lattice constant $a$ of the lattice model.
(a) Field dependence for different values of $u_{12}$. The
curvature strongly depends on $u_{12}$. (b) Fit to the neutron
scattering results\cite{Kang2003} of the $Nd_{1.85}Ce_{0.15}CuO_4$
crystal with $u_{12}=0.95$. $B_{c2}$ is about $6.2T$ in this
sample. The experimental data is obtained by subtracting the
magnetic field response along the $c$ axis by the magnetic field
response in the $ab$ plane, so that the response from the $Nd$
moment can be removed.} \label{Fig_AF_moment}
\end{figure}

\begin{figure}[h]
\begin{center}
      \includegraphics[scale=0.25]{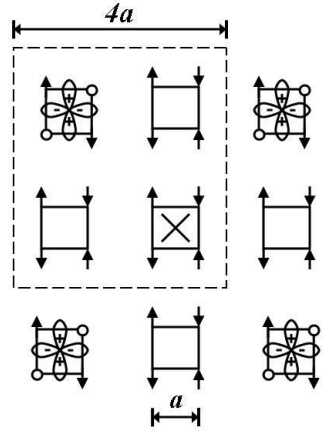}
\caption{Illustration of the $d$ wave pair density wave state at
$x=1/8$. In this state, the $d$ wave hole pairs occupy every four
non-overlapping plaquettes on the original lattice. The charge
unit cell is $4a\times 4a$. The $SO(5)$ model is defined on the
center of the non-overlapping plaquettes. Such a state could be
realized around the vortex core, whose center is depicted by the
cross. In the actual realization of this state, the hole pair can
be much more extended, and the AF ordering could be much reduced
from the classical value.} \label{PDW}
\end{center}
\end{figure}



\begin{figure}[htbp]
    \includegraphics[width=0.7 \textwidth]{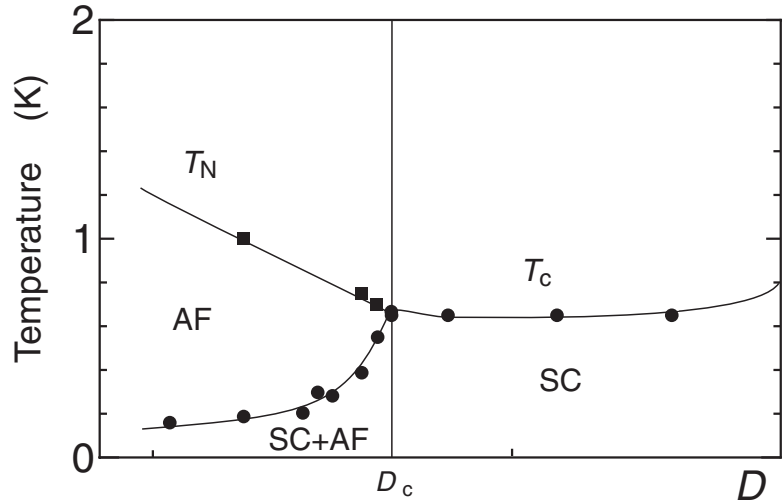}
\caption{
 The  combined phase diagram as a function of  lattice
 density
$D$ in
 $CeCu_2(Si_{1-x}Ge_x)_2$
 ($D < D_c$ ) and in
$CeCu_2Si_2$
 ($D_c \leq  D$) under
pressure $P$. Note that $D \propto  1/V$, where $V$ is the
unit-cell volume, and $D = D_{Si}[1 (V_{Ge}- V_{Si})x/V_{Ge}]$ in
the former case. Reproduced from Ref. \protect\cite{KITAOKA2002}.}
\label{kitaoka}
\end{figure}

\begin{figure}[htbp]
    \includegraphics[scale=0.35]{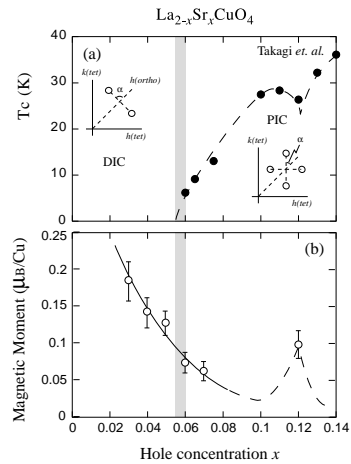}
\caption{Doping dependence of the SC transition temperature and
magnetic moment for $La_{2-x}Sr_xCuO_4$. Reproduced from Ref.
\cite{WAKIMOTO2001}}. \label{LSCOPhaseDiagram}
\end{figure}

\begin{figure}[h]
\begin{tabular}{ccc}
    \includegraphics[scale=0.4]{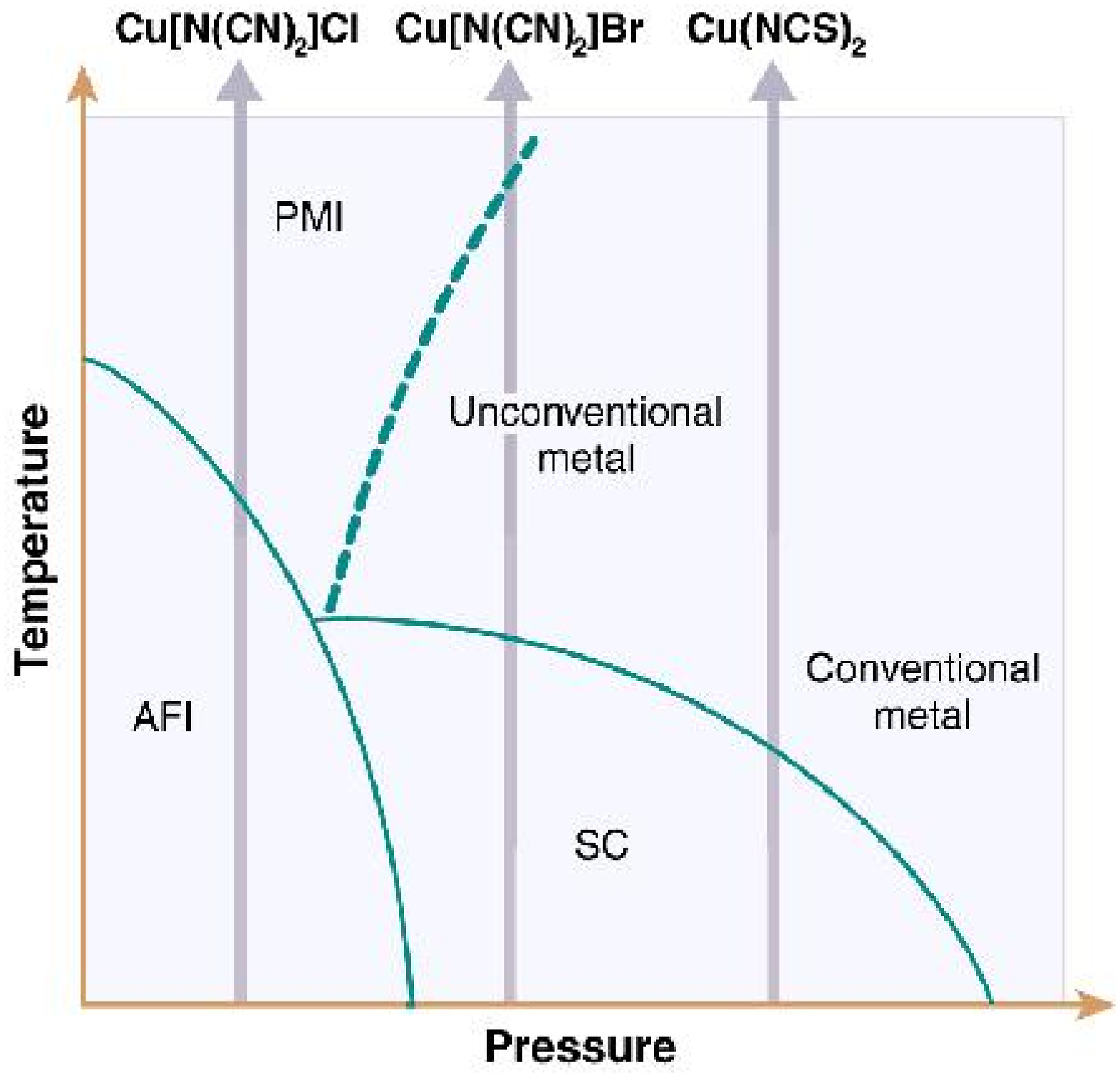}&
    \includegraphics[scale=0.5]{demler_fig49_2.eps}\\
(a)&(b)
\end{tabular}
\caption{(a) Phase diagram the two dimensional organic
superconductor BEDT salt. Reproduced from Ref.
\cite{MCKENZIE1997}. (b) Log-log plot of $T_1^{-1}$ vs $(T-T_{{\rm
c}})/T$ for (A) $\kappa$-(BEDT-TTF)$_{2}$Cu[N(CN)$_{2}$]Cl (solid
squares), and (B) deuterated
$\kappa$-(BEDT-TTF)$_{2}$Cu[N(CN)$_{2}$]Br (open squares). (Data
from \protect\cite{KAWAMOTO1995}.)} \label{exp}
\end{figure}

\begin{figure}[t]
\centerline{\epsfysize=4cm \epsfbox{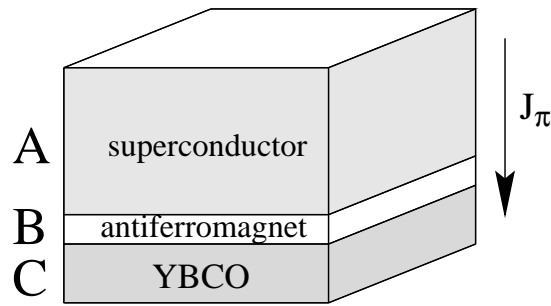} }
\caption{Setting of the tunnelling experiment for detecting the
triplet particle-particle $\pi$ mode in the normal state.}
\label{pifig1}
\end{figure}

\begin{figure}[h]
\centerline{\epsfysize=8cm \epsfbox{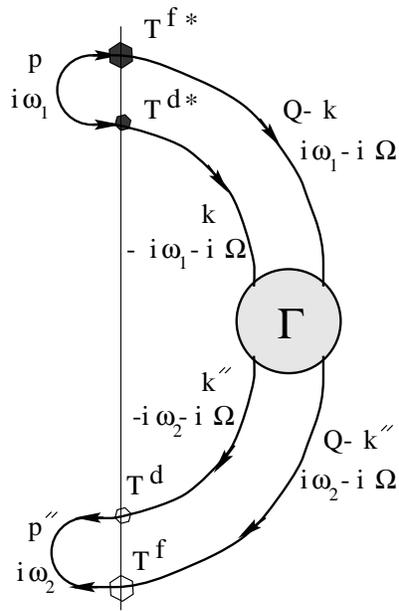} }
\caption{Second order tunnelling diagram that gives rise to the
resonant coupling of Cooper pairs and $\pi$ excitations in the
junction shown in Fig. \ref{pifig1}.} \label{pifig2}
\end{figure}
\begin{figure}[h]
\begin{center}
      \includegraphics[scale=0.8]{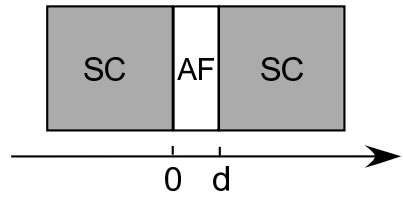}
\caption{The Superconductor-Antiferromagnet-Superconductor
(SC/AF/SC) junction} \label{fr_geometry}
\end{center}
\end{figure}
\begin{figure*}[h]
\centerline{\epsfysize=8cm \epsfbox{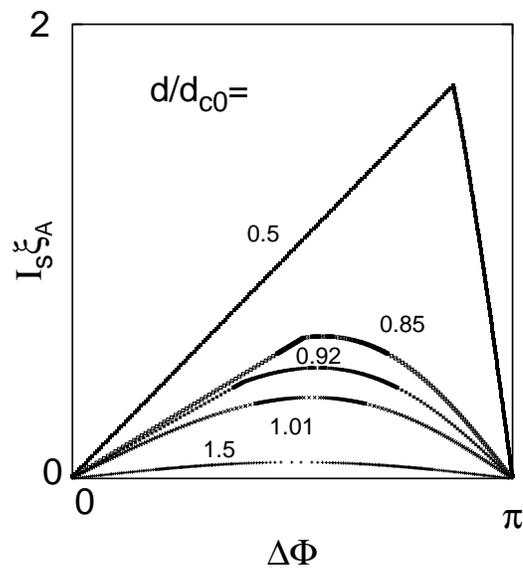}}
\caption{Predicted current-phase characteristics of a SC/AF/SC
junction with different $d/d_{c0}$.  } \label{I-fig}
\end{figure*}



\end{document}

%% file: macros.tex
\usepackage{graphicx}
\usepackage[floatfix]{epsfig}
\usepackage{dcolumn}
\usepackage{bm}

\def\inbar{\,\vrule height1.5ex width.4pt depth0pt}
\def\IR{\relax{\rm I\kern-.18em R}}
\def\IC{\relax\hbox{$\inbar\kern-.3em{\rm C}$}}

\def \be{\begin{equation}}
\def \ee{\end{equation}}
\def \bea{\begin{eqnarray}}
\def \eea{\end{eqnarray}}

\def\W{{\Omega}}

\newcommand{\ket}[1]{ |\!\; {#1} \!\;\rangle}

\def\beq{\begin{equation}}
\def\eeq{\end{equation}}
\def\eqref#1{(\ref{#1}) }

\newcommand{\helic} {\Upsilon}
\def\mathit#1{#1}

\def\bequ{\begin{equation}}
\def\eequ{\end{equation}}
\def\beqn{\begin{eqnarray}}
\def\eeqn{\end{eqnarray}}
\def\eqref#1{Eq.~(\ref{#1}) }

\def\be{\beta}
